\documentclass[aps,pra,twocolumn,showpacs,floatfix]{revtex4-1}

\usepackage{epsfig,amsmath,amssymb}
\usepackage{graphics} 
\usepackage{subfigure}
\usepackage{color}
\usepackage{multirow}
\usepackage{mathrsfs}
\usepackage{natbib}

\begin{document}

\title
{
Ground-state phases of the spin-1 $J_{1}$--$J_{2}$ Heisenberg antiferromagnet on the honeycomb lattice
}

\author
{P. H. Y. Li}
\email{peggyhyli@gmail.com}
\author
{R. F. Bishop}
\email{raymond.bishop@manchester.ac.uk}
\affiliation
{School of Physics and Astronomy, Schuster Building, The University of Manchester, Manchester, M13 9PL, UK}

\begin{abstract}
  We study the zero-temperature quantum phase diagram of a spin-1
  Heisenberg antiferromagnet on the honeycomb lattice with both
  nearest-neighbor exchange coupling $J_{1}>0$ and frustrating
  next-nearest-neighbor coupling $J_{2} \equiv \kappa J_{1} > 0$,
  using the coupled cluster method implemented to high orders of
  approximation, and based on model states with different forms of
  classical magnetic order.  For each we calculate directly in the
  bulk thermodynamic limit both ground-state low-energy parameters
  (including the energy per spin, magnetic order parameter, spin
  stiffness coefficient, and zero-field uniform transverse magnetic
  susceptibility) and their generalized susceptibilities to various
  forms of valence-bond crystalline (VBC) order, as well as the energy
  gap to the lowest-lying spin-triplet excitation.  In the range
  $0 < \kappa < 1$ we find evidence for four distinct phases.  Two of
  these are quasiclassical phases with antiferromagnetic long-range
  order, one with 2-sublattice N\'{e}el order for
  $\kappa < \kappa_{c_{1}} = 0.250(5)$, and another with 4-sublattice
  N\'{e}el-II order for $\kappa > \kappa_{c_{2}} = 0.340(5)$.  Two
  different paramagnetic phases are found to exist in the intermediate
  region.  Over the range
  $\kappa_{c_{1}} < \kappa < \kappa^{i}_{c} = 0.305(5)$ we find a
  gapless phase with no discernible magnetic order, which is a strong
  candidate for being a quantum spin liquid, while over the range
  $\kappa^{i}_{c} < \kappa < \kappa_{c_{2}}$ we find a gapped phase,
  which is most likely a lattice nematic with staggered dimer VBC
  order that breaks the lattice rotational symmetry.
\end{abstract}


\maketitle

\section{INTRODUCTION}
\label{introd_sec}
The interactions between the spins of classical spin-lattice models,
in which the sites of a specified infinite, regular, periodic lattice
are all occupied by classical spins, typically lead to magnetic
ground-state (GS) phases with perfect long-range order (LRO).  While
the actual GS phase that is realized depends on the precise values of
the exchange coupling constants (and/or any other parameters) in the
model Hamiltonian, such phases generically comprise coplanar spiral
configurations with a given ordering wave vector $\mathbf{Q}$, for the
simplest case of Bravais lattices, for example, except for some
special values of $\mathbf{Q}$ for which the GS phase may be
degenerate.  When the classical spins in such a model are replaced by
quantum spins with a specific value of the spin quantum number $s$,
quantum fluctuations can either reduce the corresponding magnetic
order parameter $M$ so that each sublattice of the quasiclassical GS
phase retains a (nonzero) extensive magnetization or, in the more
extreme case, destroy the classical LRO altogether in favor of some
alternate GS phase.

Since the classical problem corresponds to the limit
$s \rightarrow \infty$ of the respective quantum problem, the effect
of quantum fluctuations can firstly be examined in spin-wave theory
(SWT) by making power-series expansions in the parameter $1/s$.  In
the lowest-order SWT (LSWT) one typically finds that $M$ is reduced
from its classical value, but remains nonzero.  If the model
Hamiltonian is also invariant under the continuous SU(2) spin-rotation
symmetry group, as is the case for all two-body Hamiltonians in which
all interactions between each pair of spins considered are of Heisenberg exchange
type (i.e., of the type we study henceforth), it often occurs that the classical GS
phase in some region of the Hamiltonian parameter space has infinite
degeneracy.  In such cases the effect of quantum fluctuations at the
level of LSWT is often also to lift the accidental GS degeneracy,
either wholly or partially, by the order by disorder mechanism
\cite{Villain:1977_ordByDisord,
  Villain:1980_ordByDisord,Shender:1982_ordByDisord}, in favor of just
one or several of the classical infinitely degenerate family (IDF) of
states.

In order to maximize the possibility for destroying magnetic LRO we
need to introduce and optimize (geometric or dynamic) frustration
between the exchange interactions present in the model, so that not
all energy terms can be simultaneously minimized, and/or to enhance
the quantum fluctuations.  A combination of both is clearly ideal.  In
order to enhance the role of quantum fluctuations we clearly need to
be as far removed from the classical limit as possible.  In broad
terms quantum fluctuations are larger for lower values of both the
spin quantum number $s$ and the lattice spatial dimensionality $d$.
For a given value of $d$, quantum fluctuations are typically also
larger for lattices with smaller values of the coordination number
$z$.

Clearly, the quasiclassical magnetically-ordered GS phases discussed
above spontaneously break both SU(2) spin-rotation and time-reversal
symmetries.  Goldstone's theorem then implies that any such state
breaking spin-rotational symmetry must have a vanishing energy gap.
For the quasiclassical states with magnetic LRO the corresponding
gapless Goldstone modes are just the spin-wave excitations or magnons.

On the other hand, such intrinsically quantum-mechanical states as the
various forms of valence-bond crystalline (VBC) solid phases, in each
of which specific, regular periodic multiplets of the lattice spins
form spin singlets, are gapped and have zero magnetic order.  They
break neither SU(2) spin-rotation symmetry nor time-reversal symmetry,
while still breaking some lattice symmetries.  Other states, such as
multipolar or spin-nematic phases, also exist for which the SU(2)
spin-rotation symmetry is still broken, but in which time-reversal
symmetry is preserved, thereby again excluding magnetic LRO.  Finally,
there exists also the possibility, in principle of quantum spin-liquid
(QSL) phases that preserve {\it all} of the symmetries, including the
lattice symmetries.

Such featureless paramagnetic states \cite{Jiang:2016_SqLatt-honey} as
gapped QSL states, which are both fully symmetric and
unfractionalized, are particularly interesting, since their existence
for some specific broad classes of models is often strictly forbidden.
An example of such a constraint is the Lieb-Schultz-Mattis theorem
\cite{Lieb:1961_LSMH-theorem} for $d=1$ chains and its extensions to
systems with $d>1$
\cite{Hastings:2004_Lieb-LSM-Hast-theorem,Oshikawa:2000_LSMN_d-gtr-1}.
These preclude the possibility of a spin-lattice model with
half-odd-integral spin per unit cell being a short-ranged gapped
paramagnet that is fully symmetric and unfractionalized.  Such models
must be gapless, break a symmetry, or have fractionalized excitations
with corresponding topological order.  It is presently unknown what
might be the most general such restriction to preclude short-ranged,
gapped, and unfractionalized GS phases with symmetries including
global SU(2) spin-rotational symmetry, time-reversal symmetry, lattice
translational and rotational symmetries, and other possible lattice
point-group symmetries.  There are also, however, various
field-theoretical arguments that tend to disfavor their existence
(see, e.g., Refs.\
\cite{Haldane:1988_param-phases,Read:1990:param-phases,Jiang:2016_SqLatt-honey}).
For both these reasons it is of great interest to examine models where
they are not specifically excluded by any existing theorems and,
possibly, also for which idealized, candidate wave functions can be
constructed \cite{Jiang:2016_SqLatt-honey}.

The question then arises as to what are the optimal choices of
possible candidate systems that might exhibit such gapped,
featureless, paramagnetic GS phases.  The Mermin-Wagner theorem
\cite{Mermin:1966} itself disallows GS magnetic LRO in isotropic
Heisenberg systems for both the cases of $d=1$ chains, even at zero
temperature ($T=0$), and $d=2$ lattices at all nonzero temperatures
($T>0$).  Thus two-dimensional (2D) spin-lattice models at $T=0$
provide a rich hunting-ground for exotic GS phases with no classical
counterparts, and their GS quantum phase structures occupy a special
place for the study of quantum phase transitions (QPTs), as a large
amount of work in recent years attests.  Within this class the
honeycomb lattice occupies a key position for two reasons: (a) it has
the lowest coordination number ($z=3$) of all regular 2D periodic
lattices, and hence is expected, a {\it priori}, to exhibit the
largest quantum fluctuations when populated with lattice spins; and
(b) it is a non-Bravais lattice (with two sites per unit cell), to
which the Lieb-Mattis theorem \cite{Lieb:1961_LSMH-theorem} and its
relevant known extensions
\cite{Hastings:2004_Lieb-LSM-Hast-theorem,Oshikawa:2000_LSMN_d-gtr-1}
do not therefore apply.

In order to introduce frustration on the honeycomb lattice it suffices
to examine the $J_{1}$--$J_{2}$ model with antiferromagnetic (AFM)
isotropic Heisenberg exchange interactions between pairs of
nearest-neighbor (NN) and next-nearest-neighbor (NNN) spins with
exchange coupling strengths $J_{1}>0$ and $J_{2}>0$, respectively.
The $J_{1}$--$J_{2}$--$J_{3}$ model, which also includes isotropic
Heisenberg exchange interactions between pairs of
next-next-nearest-neighbor (NNNN) spins with coupling strength
$J_{3}>0$, is another possibility, especially along the line
$J_{3}=J_{2}$, which includes the point of maximum classical
frustration, at $J_{3}=J_{2}=\frac{1}{2}J_{1}$, where the three
classical phases that the model exhibits in the sector $J_{i}>0$
($i=1,2,3$) meet at a triple point, and where the classical GS phase
has macroscopic degeneracy.

The spin-$\frac{1}{2}$ $J_{1}$--$J_{2}$--$J_{3}$ honeycomb-lattice
model, or particular cases of it (e.g., when $J_{3}=J_{2}$ or
$J_{3}=0$), have been much studied using a variety of theoretical tools
\cite{Rastelli:1979_honey,Mattsson:1994_honey,Fouet:2001_honey,Mulder:2010_honey,Wang:2010_honey,Cabra:2011_honey,Ganesh:2011_honey,Clark:2011_honey,DJJF:2011_honeycomb,Reuther:2011_honey,Albuquerque:2011_honey,Mosadeq:2011_honey,Oitmaa:2011_honey,Mezzacapo:2012_honey,PHYLi:2012_honeycomb_J1neg,Bishop:2012_honeyJ1-J2,Bishop:2012_honey_circle-phase,Li:2012_honey_full,RFB:2013_hcomb_SDVBC,Ganesh:2013_honey_J1J2mod-XXX,Zhu:2013_honey_J1J2mod-XXZ,Zhang:2013_honey,Gong:2013_J1J2mod-XXX,Yu:2014_honey_J1J2mod}.
Of particular relevance for the present paper, we note that the
coupled cluster method (CCM) has been extensively employed
\cite{DJJF:2011_honeycomb,PHYLi:2012_honeycomb_J1neg,Bishop:2012_honeyJ1-J2,Bishop:2012_honey_circle-phase,Li:2012_honey_full,RFB:2013_hcomb_SDVBC}
to study the $T=0$ quantum phase structure of the model.  By contrast,
there are far fewer studies of the model in the case $s >
\frac{1}{2}$.  A specific exception is a recent study
\cite{Gong:2015_honey_J1J2mod_s1} of the $J_{1}$--$J_{2}$
honeycomb-lattice model for the case $s=1$, which used the
density-matrix renormalization group (DMRG) method.  Our specific aim
here is to extend earlier work, which applied the CCM to the
spin-$\frac{1}{2}$ version of the $J_{1}$--$J_{2}$ honeycomb-lattice
model \cite{Bishop:2012_honeyJ1-J2,RFB:2013_hcomb_SDVBC}, to its spin-1
counterpart.  While there is a broad consensus about many of the
features of the $T=0$ phase diagram of the spin-$\frac{1}{2}$ model,
there is still real uncertainty about the existence of a possible QSL
phase, as we discuss in more detail in Sec.\ \ref{model_sec} below.  A
main goal of the present work is to examine the possibility of any
similar QSL phase in the spin-1 model.

\begin{figure*}[t]
\begin{center}
\mbox{
\subfigure[]{\includegraphics[width=4.5cm]{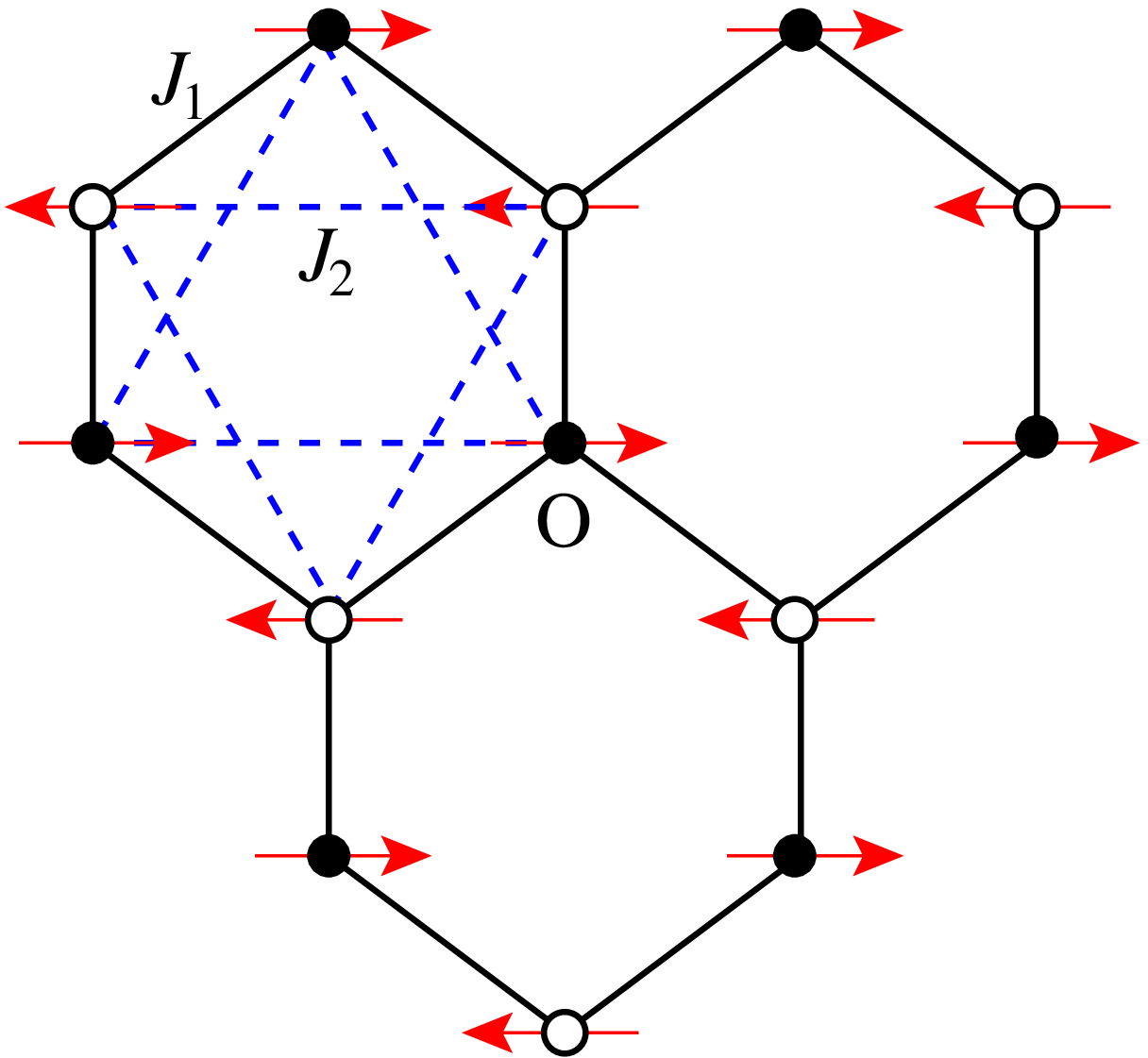}}
\quad \subfigure[]{\includegraphics[width=4.5cm]{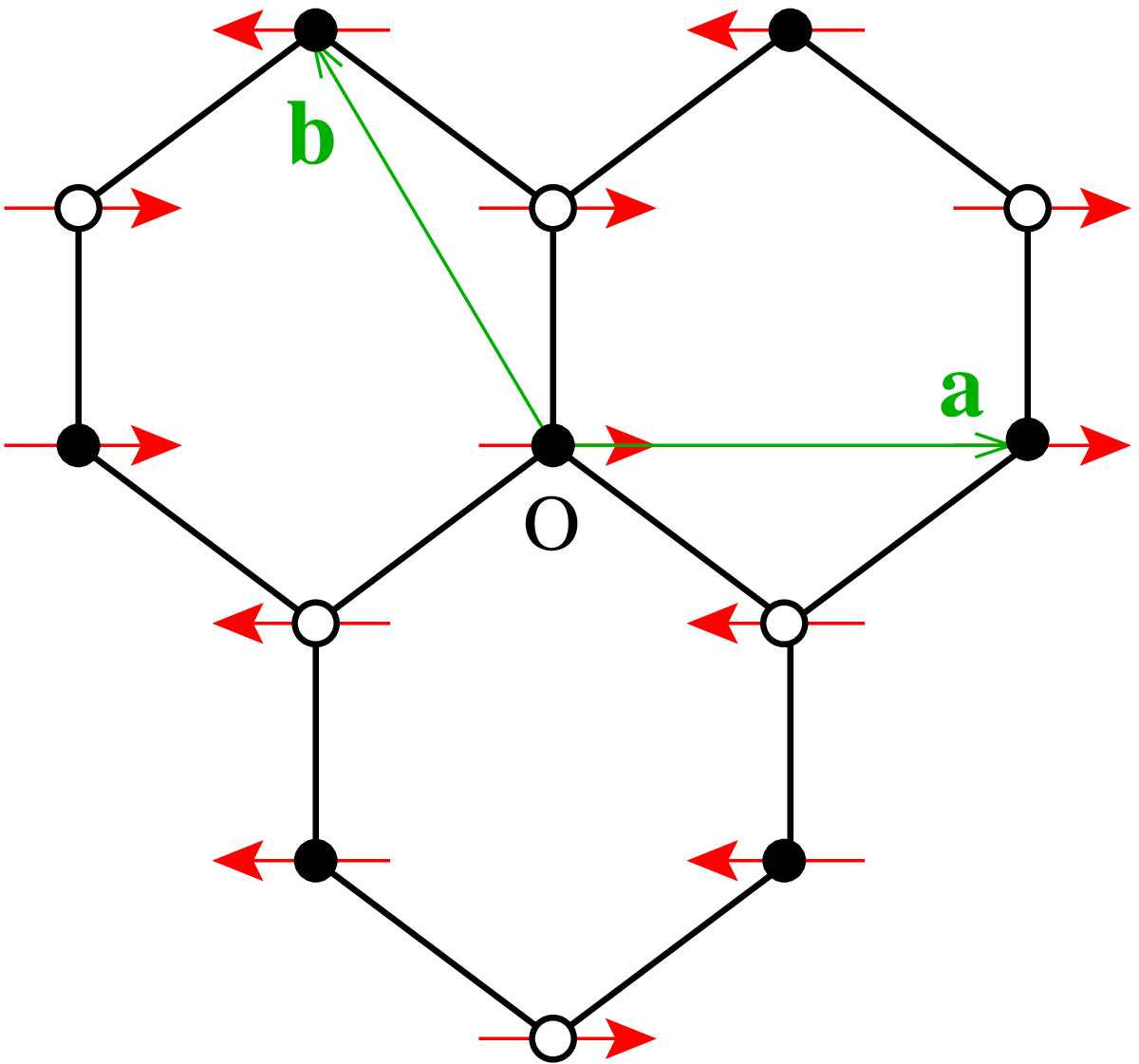}}
\quad \subfigure{\includegraphics[width=3.0cm]{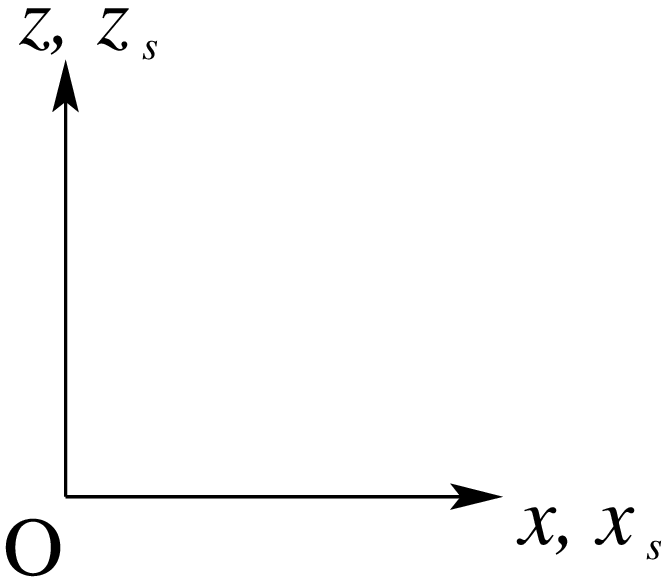}}
  }
  \caption{(Color online) The $J_{1}$--$J_{2}$ model on the honeycomb
    lattice, showing (a) the bonds ($J_{1} = $ -----; $J_{2} = $ - - -), and the N\'{e}el state, and (b) one of the three N\'{e}el-II states and the
    triangular Bravais lattice vectors $\mathbf{a}$ and $\mathbf{b}$.
    Sites on the two triangular sublattices ${\cal A}$ and ${\cal B}$
    are shown by filled and empty circles respectively, and the spins
    are represented by the (red) arrows on the lattice sites.}
\label{model_pattern}
\end{center}
\end{figure*}

Before discussing the model itself, however, it is interesting to note
that there exists a variety of quasi-2D materials that can be regarded
as experimental realizations of frustrated honeycomb-lattice systems
with AFM exchange interactions.  For example, the $s=\frac{1}{2}$
Cu$^{2+}$ ions in each of such magnetic compounds as
Na$_{3}$Cu$_{2}$SbO$_{6}$ \cite{Miura:2006_honey},
InCu$_{2/3}$V$_{1/3}$O$_{3}$ \cite{Kataev:2005_honey},
$\beta$-Cu$_{2}$V$_{2}$O$_{7}$ \cite{Tsirlin:2010_honey} and
Cu$_{5}$SbO$_{6}$ \cite{Climent:2012_honey} are situated on the sites
of weakly coupled honeycomb-lattice layers.  Also, the iridates
A$_{2}$IrO$_{3}$ (A $=$ Na, Li)
\cite{Singh:2010_honey,Liu:2011_honey,Singh:2012_honey,Choi:2012_honey}
are believed to be magnetically ordered Mott insulator materials in
which the Ir$^{4+}$ ions, which are disposed in weakly coupled
honeycomb-lattice layers, have effective $s=\frac{1}{2}$ moments.  The
two families of compounds BaM$_{2}$(XO$_{4}$)$_{2}$ (M $=$ Co, Ni; X
$=$ P, As) \cite{Regnault:1990_honey} and Cu$_{3}$M$_{2}$SbO$_{6}$ (M
$=$ Co, Ni) \cite{Roudebush:2013_honey} also comprise similar quasi-2D
materials, in each of which the magnetic M$^{2+}$ ions occupy the sites
of weakly coupled honeycomb-lattice layers.  In both of these families
the magnetic Ni$^{2+}$ ions are believed to take the high-spin value
$s=1$.  By contrast, the Co$^{2+}$ ions seem to take the low-spin
value $s=\frac{1}{2}$ in the former family BaCo$_{2}$(XO$_{4}$)$_{2}$,
and the high-spin value $s=\frac{3}{2}$ in the latter compound
Cu$_{3}$Co$_{2}$SbO$_{6}$.

In Sec.\ \ref{model_sec} we now describe the model itself and discuss
both its classical ($s \rightarrow \infty$) limit and the results
obtained to date for the extreme quantum limiting case,
$s=\frac{1}{2}$.  In Sec.\ \ref{ccm_sec} we give a brief description
of the salient features of the CCM that we use here to discuss the
$s=1$ model.  The results that we obtain are presented in Sec.\
\ref{results_sec}, and we conclude with a discussion and summary in
Sec.\ \ref{summary_sec}.

\section{THE MODEL}
\label{model_sec}
The Hamiltonian of the $J_{1}$--$J_{2}$ model studied here is given by
\begin{equation}
H = J_{1}\sum_{\langle i,j \rangle} \mathbf{s}_{i}\cdot\mathbf{s}_{j} + J_{2}\sum_{\langle\langle i,k \rangle\rangle} \mathbf{s}_{i}\cdot\mathbf{s}_{k}\,,
\label{Eq_H}
\end{equation}
where the operator
${\bf s}_{i} \equiv (s^{x}_{i}, s^{y}_{i}, s^{z}_{i})$ is the quantum
vector spin operator on lattice site $i$, with
${\bf s}^{2}_{i} = s(s+1)$, and we shall be interested here in
performing calculations for the case $s=1$.  The sums over
$\langle i,j \rangle$ and $\langle \langle i,k \rangle \rangle$ run
over all NN and NNN bonds, respectively, counting each bond once only
in each sum.  The parameters $J_{1}$ and $J_{2}$ are, respectively,
the NN and NNN exchange couplings.  We consider the case with both
bonds being AFM in nature.  Hence, with no loss of generality, we may, if we wish, put
$J_{1} \equiv 1$ to set the overall energy scale, and define
$\kappa \equiv J_{2}/J_{1}$ to be the frustration parameter.  The
honeycomb lattice is non-Bravais with two sites per unit cell.  It is
composed of two interlacing triangular Bravais sublattices ${\cal A}$
and ${\cal B}$.  Each site on a given sublattice thus has 3 NN sites
on the other sublattice and 6 NNN sites on the same sublattice.  The
lattice and the exchange bonds are shown in
Fig.~\ref{model_pattern}(a).
We define the NN lattice spacing on the
honeycomb lattice to be $d$.

The GS phase of the classical ($s \rightarrow \infty$) honeycomb
lattice $J_{1}$--$J_{2}$ model is the N\'{e}el state shown in Fig.\
\ref{model_pattern}(a) for values of the frustration parameter in the
range $0 \leq \kappa < \frac{1}{6}$.  For $\kappa > \frac{1}{6}$ the
classical spins are spirally ordered, where the model has a
one-parameter IDF of incommensurate GS phases in which the spiral wave
vector can orient in an arbitrary direction.  In LSWT spin-wave
fluctuations then lift this accidental degeneracy in favor of
particular wave vectors \cite{Mulder:2010_honey}, leading to spiral
order by disorder.  In fact, for the larger $J_{1}$--$J_{2}$--$J_{3}$
model, with all three interactions AFM in nature and
$J_{3} \equiv \lambda J_{1}$, the line segment
$\frac{1}{6} \leq \kappa \leq \frac{1}{2},\, \lambda=0$ marks the GS
$T=0$ phase boundary between two different spiral phases, known as the
spiral-I and spiral-II phases.  The $(\kappa,\,\lambda)$ points
$(\frac{1}{6},\,0)$ and $(\frac{1}{2},\,0)$ are tricritical points of
the $J_{1}$--$J_{2}$--$J_{3}$ model \cite{Fouet:2001_honey}.  The two
spiral phases meet the N\'{e}el phase at the former tricritical point,
while at the latter one they meet another classical collinear AFM
phase denoted here as the N\'{e}el-II phase and illustrated in Fig.\
\ref{model_pattern}(b).  For the $J_{1}$--$J_{2}$ model both spiral
phases are degenerate for $\frac{1}{6} < \kappa < \frac{1}{2}$, while
for $\kappa > \frac{1}{2}$ only the spiral-I phase forms the stable GS
phase.  

Both the N\'{e}el and N\'{e}el-II states comprise sets of parallel AFM
zigzag (or sawtooth) chains along one of the three equivalent
honeycomb directions.  While the NN spins on adjacent chains are also
antiparallel for the N\'{e}el state, they are parallel for the
N\'{e}el-II states.  There are thus three equivalent N\'{e}el-II
states, each of which has a 4-sublattice (i.e., a 4-site unit cell)
structure and each of which breaks the lattice rotational symmetry.
The N\'{e}el-II state is sometimes also referred to as a collinear
striped AFM phase in the literature (see, e.g., Ref.\
\cite{Gong:2015_honey_J1J2mod_s1}), for reasons which should be
apparent from Fig.\ \ref{model_pattern}(b).  However, we prefer to
preserve this name (see, e.g., Ref.\ \cite{Li:2012_honey_full}) for
yet a different classical collinear AFM state, which comprises sets of
parallel ferromagnetic zigzag chains along one of the three equivalent
honeycomb directions, with alternating chains having their spins
aligned in opposite directions.  Thus, equivalently, the striped,
N\'{e}el-II, and N\'{e}el states have, in our nomenclature, 1, 2, and
all 3 NN spins to a given spin antiparallel to it, respectively.  For
the classical $J_{1}$--$J_{2}$--$J_{3}$ model on the honeycomb lattice
with all 3 exchange bonds AFM in nature there is a third tricritical
point in the $\kappa\lambda$ plane at $(\frac{1}{2},\,\frac{1}{2})$,
at which the N\'{e}el, striped, and spiral-I phases meet
\cite{Fouet:2001_honey}.  In fact there exist two IDFs of non-coplanar
classical states, one each of which is degenerate in energy with the
N\'{e}el-II and striped states, respectively, but both thermal and quantum
fluctuations select the collinear configurations
\cite{Fouet:2001_honey}.

For the quantum models (with finite values of $s$) one expects, {\it
  a priori}, that quantum fluctuations will tend to destroy the spiral
order over a wide window of values of the frustration parameter
$\kappa$.  Indeed, this has been verified by a large variety of
calculations for the spin-$\frac{1}{2}$ case.  For example, the CCM
technique used here has shown that spiral order is destroyed over the
entire range $0 \leq \kappa \leq 1$ for the spin-$\frac{1}{2}$
$J_{1}$--$J_{2}$ model on the honeycomb lattice
\cite{Bishop:2012_honeyJ1-J2,RFB:2013_hcomb_SDVBC}.

For the spin-$\frac{1}{2}$ $J_{1}$--$J_{2}$ honeycomb-lattice
model it is by now very well established that the state with
quasiclassical N\'{e}el order is the stable GS phase for
$\kappa \lesssim 0.2$ (see, e.g., Refs.\
\cite{Albuquerque:2011_honey,Mosadeq:2011_honey,Mezzacapo:2012_honey,Bishop:2012_honeyJ1-J2,RFB:2013_hcomb_SDVBC,Ganesh:2013_honey_J1J2mod-XXX,Zhu:2013_honey_J1J2mod-XXZ,Zhang:2013_honey,Gong:2013_J1J2mod-XXX,Yu:2014_honey_J1J2mod}).
For example, the CCM technique employed here yields the value
$\kappa_{c_{1}}=0.207(3)$ for the corresponding quantum critical point
(QCP) at which N\'{e}el order melts.

It is also reasonably well established that for values of the
frustration parameter $\kappa \gtrsim 0.4$ the stable GS phase has
either the N\'{e}el-II order that occurs in the classical
$J_{1}$--$J_{2}$ model only at the isolated and highly degenerate
critical point $\kappa = \frac{1}{2}$, or has staggered dimer VBC
(SDVBC) lattice-nematic order (see, e.g., Refs.\
\cite{Fouet:2001_honey,Mulder:2010_honey,Clark:2011_honey,Reuther:2011_honey,Albuquerque:2011_honey,Mezzacapo:2012_honey,Bishop:2012_honeyJ1-J2,RFB:2013_hcomb_SDVBC,Ganesh:2013_honey_J1J2mod-XXX,Zhu:2013_honey_J1J2mod-XXZ,Zhang:2013_honey}).
The SDVBC phase is one in which the ferromagnetic bonds (i.e., the
parallel NN spin pairs) of the N\'{e}el-II states, one of which is
shown in Fig.\ \ref{model_pattern}(b), are replaced by spin-singlet
dimers.  The N\'{e}el-II and SDVBC phases thus both break the same
lattice rotational symmetry.  The CCM technique, for example, finds a
second QCP in the spin-$\frac{1}{2}$ $J_{1}$--$J_{2}$ model on the
honeycomb lattice at a value $\kappa_{c_{2}} = 0.385(10)$
\cite{Bishop:2012_honeyJ1-J2,RFB:2013_hcomb_SDVBC}, 
beyond which the stable GS phase has SDVBC order out to a third QCP at
$\kappa_{c_{3}}=0.65(5)$ \cite{RFB:2013_hcomb_SDVBC}, immediately
above which the stable GS phase has quasiclassical N\'{e}el-II order.
This value for $\kappa_{c_{2}}$ is in good agreement with the result
$\kappa_{c_{2}} \approx 0.375(25)$ from a large-scale exact
diagonalization (ED) study \cite{Albuquerque:2011_honey}, and the
estimates $\kappa_{c_{2}} \approx 0.35-0.36$ from three separate DMRG
studies
\cite{Ganesh:2013_honey_J1J2mod-XXX,Zhu:2013_honey_J1J2mod-XXZ,Gong:2013_J1J2mod-XXX}.
Whereas the ED study \cite{Albuquerque:2011_honey} finds a first-order
transition at $\kappa_{c_{2}}$ to a state that cannot be distinguished
between having N\'{e}el-II or SDVBC order, the DMRG studies
\cite{Ganesh:2013_honey_J1J2mod-XXX,Zhu:2013_honey_J1J2mod-XXZ} both
favor a transition to an SDVBC phase.  By contrast, another study using
an entangled-plaquette variational (EPV) ansatz
\cite{Mezzacapo:2012_honey}, which employs a very broad class of
entangled-plaquette states, finds that the stable GS phase for
$\kappa > \kappa_{c_{2}} \approx 0.4$ has N\'{e}el-II quasiclassical
order.  Lastly, two recent Schwinger boson mean-field (SB-MFT) studies
\cite{Zhang:2013_honey,Yu:2014_honey_J1J2mod} disagree with one
another on the nature of the GS phase for $\kappa \gtrsim 0.4$.  Zhang
and Lamas \cite{Zhang:2013_honey} find SDVBC order only in the very
narrow range $0.3732 \lesssim \kappa \lesssim 0.398$, with spiral
order for $\kappa \gtrsim 0.398$.  By contrast, Yu {\it et
  al}. \cite{Yu:2014_honey_J1J2mod} find N\'{e}el-II order for
$\kappa \gtrsim 0.43$.

The region of greatest interest, and greatest uncertainty, for the
$s=\frac{1}{2}$ model remains the region
$\kappa_{c_{1}} < \kappa < \kappa_{c_{2}}$.  For example, both SB-MFT
studies \cite{Zhang:2013_honey,Yu:2014_honey_J1J2mod} find a
transition at $\kappa_{c_{1}}$ to a QSL state.  The EPV study
\cite{Mezzacapo:2012_honey} also favors a disordered QSL phase.  On
the other hand, it is quite well established both from ED
\cite{Fouet:2001_honey,Albuquerque:2011_honey} and CCM
\cite{DJJF:2011_honeycomb,Li:2012_honey_full} studies that the
spin-$\frac{1}{2}$ honeycomb-lattice model exhibits a quantum
paramagnetic phase with strong plaquette VBC (PVBC) order in the
intermediate regime in the presence of additional AFM NNNN coupling,
$J_{3}>0$.  However, whether this PVBC order is maintained
as $J_{3} \rightarrow 0$ is more open to doubt, with various scenarios
being possible.  These include both PVBC and QSL states, as well as a
QCP between the N\'{e}el and PVBC phases.  In this context it is
particularly interesting to review the results in the intermediate
regime from applying the more systematic, less inherently biased, and
potentially more accurate methods that have been applied to the
spin-$\frac{1}{2}$ honeycomb-lattice $J_{1}$--$J_{2}$ model, {\it
  viz.}, the DMRG method
\cite{Ganesh:2013_honey_J1J2mod-XXX,Zhu:2013_honey_J1J2mod-XXZ,Gong:2013_J1J2mod-XXX}
and the CCM \cite{Bishop:2012_honeyJ1-J2,RFB:2013_hcomb_SDVBC}.

Of the DMRG studies, Ganesh {\it et al}.\
\cite{Ganesh:2013_honey_J1J2mod-XXX} found a PVBC phase over the region
$0.22 \lesssim \kappa \lesssim 0.35$, with both transitions being
continuous and thus indicative of deconfined quantum criticality
\cite{Senthil:2004_Science_deconfinedQC,Senthil:2004_PRB_deconfinedQC}.
On the other hand, Zhu {\it et al}.\ \cite{Zhu:2013_honey_J1J2mod-XXZ}
found N\'{e}el order to vanish at $\kappa \approx 0.26$, and they
suggested that in the region $0.26 \lesssim \kappa \lesssim 0.36$ the
system either has weak PVBC order or is quantum critical.  The narrow
window $0.22 \lesssim \kappa \lesssim 0.26$ in which the discrepancy between
these two DMRG studies occurs was the particular focus of a third DMRG
study by Gong {\it et al}.\ \cite{Gong:2013_J1J2mod-XXX}.  They found
N\'{e}el order to vanish at a value $\kappa \approx 0.22$, with a
possible PVBC ordering in the region
$0.25 \lesssim \kappa \lesssim 0.35$.  They found that both magnetic
(spin) and dimer orderings vanish in the thermodynamic limit in the narrow range $0.22 \lesssim \kappa \lesssim 0.25$,
consistent with a possible QSL phase of the sort that other studies
\cite{Wang:2010_honey,Clark:2011_honey,Mezzacapo:2012_honey,Yu:2014_honey_J1J2mod} favor.  The
CCM studies \cite{Bishop:2012_honeyJ1-J2,RFB:2013_hcomb_SDVBC} find a
paramagnetic phase in the region
$\kappa_{c_{1}} < \kappa < \kappa_{c_{2}}$, with the transition at
$\kappa_{c_{1}} = 0.207(3)$ of the continuous deconfined type, while
that at $\kappa_{c_{2}}=0.385(10)$ is of first-order type.  While
N\'{e}el order melts at $\kappa_{c_{1}}$, it was found
\cite{Bishop:2012_honeyJ1-J2} that PVBC order did not set in until the
slightly higher value $\kappa \approx 0.24$, and the CCM study also
thereby indicated a possible QSL phase in the narrow window
$0.21 \lesssim \kappa \lesssim 0.24$, in broad agreement with the
latest large-scale DMRG study \cite{Gong:2013_J1J2mod-XXX}.

In view of this very interesting situation for the spin-$\frac{1}{2}$
$J_{1}$--$J_{2}$ model on the honeycomb lattice in the paramagnetic
region $\kappa > \kappa_{c_{1}}$, beyond which N\'{e}el order melts,
especially with regard to the possible existence of a (likely, very
narrow) sub-regime where a QSL forms the stable GS phase, it is
clearly of great interest to study its spin-1 counterpart.  To that
end we employ here the same CCM technique as has been used for the
spin-$\frac{1}{2}$ model
\cite{Bishop:2012_honeyJ1-J2,RFB:2013_hcomb_SDVBC} to considerable
effect, as discussed above.  We will calculate a complete set of
low-energy parameters, including the GS energy per spin $E/N$, the
magnetic order parameter (i.e., the relevant sublattice magnetization)
$M$, the spin stiffness $\rho_{s}$, and the zero-field (uniform,
transverse) magnetic susceptibility $\chi$, in order to build up as much
information as possible about the $T=0$ quantum phase diagram of the
model.  We also calculate the triplet spin gap $\Delta$.  Before presenting our results
in Sec.\ \ref{results_sec} we first give a brief discussion of the key
ingredients of the CCM itself, including the reference states we
employ to calculate the various low-energy parameters.

\section{THE COUPLED CLUSTER METHOD}
\label{ccm_sec}
We briefly review the most important elements of the CCM, and refer
the interested reader to the extensive literature (and see, e.g.,
Refs.\
\cite{Kummel:1978_ccm,Bishop:1978_ccm,Bishop:1982_ccm,Arponen:1983_ccm,Bishop:1987_ccm,Bartlett:1989_ccm,Arponen:1991_ccm,Bishop:1991_TheorChimActa_QMBT,Bishop:1998_QMBT_coll,Zeng:1998_SqLatt_TrianLatt,Fa:2004_QM-coll}
and references cited therein) for further details.  As we shall see,
the method is size-extensive at all levels of implementation, and
hence directly provides results in the (thermodynamic)
infinite-lattice limit, $N \rightarrow \infty$.  The first step is always
the choice of a suitable (normalized) many-body reference state (or
model state) $|\Phi\rangle$, with respect to which the exact many-body
GS wave function $|\Psi\rangle$ may be parametrized via a systematic
scheme to incorporate the correlations, which we describe below.  We
use here, for example, the quasiclassical N\'{e}el and N\'{e}el-II
collinear AFM states shown in Figs.\ \ref{model_pattern}(a) and
\ref{model_pattern}(b), respectively, as suitable model states, among
others described below too.  Broadly speaking, the role of any CCM
model state $|\Phi\rangle$ is that of a generalized vacuum state.  The
precise properties needed of a model state are described in detail
below.

The exact ket GS wave function $|\Psi\rangle$ and its bra counterpart
$\langle\tilde{\Psi}|$ satisfy the normalization
conditions,
\begin{equation}
\langle\tilde{\Psi}|\Psi\rangle = \langle{\Phi}|\Psi\rangle =
\langle{\Phi}|\Phi\rangle \equiv 1\,.   \label{norm_conditions}
\end{equation}
They satisfy their respective GS Schr\"{o}dinger equations,
\begin{equation}
H|\Psi\rangle=E|\Psi\rangle\,; \quad \langle\tilde{\Psi}|H=E\langle\Psi|\,.  \label{GS_schrodinger_eq}
\end{equation}
In terms of a suitably chosen model state $|\Phi\rangle$ they are parametrized within the CCM by the characteristic exponentiated forms,
\begin{equation}
|\Psi\rangle=e^{S}|\Phi\rangle\,; \quad \langle\tilde{\Psi}|=\langle\Phi|\tilde{S}e^{-S}\,.  \label{exp_para}
\end{equation}
Hermiticity clearly implies the explicit relation,
\begin{equation}
\langle\Phi|\tilde{S} = \frac{\langle\Phi|e^{S^{\dagger}}e^{S}}{\langle\Phi|e^{S^{\dagger}}e^{S}|\Phi\rangle}\,,  \label{correlation-opererators-relationship}
\end{equation}
between the two CCM GS correlation operators $\tilde{S}$ and $S$.  

Nevertheless, another key feature of the CCM is that the constraint implied by Eq.\ (\ref{correlation-opererators-relationship}) is {\it not} explicitly imposed.  Instead, the 
two correlation operators are 
formally decomposed independently as the two sums,
\begin{equation}
S=\sum_{I\neq 0}{\cal S}_{I}C^{+}_{I}\,; \quad \tilde{S}=1+\sum_{I\neq 0}\tilde{{\cal S}}_{I}C^{-}_{I}\,,  \label{correlation_oper}
\end{equation}
in which $C^{+}_{0}\equiv 1$ is the identity operator in the
corresponding many-body Hilbert (or Fock) space, and where $I$ is a
set index that captures a complete set of single-body configurations
for all $N$ particles.  More specifically, what is mathematically
required of $|\Phi\rangle$ is that it is a cyclic (or fiducial) vector
with respect to a complete set of multiconfigurational (many-body)
creation operators $\{C^{+}_{I}\}$, all of which, very importantly,
are chosen so as to mutually commute between themselves,
\begin{equation}
[C^{+}_{I},C^{+}_{J}]=0\,, \quad \forall I,J \neq 0\,.   \label{commute-relation-create-destruct-oper}
\end{equation}
Hence, the set of states $\{C^{+}_{I}|\Phi\rangle\}$ forms a complete basis for the ket-state Hilbert space.  Furthermore, the state $|\Phi\rangle$ also has the property that it is a generalized vacuum state with respect to the set of operators $\{C^{-}_{I}\}$, so that
\begin{equation}
\langle\Phi|C^{+}_{I} = 0 = C^{-}_{I}|\Phi\rangle\,, \quad \forall I
\neq 0\,,  \label{creat-destruct-operators-relationship}
\end{equation}
where the corresponding multiconfigurational destruction operators, $C^{-}_{I} \equiv
(C^{+}_{I})^{\dagger}$, similarly span the bra-state Hilbert space in the sense that the set of states $\{\langle\Phi|C^{-}_{I}\}$ forms a complete basis for it.

These rather general parametrizations encapsulated in Eqs.\
(\ref{exp_para}),
(\ref{correlation_oper})--(\ref{creat-destruct-operators-relationship})
form the core of the CCM, and have several immediate consequences.  At
first sight it might appear to be a drawback of the method that
Hermiticity is not imposed via Eq.\
(\ref{correlation-opererators-relationship}).  While the exact CCM
correlation operators of Eq.\ (\ref{correlation_oper}) will surely
fulfill Eq.\ (\ref{correlation-opererators-relationship}) exactly,
when approximations are made in practice (e.g., by truncating the sums
over configurations $I$ in Eq.\ (\ref{correlation_oper}) to some
suitable, manageable subset) the Hermiticity relation may only be
satisfied approximately.  In turn this will have the consequence that
the GS energy estimates obtained at any approximate level of
implementation of the CCM, as described more fully below, do not
automatically provide strict upper bounds to the true GS energy.  We
return to this point in Sec.\ \ref{summary_sec} after presenting our
results.  Nevertheless, this potential disadvantage is almost always
far outweighed in practice by several advantages that similarly flow
from the CCM parametrization scheme.

One very important such advantage is that the scheme itself implies
that the Goldstone linked-cluster theorem will always be exactly
preserved, even when the sums in Eq.\ (\ref{correlation_oper}) are
arbitrarily truncated, as we demonstrate explicitly below.  As an
immediate consequence, the CCM is thus size-extensive at any such
approximate level of implementation.  Hence, all thermodynamically
extensive variables, such as the GS energy $E$, scale linearly with
$N$ at arbitrary levels of approximation, thereby allowing us to work
from the outset in the thermodynamic limit ($N \rightarrow \infty$),
and obviating the need for any finite-size scaling of the numerical
results that is required in most alternative techniques.  A second key
advantage of the CCM, which similarly stems as an immediate
consequence of the exponentiated parametrization scheme, is that it
also exactly preserves the very important Hellmann-Feynman theorem at
all similar levels of implementation as above.

Clearly, all GS information of the system is obtainable from the CCM
$c$-number correlation coefficients
$\{{\cal S}_{I},\tilde{{\cal S}}_{I}\}$, which may themselves now
formally be calculated by minimizing the GS energy functional,
\begin{equation}
\bar{H}=\bar{H}[{\cal S}_{I},{\tilde{\cal S}_{I}}] \equiv
\langle\Phi|\tilde{S}e^{-S}He^{S}|\Phi\rangle\,,  \label{eq_GS_E_xpect_funct}
\end{equation}
from Eq.\ (\ref{exp_para}), with respect to each of the parameters
$\{{\cal S}_{I},{\tilde{\cal S}}_{I}\,; \forall I \neq 0\}$.
Variation of $\bar{H}$ with respect to the coefficient ${\tilde{\cal S}}_{I}$ thus
yields the set of conditions,
\begin{equation}
\langle\Phi|C^{-}_{I}e^{-S}He^{S}|\Phi\rangle = 0\,, \quad \forall I \neq 0\,,  \label{ket_eq}
\end{equation}
while variation with respect to the coefficient ${\cal S}_{I}$ yields the 
corresponding set of conditions,
\begin{equation}
\langle\Phi|\tilde{S}e^{-S}[H,C^{+}_{I}]e^{S}|\Phi\rangle=0\,, \quad \forall I \neq 0\,.  \label{bra_eq}
\end{equation}
Equation (\ref{ket_eq}) is a coupled set of nonlinear equations for
the set of GS ket coefficients $\{{\cal S}_{I},\,\forall I \neq 0\}$,
while Eq.\ (\ref{bra_eq}) is a coupled set of linear equations
for the set of GS bra coefficients
$\{{\tilde{\cal S}}_{I},\,\forall I \neq 0\}$ once the ket
coefficients $\{{\cal S}_{I},\,\forall I \neq 0\}$ are used as input,
having been found first from solving Eq.\ (\ref{ket_eq}).  Both Eqs.\
(\ref{ket_eq}) and (\ref{bra_eq}) have the same number of equations as
unknown parameters to be solved for.

The GS energy $E$ is now simply the value of $\bar{H}$ from Eq.\ (\ref{eq_GS_E_xpect_funct}) at the extremum obtained from Eqs.\
(\ref{ket_eq}) and (\ref{bra_eq}),
\begin{equation}
E=\langle\Phi|e^{-S}He^{S}|\Phi\rangle=\langle\Phi|He^{S}|\Phi\rangle\,. \label{eq_GS_E}
\end{equation}
While $E$, uniquely, is thus given in terms of the ket-state
correlation coefficients $\{{\cal S}_{I}\}$ alone, the GS expectation
value of any other physical operator (e.g., the sublattice
magnetization $M$) will require a knowledge of both sets
$\{{\cal S}_{I}\}$ and $\{\tilde{{\cal S}_{I}}\}$.  The use of Eq.\
(\ref{eq_GS_E}) in Eq.\ (\ref{bra_eq}) also yields the equivalent set
of generalized linear eigenvalue equations,
\begin{equation}
\langle\Phi|\tilde{S}(e^{-S}He^{S}-E)C^{+}_{I}|\Phi\rangle=0\,, \quad \forall I \neq 0\,,  \label{bra_eq_alternative}
\end{equation}
for the GS bra coefficients $\{\tilde{\cal S}_{I},\,\forall I \neq 0\}$.

Within the CCM framework, excited-state (ES) wave functions are parametrized as
\begin{equation}
|\Psi_{e}\rangle = X^{e}e^{S}|\Phi\rangle\,,   \label{xcited_wave_funct}
\end{equation} 
where the linear excitation operator $X^{e}$ is expanded as
\begin{equation}
X^{e}=\sum_{I\neq 0}{\cal X}^{e}_{I}C^{+}_{I}\,,  \label{xcited_operator}
\end{equation} 
by analogy to Eq.\ (\ref{correlation_oper}).  A simple 
combination of the GS ket-state Schr\"{o}dinger equation (\ref{GS_schrodinger_eq}) with its ES counterpart,
\begin{equation}
H|\Psi_{e}\rangle=E_{e}|\Psi_{e}\rangle\,, \label{schrodinger_eq_xcited}
\end{equation}
readily yields the equation,
\begin{equation}
e^{-S}[H, X^{e}]e^{S}|\Phi\rangle = \Delta_{e}X^{e}|\Phi\rangle\,,  \label{eq_xcited}
\end{equation}
by making use of the simple commutativity relation, $[X^{e},S]=0$, which follows trivially from their definitions in Eqs.\ (\ref{correlation_oper}) and (\ref{xcited_operator}), and from Eq.\ (\ref{commute-relation-create-destruct-oper}), and
where $\Delta_{e}$ is the excitation energy,
\begin{equation}
\Delta_{e} \equiv E_{e}-E\,.
\end{equation}
By taking the overlap of Eq.\ (\ref{eq_xcited}) with $\langle
\Phi|C^{-}_{I}$, we find the set of equations,
\begin{equation}
\langle\Phi|C^{-}_{I}[e^{-S}He^{S},X^{e}]|\Phi\rangle = \Delta_{e} {{\cal X}}^{e}_{I}\,, \quad \forall I \neq 0\,,  \label{ket_eq_xcited}
\end{equation}
where we have made use of the fact that the set of states $\{C^{+}_{J}|\Phi\rangle\}$ is
(or may be constructed to be) orthonormalized, 
\begin{equation}
\langle \Phi|C^{-}_{I}C^{+}_{J}|\Phi\rangle = \delta\,(I,J)\,, \quad \forall I,J\,.
\end{equation}
These generalized eigenvalue equations (\ref{ket_eq_xcited}) are
then solved for the set of ket-state ES correlation coefficients $\{{\cal X}^{e}_{I}\}$ and the excitation energy $\Delta^{e}$.

Up to this point the CCM procedure is exact.  Nevertheless, one may
wonder whether it is necessary in practice to truncate the
infinite-series expansions for the ubiquitous exponential terms
$e^{\pm S}$.  However, we note that these exponentiated forms of the
operator $S$ occur in all of the equations to solve or compute [e.g.,
Eqs.\ (\ref{ket_eq})--(\ref{bra_eq_alternative}), (\ref{ket_eq_xcited})] only in the combination
of a similarity transformation of the Hamiltonian of the system,
$e^{-S}He^{S}$.  This may be expanded as the well-known nested
commutator series,
\begin{equation}
e^{-S}He^{S} = \sum^{\infty}_{n=0}\frac{1}{n!}[H,S]_{n}\,,  \label{eq_expon_nested_commutator}
\end{equation}
where $[H,S]_{n}$, defined
iteratively as
\begin{equation}
[H,S]_{n}=[[H,S]_{n-1},S]\,; \quad [H,S]_{0}=H\,,
\end{equation}
is an $n$-fold nested commutator.  Yet another key feature of the CCM
parametrization of Eq.\ (\ref{correlation_oper}) is that this
otherwise infinite sum in Eq.\ (\ref{eq_expon_nested_commutator}) will
now (usually, as here) terminate at some low finite order, due to the
mutual commutativity relation of Eq.\
(\ref{commute-relation-create-destruct-oper}) and the fact that $H$ is
(usually, as here) of finite order in the corresponding set of
relevant single-particle operators, as we now explain.

In general, if $H$ involves up to $m$-body interaction terms, its
second-quantized form will contain products of up to $2m$ single-body
creation and destruction operators, and the sum in Eq.\
(\ref{eq_expon_nested_commutator}) then terminates at the term with
$n=2m$.  Similarly, in our present case where the Hamiltonian of Eq.\
(\ref{Eq_H}) is bilinear in the SU(2) operators it is easy to see from
the SU(2) commutation relations that Eq.\
(\ref{eq_expon_nested_commutator}) terminates exactly at the term with
$n=2$ when $C^{+}_{I}$ comprises a product of single spin-raising
operators, $s^{+}_{k} \equiv s^{x}_{k} + is^{y}_{k}$, on various sites
$k$, as we discuss below.  We also observe that the mutual
commutativity requirement of Eq.\
(\ref{commute-relation-create-destruct-oper}) on all the operators
$\{C^{+}_{I},\,\forall I \neq 0\}$ that comprise the decomposition of
the correlation operators $S$ in Eq.\ (\ref{correlation_oper}), has
the immediate consequence that all nonzero terms in the expansion in
Eq.\ (\ref{eq_expon_nested_commutator}) must be linked to the
Hamiltonian.  It is simply not possible to generate unlinked terms in
this way, and hence the Goldstone linked-cluster theorem (and its corollary of
size-extensivity) is preserved even if truncations are made for the
expansions in Eq.\ (\ref{correlation_oper}) of the CCM correlation
operators.

Thus, finally, to implement the CCM in practice, the sole
approximation made is to restrict the set of multiconfigurational
set-indices $\{I\}$ that are retained in the expansions of the
correlation operators $\{S,\tilde{S},X^{e}\}$ in Eqs.\
(\ref{correlation_oper}) and (\ref{xcited_operator}) to some manageable (infinite or finite)
subset.  Clearly, the related choices for both the appropriate model state
$|\Phi\rangle$ and the appropriate truncation scheme should be based
on physical grounds, and we turn now to how such choices are made for
spin-lattice models in general.

The simplest broad class of model states for quantum magnets comprises
independent-spin product states, in each of which the spin projection
of every spin, along some specified quantization axis on each lattice
site, is independently specified.  The two collinear
N\'{e}el and N\'{e}el-II AFM states shown, respectively, in Figs.\
\ref{model_pattern}(a) and \ref{model_pattern}(b) are clearly of this
type.  More generally, so are all such similar quasiclassical states
with perfect magnetic LRO.  It is extremely convenient to put all such
states on the same footing, and thereby treat them universally.  One
simple way to do so is to make a passive rotation of each spin
independently (i.e., by choosing local spin quantization axes on each lattice 
site independently), so that every spin on every site then points,
say, downwards, (i.e., along the negative $z_{s}$ direction of the
axes shown in Fig.\ \ref{model_pattern}) in its own local spin-coordinate
frame.  The basic SU(2) spin commutation relations are, of course,
preserved under such unitary transformations.  In this way all lattice
sites become equivalent to one another, whatever the form of the
independent-spin product, quasiclassical, model state $|\Phi\rangle$
that is chosen as the CCM reference state.  All such states take the
universal form
$|\Phi\rangle=|$$\downarrow\downarrow\downarrow\cdots\downarrow\rangle$
in their own local spin-coordinate frames.  Once such frames have been
chosen, all that is needed is to re-express the Hamiltonian $H$
in terms of them.

It is now simple also to see how to make all such states
$|\Phi\rangle$
fiducial vectors with respect to a suitable set of mutually commuting
many-body creation operators $\{C^{+}_{I}\}$.
Thus, we may simply construct $C^{+}_{I}$
as a product of single-spin raising operators, $C^{+}_{I}
\rightarrow s^{+}_{k_{1}}s^{+}_{k_{2}}\cdots s^{+}_{k_{n}};\;
n=1,2,\cdots , 2sN$, with the set-index
$I$ now becoming a collection of lattice-site indices, $I
\rightarrow \{k_{1},k_{2},\cdots , k_{n};\; n=1,2,\cdots ,
2sN\}$, wherein each individual site index may appear no more than
$2s$ times.

Within the CCM framework so described for quantum spin-lattice models,
a very general and systematic hierarchy of approximations, called the
SUB$n$--$m$ scheme, has been extensively applied to a wide variety of
systems, ranging from unfrustrated to highly frustrated models, with
considerable success.  For specified values of the pair of
positive-integral truncation indices $n$ and $m$, the CCM SUB$n$--$m$
approximation retains only those multi-spin configurations $I$ above,
which involve no more than $n$ spin flips that span a range of up to a
maximum of $m$ contiguous sites.  In this context a single spin flip
requires the action of a spin-raising operator $s^{+}_{k}$ acting
once, and a set of lattice sites is defined to be contiguous if each
site in the set is NN (in the specified geometry) to at least one
other in the set.  Clearly, the SUB$n$--$m$ approximation becomes
exact as both truncation indices $n$ and $m$ become indefinitely
large.  Different sub-schemes can also be specified according to how
each truncation index approaches the exact infinite limit.

A very extensively used approximation scheme is the localized
(lattice-animal-based subsystem) LSUB$m$ scheme
\cite{Zeng:1998_SqLatt_TrianLatt,Fa:2004_QM-coll}.  At the $m$th level
of approximation, this scheme is defined to retain all clusters of
spins described by multispin-flip configurations $\{I\}$ in the sums
in Eqs.\ (\ref{correlation_oper}) and (\ref{xcited_operator}) that span $m$ or fewer contiguous
lattice sites.  The configurations retained are thus defined on all
possible lattice animals (or polyominos, equivalently, in the usual
graph-theoretic sense) of maximal size $m$.  Clearly, the LSUB$m$
scheme is equivalent to the previously defined SUB$n$--$m$ scheme in
the case $n=2sm$, i.e., LSUB$m \equiv$ SUB$2sm$--$m$.  It is precisely
the LSUB$m$ scheme that we used, for example, in our earlier studies
\cite{Bishop:2012_honeyJ1-J2,RFB:2013_hcomb_SDVBC} of the
spin-$\frac{1}{2}$, honeycomb-lattice $J_{1}$--$J_{2}$ model.

The number $N_{f}=N_{f}(m)$ of fundamental multispin-flip
configurations that are defined to be distinct under the symmetries of
the lattice and the particular model state $|\Phi\rangle$ being
employed (i.e., the effective size of the index set $\{I\}$), and
which are retained at a given $m$th level of LSUB$m$ approximation, is
clearly lowest for $s=\frac{1}{2}$.  Since $N_{f}(m)$ rises sharply
for a given truncation index $m$ as $s$ is increased, and also since
$N_{f}(m)$ typically increases faster than exponentially for a given
value of $s$ as $m$ is increased, the alternative SUB$n$--$n$ scheme
is usually preferred for models with $s >\frac{1}{2}$, as here.
Clearly, SUB$n$--$n$ $\equiv$ LSUB$n$ only for $s=\frac{1}{2}$,
whereas for $s > \frac{1}{2}$ we have SUB$n$--$n$ $\subset$ LSUB$n$.
For most of the calculations performed here, including all of those
based on the N\'{e}el and N\'{e}el-II states as CCM model states, we
employ the SUB$n$--$n$ scheme up to the very high order $n = 10$.

We note that the multispin-flip cluster configurations $\{I\}$ in the
expansion of Eq.\ (\ref{xcited_operator}) for the excitation operator
$X^{e}$ are different to those in the corresponding expansion of Eq.\
(\ref{correlation_oper}) for the GS correlations operators $S$ and $\tilde{S}$.
They are also different for each model state.  Thus, for the ES
calculations for the triplet spin gap $\Delta$ we restrict ourselves to configurations $I$
that change the $z$ component of total spin, $S^{z}$, by one unit, whereas for the GS
calculations we restrict ourselves to those with $S^{z}=0$.  To ensure
comparable accuracy for both the GS and ES calculations, however, we
use the SUB$n$--$n$ approximation in both cases.  Once again, even
though the number of fundamental configurations, $N_{f}(n)$, at a
given $n$th level of SUB$n$--$n$ approximation, is different for the
ES case than for the GS case using the same CCM model state
$|\Phi\rangle$, our calculations here for $\Delta$ are also done up to
the very high order $n=10$, for both the N\'{e}el and N\'{e}el-II
choices of model state.  For example, for the N\'e{e}l model state in
the present spin-1 model, we have $N_{f}(10)=219521$ for the GS case
and $N_{f}(10)=244533$ for the triplet spin-gap ES case.  Corresponding
numbers for the N\'{e}el-II model state are $N_{f}(10)=630130$ and
$N_{f}(10)=710533$ for the GS and triplet spin-gap ES cases,
respectively.  Clearly, with such large numbers of equations to derive
and solve \cite{Zeng:1998_SqLatt_TrianLatt}, one needs both massive parallelization
and supercomputing resources, as well as purpose-built computer-algebra
packages for the derivation of the equations \cite{ccm_code}.

\begin{figure}[!tb]
\includegraphics[width=4.5cm]{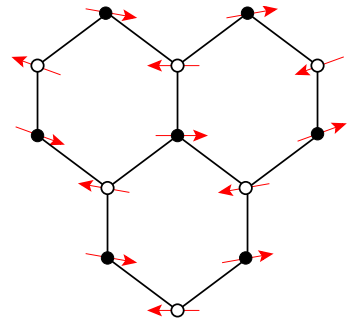}
\caption{(Color online) The twisted reference state for the calculation of the spin stiffness coefficient, $\rho_{s}$, for the $J_{1}$--$J_{2}$ honeycomb model.  The twist is applied in the $x$ direction to 
  the N\'{e}el state shown in Fig.\ \ref{model_pattern}(a).
  The spins on lattice sites \textbullet \hspace{0.01cm} are represented by the (red) arrows.}
\label{pattern_sStiff}
\end{figure}

We are also interested here in calculating other low-energy parameters
of the system, namely the spin stiffness coefficient $\rho_{s}$ and the zero-field
(uniform, transverse) magnetic susceptibility $\chi$, for which
suitably twisted and canted quasiclassical states, respectively, are
also required as CCM model states, as we now describe.  For example,
the spin stiffness (or helicity modulus) $\rho_{s}$ of a spin-lattice
system provides a quantitative measure of the energy required to
rotate the order parameter of a magnetically ordered state by an
(infinitesimal) angle $\theta$ per unit length in a specified
direction.  Thus if the GS energy as a function of the imposed twist
is $E(\theta)$, and $N (\rightarrow \infty)$ is the number of lattice sites, we have
\begin{equation}
\frac{E(\theta)}{N}=\frac{E}{N} + \frac{1}{2}\rho_{s}\theta^{2} + O(\theta^{4})\,,  \label{eq_GS-E_theta}
\end{equation}
where $E \equiv E(\theta=0)$.  We note that $\theta$ has the
dimensions of an inverse length.  In the thermodynamic limit
($N \rightarrow \infty$) considered here a nonzero (positive) value of
$\rho_{s}$ implies the stability of the magnetic LRO of the state
in question.  Exceptionally, for the N\'{e}el state (illustrated in
Fig.\ \ref{model_pattern}(a) with a staggered magnetization in the
$x_{s}$ direction), the value of $\rho_{s}$ is completely independent
of the applied twist direction, since its ordering wave vector takes
the value $\mathbf{Q}=(0,0)$ in the $xz$ plane shown in Fig.\
\ref{model_pattern}.  We show in Fig.\ \ref{pattern_sStiff} the twist
applied in the $x$ direction to the unperturbed N\'{e}el state of
Fig.\ \ref{model_pattern}(a), and it is just the twisted state shown
in Fig.\ \ref{pattern_sStiff} that we now use as our CCM model state
to calculate $\rho_{s}$ for the N\'{e}el GS phase.
The definition of Eq.\ (\ref{eq_GS-E_theta}) readily yields that the corresponding classical ($s \rightarrow \infty$) value of $\rho_{s}$ for the  N\'{e}el state of the $J_{1}$--$J_{2}$ model on the honeycomb lattice is
\begin{equation}
\rho^{{\rm N\acute{e}el}}_{s;\,{\rm cl}}=\frac{3}{4}(1-6\kappa)J_{1}d^{2}s^{2}\,, \label{sStiff_neel_classical}
\end{equation}
where $d$ is the honeycomb lattice spacing.  Unsurprisingly,
$\rho^{{\rm N\acute{e}el}}_{s;\,{\rm cl}} \rightarrow 0$ at precisely
the point $\kappa=\frac{1}{6}$ where the classical N\'{e}el LRO
vanishes, and we have a continuous transition to a stable GS phase with
spiral order.

\begin{figure*}[t]
\begin{center}
\mbox{
\subfigure[]{\includegraphics[width=4.5cm]{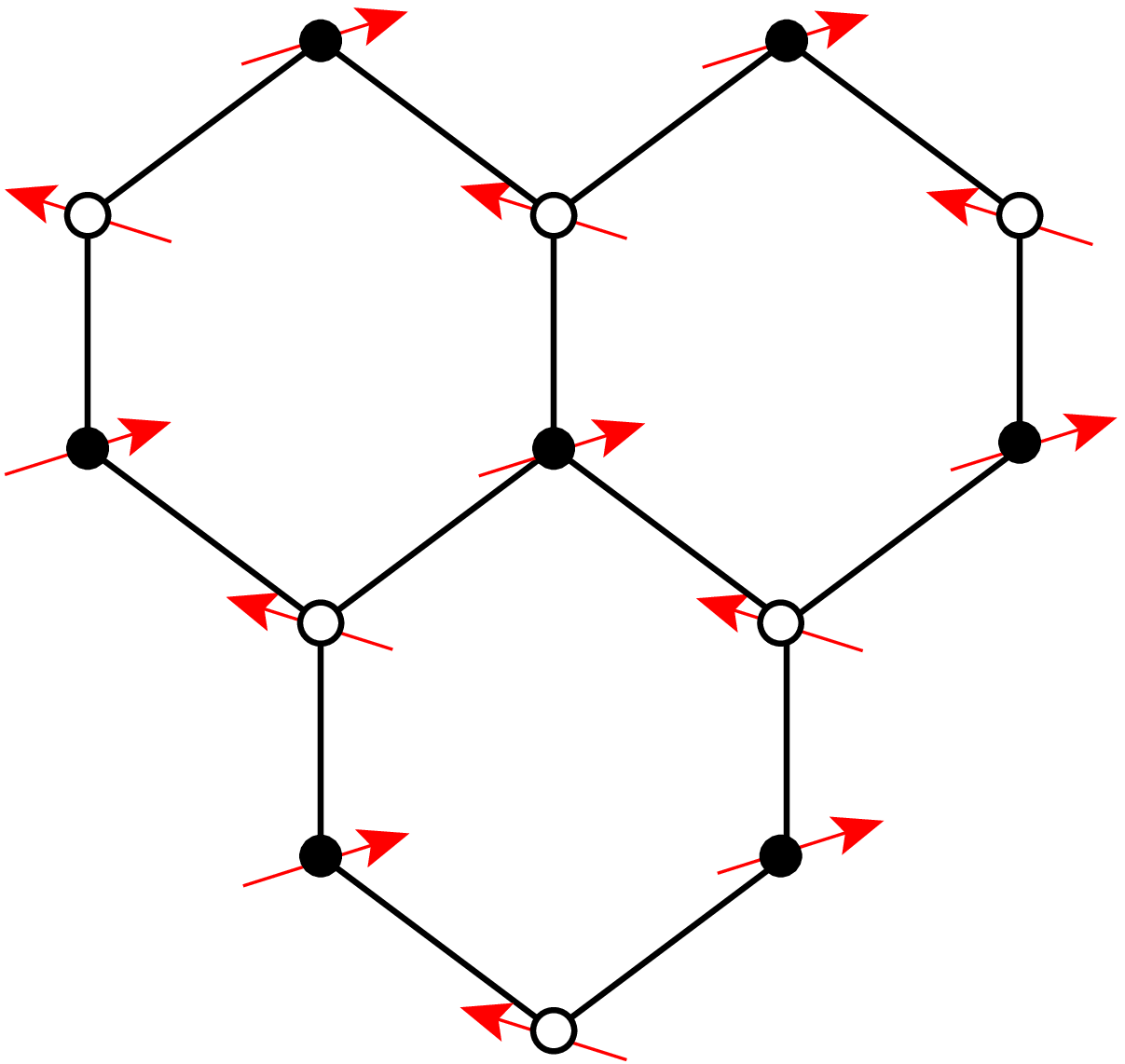}}
\quad \subfigure[]{\includegraphics[width=4.5cm]{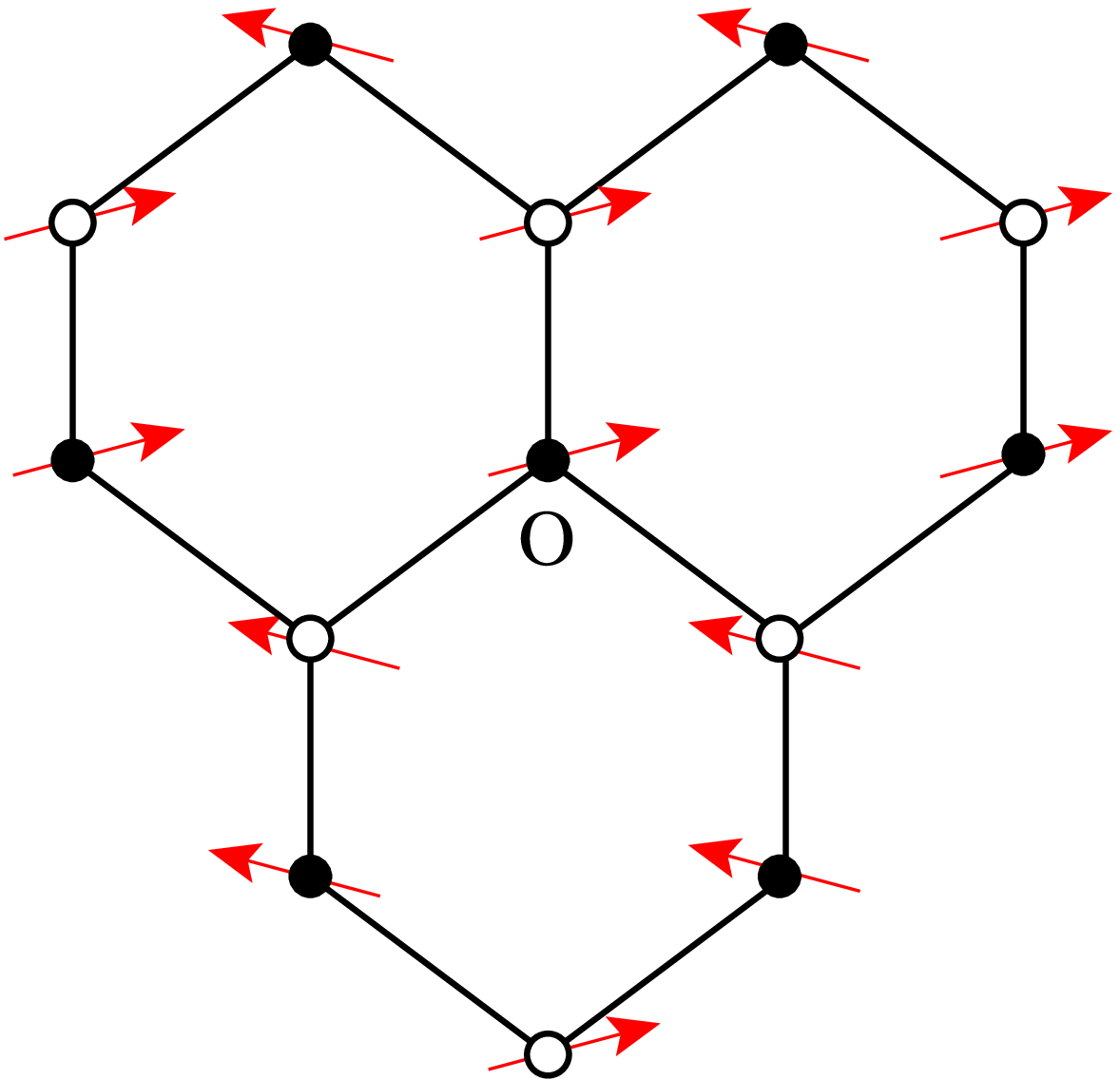}}
  }
  \caption{(Color online) The canted reference states for the calculation of the zero-field magnetic susceptibility, $\chi$, for the $J_{1}$--$J_{2}$ honeycomb model.  The external magnetic field is applied in the $z_{s}$ direction to 
  the (a) N\'{e}el and (b) N\'{e}el-II states shown in Figs.\ \ref{model_pattern}(a) and \ref{model_pattern}(b), respectively.
  The spins on lattice sites \textbullet \hspace{0.01cm} are represented by the (red) arrows.}
\label{pattern_M_ExtField}
\end{center}
\end{figure*}

Suppose we now place our unperturbed system in an external transverse magnetic field $\mathbf{h}$, in order to calculate its zero-field magnetic susceptibility $\chi$.  For the two collinear, quasiclassical AFM states shown in Fig.\ \ref{model_pattern}, both of which have spins aligned along the $x_{s}$ axis, the field is applied in the $z_{s}$ direction, $\mathbf{h} = h\hat{z}_{s}$.  In units where the 
gyromagnetic ratio $g\mu_{B}/\hbar=1$, the Hamiltonian $H=H(h=0)$ of Eq.\
(\ref{Eq_H}) then becomes
\begin{equation}
H(h)=H(0) + h\sum_{l} s^{z}_{l}\,,  \label{eq_H-h}
\end{equation}
In the presence of the magnetic field the spins will cant at an angle $\alpha$ to
the $x_{s}$ axis with respect to their zero-field configurations, as
shown in Figs.\ \ref{pattern_M_ExtField}(a) and \ref{pattern_M_ExtField}(b) for the two quasiclassical AFM states
shown in Figs.\ \ref{model_pattern}(a) and \ref{model_pattern}(b), respectively.  
For both states the classical ($s \rightarrow
\infty$) value of $\alpha$ is readily found 
by minimizing the classical energy, $E = E(h)$, corresponding to Eq.\
(\ref{eq_H-h}), with respect to the cant angle $\alpha$.  As usual, the uniform (transverse) magnetic
susceptibility is then defined as
\begin{equation}
\chi({h})=-\frac{1}{N}\frac{{\rm d}^{2}E}{{\rm d}h^{2}}\,.
\end{equation}
Its zero-field limit is then the respective low-energy parameter, $\chi \equiv \chi(0)$, and the corresponding
analog of Eq.\ (\ref{eq_GS-E_theta}) is hence,
\begin{equation}
\frac{E(h)}{N}=\frac{E}{N}-\frac{1}{2}\chi h^{2} + O(h^{4})\,.
\end{equation}
where $E \equiv E(h=0)$.  The two canted states shown in Figs.\ \ref{pattern_M_ExtField}(a) and \ref{pattern_M_ExtField}(b) are just the CCM model states that we use to calculate $\chi$ for the N\'{e}el and N\'{e}el-II GS phases, respectively.  Simple calculations show that the corresponding classical ($s \rightarrow \infty$) values for the $J_{1}$--$J_{2}$ model on the honeycomb lattice are
\begin{equation}
\chi^{{\rm N\acute{e}el}}_{{\rm cl}}=\frac{1}{6J_{1}}\,, \label{chi_neel_classical}
\end{equation}
and
\begin{equation}
\chi^{{\rm N\acute{e}el-II}}_{{\rm cl}}=\frac{1}{4(1+2\kappa)J_{1}}\,,  \label{chi_neel2_classical}
\end{equation}
independent of $s$ in each case classically.

Unlike the classical spin stiffness coefficient, the classical zero-field susceptibility coefficient is not expected to go to zero at a phase transition point where the corresponding magnetic LRO vanishes.  However, the two classical values in Eqs.\ (\ref{chi_neel_classical}) and (\ref{chi_neel2_classical}) do meet (at a finite value) at the point $\kappa =\frac{1}{4}$.  This is precisely the point at which the corresponding classical energy curves, 
\begin{equation}
\frac{E^{{\rm N\acute{e}el}}_{{\rm cl}}}{N}=-\frac{3}{2}(1-2\kappa)J_{1}\,, \label{E_neel_classical}
\end{equation}
and
\begin{equation}
\frac{E^{{\rm N\acute{e}el-II}}_{{\rm cl}}}{N}=-\frac{1}{2}(1+2\kappa)J_{1}\,, \label{E_neel2_classical}
\end{equation}
cross (or, more precisely, would cross if these two GS phases were in competition with one another in the classical model).  We note, however, that, unlike in the classical limit, $\chi$ {\it can} become zero in a quantum case (i.e., for a finite value of $s$) at a QCP, where it provides a clear signal of a spin gap opening \cite{Mila:2000_M-Xcpty_spinGap,Bernu:2015_M-Xcpty_spinGap} (i.e., a transition to a gapped state).

While the CCM does not involve any finite-size scaling of its results, as a final step we do need to extrapolate our SUB$n$--$n$ sequences of approximants for any GS or ES calculated physical parameter to the limit $n \rightarrow \infty$ where, by construction, the method becomes exact.
Although no exact such extrapolation rules are known, by now there exists a large body of heuristic work on a very wide variety of spin-lattice models that has led to empirical schemes for many physical quantities.  Thus, for example, a very well tested and highly accurate extrapolation scheme for the GS energy per spin has been shown to be 
(and see, e.g., Refs.\
\cite{Fa:2004_QM-coll,DJJF:2011_honeycomb,Bishop:2000_XXZ,Kruger:2000_JJprime,Fa:2001_SqLatt_s1,Darradi:2005_Shastry-Sutherland,Darradi:2008_J1J2mod,Bi:2008_EPL_J1J1primeJ2_s1,Bi:2008_JPCM_J1xxzJ2xxz_s1,Bi:2009_SqTriangle,Bishop:2010_UJack,Bishop:2010_KagomeSq,Bishop:2011_UJack_GrtSpins,PHYLi:2012_SqTriangle_grtSpins,PHYLi:2012_honeycomb_J1neg,Bishop:2012_honeyJ1-J2,Bishop:2012_honey_circle-phase,Li:2012_honey_full,Li:2012_anisotropic_kagomeSq,RFB:2013_hcomb_SDVBC})
\begin{equation}
\frac{E(n)}{N} = e_{0}+e_{1}n^{-2}+e_{2}n^{-4}\,.     \label{extrapo_E}
\end{equation}

As expected, the convergence as a function of the SUB$n$--$n$ truncation index $n$ of all other GS physical parameters is slower than for the energy.  A typical example is the magnetic order parameter $M$ (i.e., the appropriate sublattice magnetization), which takes the simple form,
\begin{equation}
M = -\frac{1}{N}\sum^{N}_{k=1}\langle\Phi|\tilde{S}
  e^{-S}s^{z}_{k}e^{S}|\Phi\rangle\,.   \label{M_eq}
\end{equation}
in terms of the local rotated
spin-coordinate frames discussed above.  For unfrustrated or only very mildly frustrated systems, an extrapolation scheme for $M(n)$ with leading power $1/n$ (rather than $1/n^{2}$, as for the GS energy),
\begin{equation}
M(n) = m_{0}+m_{1}n^{-1}+m_{2}n^{-2}\,,   \label{M_extrapo_standard}
\end{equation}
has been found (and see, e.g., Refs.\
\cite{Bishop:2000_XXZ,Kruger:2000_JJprime,Fa:2001_SqLatt_s1,Darradi:2005_Shastry-Sutherland,Bi:2009_SqTriangle,Bishop:2010_UJack,Bishop:2010_KagomeSq,Bishop:2011_UJack_GrtSpins,PHYLi:2012_SqTriangle_grtSpins,PHYLi:2012_honeycomb_J1neg,Bishop:2012_honeyJ1-J2,Bishop:2012_honey_circle-phase,RFB:2013_hcomb_SDVBC}) to fit the CCM results extremely well.  By contrast, for highly frustrated systems, a scheme which has been found to be more appropriate and to fit the CCM results very closely in a wide variety of earlier studies (and see, e.g., Refs.\ \cite{DJJF:2011_honeycomb,PHYLi:2012_honeycomb_J1neg,Bishop:2012_honeyJ1-J2,Bishop:2012_honey_circle-phase,Li:2012_honey_full,RFB:2013_hcomb_SDVBC,Darradi:2008_J1J2mod,Bi:2008_EPL_J1J1primeJ2_s1,Bi:2008_JPCM_J1xxzJ2xxz_s1,Li:2012_anisotropic_kagomeSq} is one with a leading power $1/n^{1/2}$,
\begin{equation}
M(n) = \mu_{0}+\mu_{1}n^{-1/2}+\mu_{2}n^{-3/2}\,.   \label{M_extrapo_frustrated}
\end{equation}
This latter scheme is particularly appropriate for systems with an order-disorder transition, or for systems which are either close to a QCP or for which $M$ is close to zero.

CCM SUB$n$--$n$ extrapolation schemes with a leading power $1/n$ have also been shown to fit the results very well for the corresponding approximants for each of the spin
gap $\Delta(n)$ (and see, e.g., Refs.\ \cite{Kruger:2000_JJprime,Richter:2015_ccm_J1J2sq_spinGap,Bishop:2015_honey_low-E-param,Bishop:2015_J1J2-triang_spinGap}),
\begin{equation}
\Delta(n) = d_{0}+d_{1}n^{-1}+d_{2}n^{-2}\,,   \label{Eq_spin_gap}
\end{equation}
the spin stiffness coefficient
$\rho_{s}(n)$ (and see, e.g., Refs.\ \cite{Darradi:2008_J1J2mod,Bishop:2015_honey_low-E-param,SEKruger:2006_spinStiff,Gotze:2016_triang,Bishop:2016_honey_grtSpins}),
\begin{equation}
\rho_{s}(n) = s_{0}+s_{1}n^{-1}+s_{2}n^{-2}\,,   \label{Eq_sstiff}
\end{equation}
and the zero-field magnetic susceptibility, $\chi(n)$ (and see, e.g., Refs.\
\cite{Bishop:2015_honey_low-E-param,Gotze:2016_triang,Bishop:2016_honey_grtSpins,Farnell:2009_Xcpty_ExtMagField}),
\begin{equation}
\chi(n) = x_{0}+x_{1}n^{-1}+x_{2}n^{-2}\,,   \label{Eq_X}
\end{equation}
from which we obtain, respectively, the extrapolated values $\Delta \equiv \Delta(\infty)=d_{0}$, $\rho_{s} \equiv \rho_{s}(\infty)=s_{0}$, and $\chi \equiv \chi(\infty)=x_{0}$.

We note that each of the extrapolation schemes of Eqs.\
(\ref{extrapo_E}) and (\ref{M_extrapo_standard})--(\ref{Eq_X})
contains three fitting parameters.  Clearly, in order to obtain stable
and robust fits to such schemes it is preferable to use at least four
SUB$n$--$n$ data points as input.  However, occasionally
this is either impracticable or inappropriate, for various reasons we
describe.  In such cases it is often then preferable to utilize, for
the particular GS physical parameter $P$ involved, a completely
unbiased extrapolation scheme for the CCM SUB$n$--$n$ approximants
$P(n)$ of the form,
\begin{equation}
P(n) = p_{0}+p_{1}n^{-\nu}\,,   \label{Eq_exponFit}
\end{equation}
in which the leading exponent $\nu$ is itself a free fitting
parameter, along with $p_{0}$ and $p_{1}$.  Naturally, it is always
possible to perform such a fit first for any GS quantity, even when we
have four or more data points to utilize, in order to check the value
of the leading exponent, before using one of the afore-mentioned
schemes of Eqs.\ (\ref{extrapo_E}) and
(\ref{M_extrapo_standard})--(\ref{Eq_X}).

\section{RESULTS}
\label{results_sec}
\begin{figure}
  \includegraphics[width=6.2cm,angle=270]{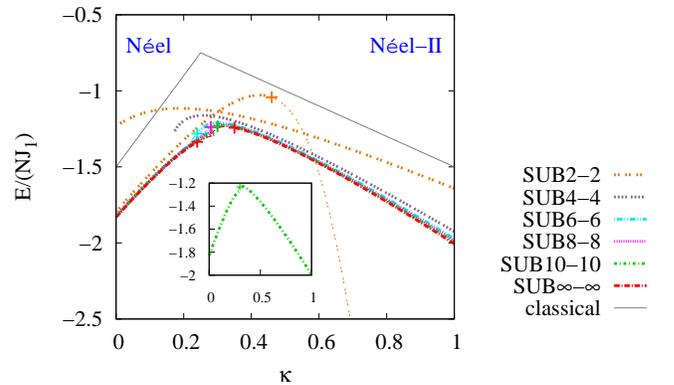}
\caption{(Color online) CCM results for the GS energy per spin $E/N$ (in units of $J_{1}$) versus the frustration parameter $\kappa \equiv J_{2}/J_{1}$, for
  the spin-1 $J_{1}$--$J_{2}$ model on the honeycomb lattice (with $J_{1}>0$).  
  Results based on both the N\'{e}el and N\'{e}el-II states as CCM model states are shown in SUB$n$--$n$
  approximations with $n=\{2,4,6,8,10\}$, together with the corresponding
  SUB$\infty$--$\infty$ extrapolations using Eq.\ (\ref{extrapo_E}), with the respective data set $n=\{4,6,8,10\}$.  The plus (+) symbols on the curves mark the
  points where the respective solutions have $M \rightarrow 0$, and those portions of the curves shown with thinner lines beyond the plus (+) symbols indicate unphysical regions where $M<0$.  For comparison we also show the corresponding
  classical curves from Eqs.\ (\ref{E_neel_classical}) and
  (\ref{E_neel2_classical}) for the value $s=1$.}
\label{E_s1}
\end{figure}

We will return in Sec.\ \ref{summary_sec}, after first presenting our
results, to the question of the role, in practice, of the CCM model
state, and the related question of whether or not our results depend
on the specific choices made.  Firstly, however, we show in Fig.\
\ref{E_s1} our results for the GS energy per spin, $E/N$, for the
model, using both the N\'{e}el and N\'{e}el-II AFM states as CCM model
states.
In each case we display the respective results from SUB$n$--$n$
approximations with $n=\{2,4,6,8,10\}$, together with the corresponding
SUB$\infty$--$\infty$ extrapolations based on the data sets with
$n=\{4,6,8,10\}$ as input to Eq.\ (\ref{extrapo_E}), to find the
corresponding ($n \rightarrow \infty$) extrapolated values $e_{0}$.
It is clear from Fig.\ \ref{E_s1} that for the SUB$n$--$n$
calculations based on both quasiclassical AFM states, convergence is
rapid as the truncation index $n$ is increased.  We also note that
each of the energy curves terminates at some critical value of the
frustration parameter that depends both on the model state chosen and
the SUB$n$--$n$ approximation used.  For the curves based on the
N\'{e}el model state there is an upper critical value, and for those
based on the N\'{e}el-II model state a lower critical value.  In both
cases, for values of $\kappa$ beyond the respective critical value no
real solution can be found to the corresponding CCM equations
(\ref{ket_eq}).

Such CCM termination points of the coupled sets of SUB$n$--$n$
equations are both common in practice and well understood (see, e.g.,
Refs.\
\cite{Bishop:2012_honeyJ1-J2,Fa:2004_QM-coll,Bi:2009_SqTriangle}).
They are always simply a consequence of the corresponding QCP that
exists in the model under study, and which marks the melting of the
respective form of magnetic LRO corresponding to the particular model
state used.  For a given (finite) value of the SUB$n$--$n$ truncation
index $n$ and for a given phase, Fig.\ \ref{E_s1} shows that the CCM
solutions extend beyond the actual (SUB$\infty$--$\infty$) QCP into the
unphysical regime beyond the QCP, as is usually the case in other
models studied.  The extent of the unphysical regime diminishes as the
truncation index $n$ increases, and ultimately vanishes in the exact
($n \rightarrow \infty$) limit.

Figure \ref{E_s1} demonstrates clear preliminary evidence for the
existence of an intermediate phase between the phases with N\'{e}el and
N\'{e}el-II magnetic LRO for the $s=1$ $J_{1}$--$J_{2}$ model, just as
for its $s=\frac{1}{2}$ counterpart.  For example, Fig.\ \ref{E_s1}
shows that, whereas the SUB$n$--$n$ GS energy results based on the
N\'{e}el and N\'{e}el-II model states for a given value of $n$ cross
one another for values $n \leq 8$, before their respective termination
points, this is no longer the case for $n=10$, as the inset shows
clearly, where the upper termination point for the N\'{e}el phase and
the lower termination point for the N\'{e}el-II phase are both around
$\kappa \approx 0.28$.  Presumably, if we could perform CCM
SUB$n$--$n$ calculations with $n > 10$ a gap would appear between the
respective termination points.

\begin{figure}
\includegraphics[width=6.2cm,angle=270]{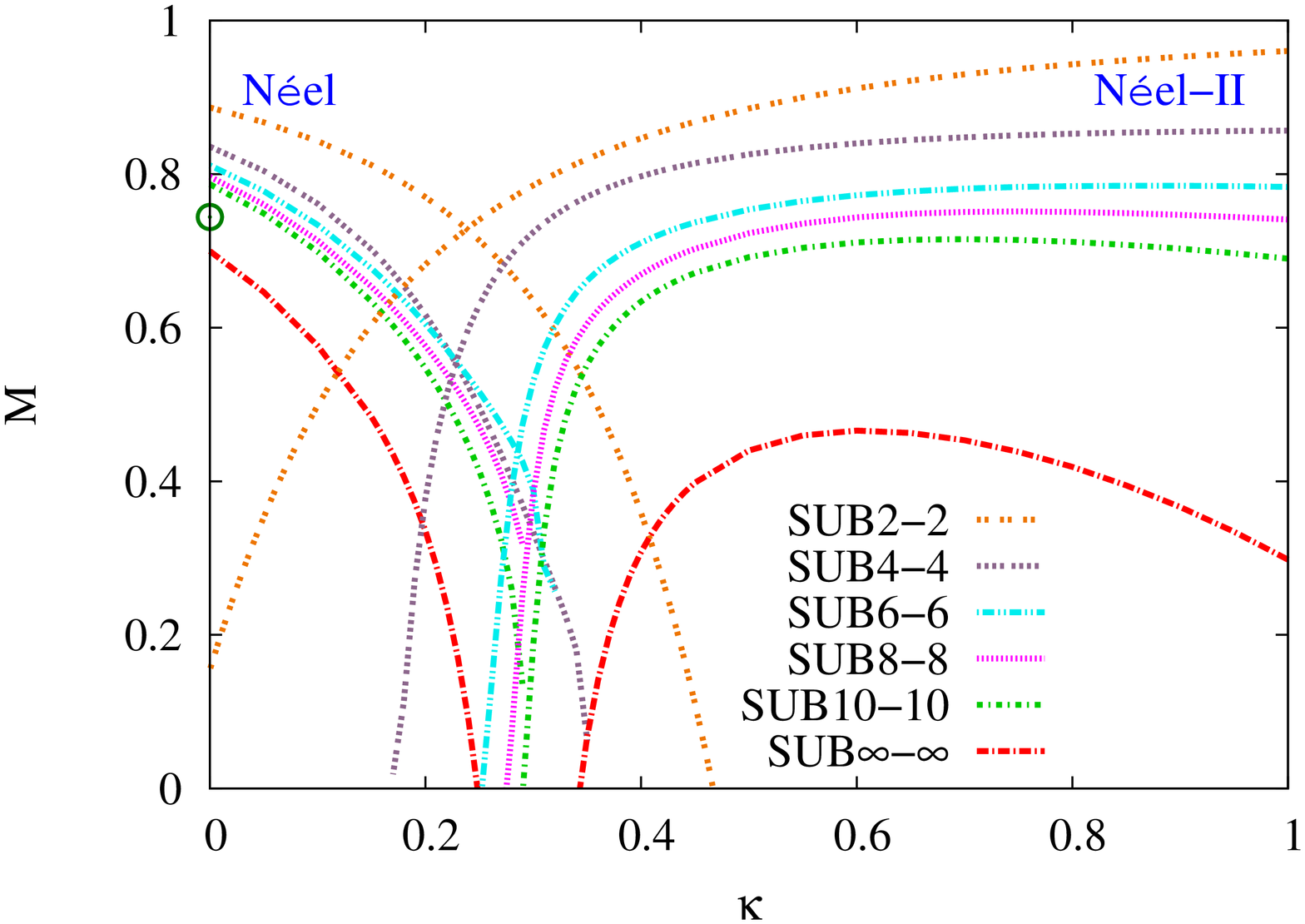}
\caption{(Color online) CCM results for the GS magnetic order parameter
  $M$ versus the frustration parameter $\kappa \equiv J_{2}/J_{1}$, for
  the spin-1 $J_{1}$--$J_{2}$ model on the honeycomb lattice (with $J_{1}>0$).  
  Results based on both the N\'{e}el and N\'{e}el-II states as CCM model states are shown in SUB$n$--$n$
  approximations with $n=\{2,4,6,8,10\}$, together with the corresponding
  SUB$\infty$--$\infty$ extrapolations using Eq.\ (\ref{M_extrapo_frustrated}), with the respective data set $n=\{2,6,10\}$.  Rather than crowd the figure with additional full curves based on (the largely inappropriate) Eq.\ (\ref{M_extrapo_standard}), we show with the circle ($\bigcirc$) symbol the corresponding extrapolated value  based on the N\'{e}el state using Eq.\ (\ref{M_extrapo_standard}), with data set $n=\{2,6,10\}$, for the case $\kappa=0$ only, where this extrapolation scheme {\it is} the appropriate one.}
\label{M_s1}
\end{figure}

More detailed evidence for the regions of stability of the
quasiclassical AFM phases with magnetic LRO, and for any intermediate
phase, can clearly be obtained from the GS magnetic order parameter
$M$ of Eq.\ (\ref{M_eq}).  Thus, in Fig.\ \ref{M_s1} we show our
analogous SUB$n$--$n$ results for $M$ to those shown in Fig.\
\ref{E_s1} for $E/N$.
Firstly, we note that, exactly as expected, the SUB$n$--$n$ sequences
of approximations for $M$ based on both model states converge
appreciably more slowly than for $E/N$.  Secondly, it is very
interesting to note from Fig.\ \ref{M_s1} that many of the SUB$n$--$n$
curves for $M$, particularly all of those based on the N\'{e}el-II
model state, become zero {\it before} the respective termination
point.  These points have been denoted by plus ($+$) symbols on the
corresponding energy curves in Fig.\ \ref{E_s1}, wherein the
respective portions of the curves beyond these points where $M < 0$
are shown by thinner lines to denote that these regions are
unphysical.

A third, more subtle, point can also be seen from Fig.\ \ref{M_s1}.
Thus, over both the entire region of the N\'{e}el-II curves and the
region of the N\'{e}el curves away from the immediate vicinity of the
unfrustrated $\kappa=0$ limit, there is a marked $(4m-2)/4m$
staggering effect for the raw SUB$n$--$n$ results.  For example, for
the N\'{e}el curves, the staggering effect is strong enough that
the two SUB$n$--$n$ curves for $M$ with $n=4,6$ even cross one another
at a value $\kappa \approx 0.2$.  Thus, the two SUB$n$--$n$ sequences
of corresponding values for $M$ at a fixed value of $\kappa$, i.e., for
$n=(4m-2)$ on the one hand and $n=4m$ on the other, tend to converge
quite differently from one another, for all positive integral values
of $m$.  Conversely, each of these sequences separately seems to
converge monotonically.  Such staggered (or non-monotonic) CCM
SUB$n$--$n$ sequences have also been observed in other models.  For
example, for the spin-$\frac{1}{2}$ $J_{1}$--$J_{2}$ model on the
triangular lattice, it has been shown explicitly
\cite{Li:2015_j1j2-triang} that there exists corresponding $(2m-1)/2m$
(i.e., odd/even) staggering in the CCM SUB$n$--$n$ sequences based on
both the 3-sublattice 120$^{\circ}$ N\'{e}el state and the
2-sublattice AFM striped state as model states.  In view of the fact
that the present honeycomb lattice comprises two interlocking
triangular Bravais sublattices, it is probable that the staggering
effects observed in the $J_{1}$--$J_{2}$ models on the honeycomb and
triangular lattices are related, and consistent with one another.

It is interesting to note that a closer inspection of the CCM
SUB$n$--$n$ results for $M$ for the spin-$\frac{1}{2}$ version of the
present honeycomb-lattice $J_{1}$--$J_{2}$ model in Ref.\
\cite{RFB:2013_hcomb_SDVBC} also reveals a similar $(4m-2)/4m$
staggering to that observed here.  In the light of the current
results, one can see that that staggering effect was overlooked there,
but was rather interpreted as showing that the SUB6--6 result was
anomalous by comparison with the SUB$n$--$n$ approximants,
$n=\{8,10,12\}$.

In view of the staggering of the results our extrapolated
SUB$\infty$--$\infty$ results for $M$, which are shown in Fig.\
\ref{M_s1} are based on the SUB$n$--$n$ data set $n=\{2,6,10\}$ as
input to the scheme of Eq.\ (\ref{M_extrapo_frustrated}), which is the
appropriate scheme when $\kappa$ is appreciable, and especially near
any QCPs at which $M$ vanishes.  The corresponding values where
$M \rightarrow 0$ for the extrapolated curves are thus also those
shown in Fig.\ \ref{E_s1} on the extrapolated GS energy per spin
curves by the plus ($+$) symbols.  They hence provide us with our best
estimates so far for the two QCPs, viz., $\kappa_{c_{1}}$ above which
N\'{e}el order melts, and $\kappa_{c_{2}}$ below which N\'{e}el-II
order similarly melts.  Using the extrapolation scheme of Eq.\
(\ref{M_extrapo_frustrated}) and the input SUB$n$--$n$ data sets
$n=\{2,6,10\}$ for the two model states provides the values $\kappa_{c_{1}} \approx 0.248$ and
$\kappa_{c_{2}} \approx 0.343$, with clear evidence for an
intermediate state in the range
$\kappa_{c_{1}} < \kappa < \kappa_{c_{2}}$.

Before we proceed to investigate the intermediate regime further it is
worth pausing for a moment to consider the accuracy of our results
using the necessarily restricted set of SUB$n$--$n$ data points with
$n=\{2,6,10\}$ only.  To do so it is perhaps sufficient to consider
the case $\kappa=0$ only, where the extrapolation scheme of Eq.\
(\ref{M_extrapo_standard}) is the apposite one for $M$.  Precisely in
the unfrustrated limit, $\kappa=0$, the $(4m-2)/4m$ staggering effect,
seen in the raw SUB$n$--$n$ results $M(n)$ for values of $\kappa$
appreciably far from $\kappa=0$, essentially disappears.  Thus,
precisely at $\kappa=0$ we may also use the SUB$n$--$n$ data set
$n=\{4,6,8,10\}$, which, {\it a priori}, should provide a much more
robust result than using the restricted set $n=\{2,6,10\}$.  For the
GS energy per spin we obtain values $E(\kappa=0)/N \approx -1.83063$
using the restricted set $n=\{2,6,10\}$, which may be compared with
the corresponding value $E(\kappa=0)/N \approx -1.83061$ using the set
$n=\{4,6,8,10\}$ (which is also the value quoted in Ref.\
\cite{Bishop:2016_honey_grtSpins}).  Similarly, for the magnetic order
parameter we find, using Eq.\ (\ref{M_extrapo_standard}),
$M(\kappa=0) \approx 0.7441$ using the restricted set $n=\{2,6,10\}$,
and $M(\kappa=0) \approx 0.7412$ \cite{Bishop:2016_honey_grtSpins}
using the set $n=\{4,6,8,10\}$.  Clearly, the very close agreement
between the respective pairs of values lends considerable credence to
our results for $M$ at arbitrary values of $\kappa$, and also to the values for the
QCPs, $\kappa_{c_{1}}$ and $\kappa_{c_{2}}$, so obtained.

\begin{figure}
\includegraphics[width=6.2cm,angle=270]{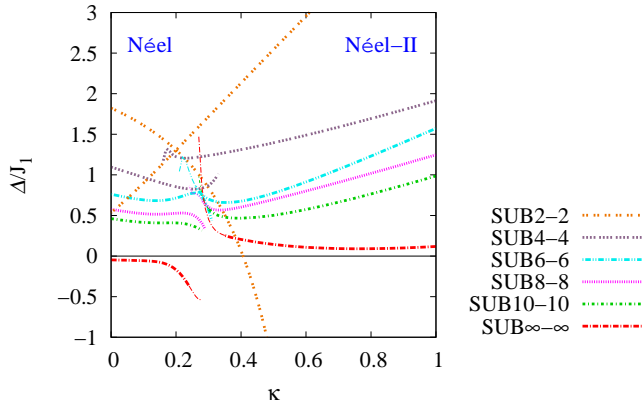}
\caption{(Color online) CCM results for spin gap $\Delta$ (in units of $J_{1}$) versus the frustration parameter $\kappa
  \equiv J_{2}/J_{1}$, for the spin-1 $J_{1}$--$J_{2}$
  model on the honeycomb lattice (with $J_{1}>0$).  Results based on both the N\'{e}el and N\'{e}el-II states as the CCM model
  states are shown in SUB$n$--$n$ approximations with $n=\{2,4,6,8,10\}$, together with the corresponding SUB$\infty$--$\infty$ extrapolation using Eq.\ 
  (\ref{Eq_spin_gap}), with the data sets $n=\{2,6,10\}$.  Those portions of the curves with thinner lines indicate the
  respective unphysical regions where $M<0$.}
\label{E_gap_s1}
\end{figure}

We turn our attention next to the triplet spin gap $\Delta$.  In Fig.\
\ref{E_gap_s1} we show our SUB$n$--$n$ approximants with
$n=\{2,4,6,8,10\}$, again based on both the N\'{e}el and N\'{e}el-II
quasiclassical AFM states as CCM model states.  A similar $(4m-2)/4m$
staggering of the curves to that seen in Fig.\ \ref{M_s1} for the
magnetic order parameter $M$ is also clearly visible in Fig.\
\ref{E_gap_s1} for the corresponding curves for the spin gap $\Delta$.
We thus again show the corresponding SUB$\infty$--$\infty$
extrapolations, based now on Eq.\ (\ref{Eq_spin_gap}) to obtain the
value $d_{0}$, using the two restricted sets of SUB$n$--$n$
approximants with $n=\{2,6,10\}$ as input data.  As is completely to
be expected from states with magnetic LRO, and hence with gapless
Goldstone magnon modes, our results are compatible, within very small
numerical errors, with $\Delta=0$ in the two regions
$\kappa < \kappa_{c_{1}}$ and $\kappa > \kappa_{c_{2}}$ with the
values $\kappa_{c_{1}}$ and $\kappa_{c_{2}}$ as determined above from
the points at which $M \rightarrow 0$.

However, the most striking feature of Fig.\ \ref{E_gap_s1} is the
quite different behavior of the respective curves for $\Delta$ at the
two critical points $\kappa_{c_{1}}$ and $\kappa_{c_{2}}$.  Thus, for
example, near the lower termination point $\kappa_{c_{2}}$ of the
N\'{e}el-II phase all of the SUB$n$--$n$ curves (with $n > 2$), as
well as the SUB$\infty$--$\infty$ extrapolant shown, exhibit a clear
tendency for a gapped state to open up below the QCP at
$\kappa_{c_{2}}$.  By contrast, near the upper termination point
$\kappa_{c_{1}}$ of the N\'{e}el phase there is no such obvious
tendency for a gapped state to appear immediately beyond the QCP at
$\kappa_{c_{1}}$.  The preliminary evidence from our spin gap results
is thus that in the intermediate region
$\kappa_{c_{1}} < \kappa < \kappa_{c_{2}}$ there are (at least) two
different GS phases.  The transition at $\kappa_{c_{1}}$ appears to be
from a N\'{e}el-ordered state to a gapless state, while that at
$\kappa_{c_{2}}$ appears to be between a gapped state and a state with
N\'{e}el-II magnetic LRO.  The precise numerical values for
$\kappa_{c_{1}}$ and $\kappa_{c_{2}}$ are difficult to determine from
the results for $\Delta$ but, from Fig.\ \ref{E_gap_s1}, they are
clearly compatible with those found earlier from the results for the
order parameter $M$.

\begin{figure}
\includegraphics[angle=270,width=7.7cm]{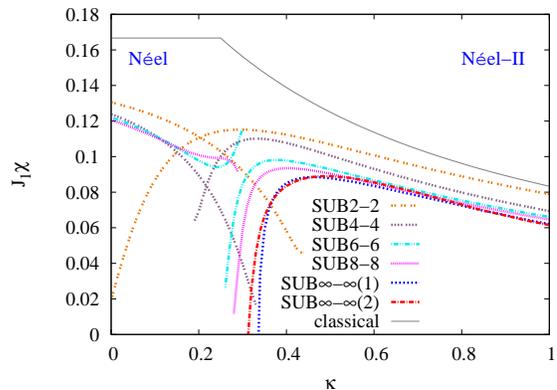}
\caption{(Color online) CCM results for the zero-field magnetic
  susceptibility $\chi$ (in units of $J_{1}^{-1}$, and where the gyromagnetic ratio $g\mu_{B}/\hbar=1$) versus the frustration parameter $\kappa \equiv
  J_{2}/J_{1}$, for the spin-1
  $J_{1}$--$J_{2}$ model on the honeycomb lattice (with
  $J_{1}>0$).  Results based on both the canted N\'{e}el and
  canted N\'{e}el-II states as CCM model states are shown in SUB$n$--$n$ approximations with $n=\{2,4,6,8\}$, together with the corresponding SUB$\infty$--$\infty(1)$ and SUB$\infty$--$\infty(2)$ N\'{e}el-II extrapolations using Eqs.\ (\ref{Eq_exponFit})
  and (\ref{Eq_X}), respectively, with the data set $n=\{4,6,8\}$.  The
  classical results from Eqs.\ (\ref{chi_neel_classical}) and
  (\ref{chi_neel2_classical}) are also shown for comparison.}
\label{M_Xcpty_s1}
\end{figure}

In order to try to corroborate the above findings from the spin gap
$\Delta$, we now turn our attention to the zero-field magnetic
susceptibility $\chi$.  To calculate $\chi$ within the CCM we now use
the canted N\'{e}el and N\'{e}el-II states shown in Figs.\
\ref{pattern_M_ExtField}(a) and \ref{pattern_M_ExtField}(b),
respectively, as model states.  In view of the lower symmetries of
these states in comparison with their zero-field counterparts in
Figs.\ \ref{model_pattern}(a) and \ref{model_pattern}(b),
respectively, the numbers $N_{f}(n)$ of fundamental CCM configurations
at a given SUB$n$--$n$ level of approximation are considerably greater
for $\chi$ than those for the GS parameters $E/N$ and $M$.  For example,
for the canted N\'{e}el (canted N\'{e}el-II) model state for the
present spin-1 model, we have $N_{f}(8)=59517$ $(177358)$ at the
respective SUB$8$--$8$ levels.  Thus, whereas for the zero-field
N\'{e}el and N\'{e}el-II model states we are able to perform
SUB$n$--$n$ calculations for the parameters $E/N$ and $M$ with
$n \leq 10$, for the corresponding canted states we are only able to
perform SUB$n$--$n$ calculations for $\chi$ with $n \leq 8$.  Figure
\ref{M_Xcpty_s1} displays the corresponding results obtained.

Once again we see that there is a marked difference in the behavior of
the curves near the two QCPs at $\kappa_{c_{1}}$ and $\kappa_{c_{2}}$.
Firstly, on the N\'{e}el-II side, each of the SUB$n$--$n$ curves
exhibits a sharp downturn near its respective termination point.
Although there appears to be some possible $(4m-2)/4m$ staggering of
the results around the region close to $\kappa_{c_{2}}$, it is both
much less marked for higher values of $\kappa$ and much less evident
than for the corresponding results for either $M$ or $\Delta$.  Thus,
in Fig.\ \ref{M_Xcpty_s1}, we also show for the N\'{e}el-II results
extrapolations using the data set $n=\{4,6,8\}$.  For comparison
purposes we show separate extrapolations based on each of Eqs.\
(\ref{Eq_X}) and (\ref{Eq_exponFit}).  The two extrapolations are
clearly in very close agreement with one another except in a very
small region near the QCP at $\kappa_{c_{2}}$, which is where the
exponent $\nu$ in the fit of the form of Eq.\ (\ref{Eq_X}) also
differs appreciably from the value 1, which is appropriate for Eq.\
(\ref{Eq_exponFit}).  Thus, in the critical region where $\chi$
becomes small the SUB$\infty$--$\infty(1)$ fit based on Eq.\
(\ref{Eq_exponFit}) must clearly be preferred, and the value of
$\kappa$ at which $\chi \rightarrow 0$ for this fit is
$\kappa \approx 0.337$.  This value is in excellent agreement with the
corresponding value $\kappa_{c_{2}} \approx 0.343$ at which
$M^{{\rm N\acute{e}el-II}} \rightarrow 0$ from Fig.\ \ref{M_s1} (Note
that even the corresponding value of $\kappa \approx 0.313$ obtained
from the less justified SUB$\infty$--$\infty(2)$ fit in Fig.\
\ref{M_Xcpty_s1} is remarkably close to the value of $\kappa_{c_{2}}$
obtained from the vanishing of the N\'{e}el-II LRO.)  As we have noted
before, the vanishing of $\chi$ in a quantum spin model is a very
clear signal of a transition at that point to a gapped state
\cite{Mila:2000_M-Xcpty_spinGap,Bernu:2015_M-Xcpty_spinGap}.  Hence,
the results for $\chi$ lend weight to our corresponding results for
$\Delta$ that the transition at $\kappa_{c_{2}} \approx 0.34$ is
between a state with N\'{e}el-II magnetic LRO and a gapped state.

Secondly, by contrast with the N\'{e}el-II results for $\chi$, the
N\'{e}el results behave markedly differently near their respective
termination points in Fig.\ \ref{M_Xcpty_s1}.
Thus, on the N\'{e}el side, whereas the lower-order SUB$n$--$n$ curves
with $n=\{2,4\}$ exhibit a downturn near their termination points,
this feature disappears for the higher-order counterparts,
$n=\{6,8\}$.  Hence, the results shown in Fig.\ \ref{M_Xcpty_s1} for
the N\'{e}el side cannot readily be extrapolated.  What is indubitably
clear, however, from the most accurate results with the higher values
of the SUB$n$--$n$ truncation index $n$, is that there is no tendency
at all for $\chi$ to vanish as the N\'{e}el order melts, thereby
providing further evidence for the QPT at $\kappa_{c_{1}}$ being to a
gapless state.  Furthermore, the SUB$n$--$n$ results with $n=\{6,8\}$,
based on the N\'{e}el state as CCM model state, are completely
compatible with the same value $\kappa_{c_{1}} \approx 0.25$ as found
from the vanishing of the N\'{e}el magnetic LRO parameter $M$.

\begin{figure}
\includegraphics[angle=270,width=7.7cm]{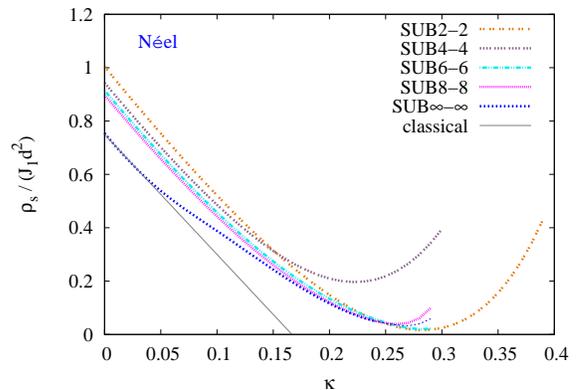}
\caption{(Color online) CCM results for the spin stiffness coefficient
  $\rho_{s}$ (in units of $J_{1}d^{2}$) versus the frustration
  parameter $\kappa \equiv J_{2}/J_{1}$, for the spin-1
  $J_{1}$--$J_{2}$ model on the honeycomb lattice (with $J_{1}>0$).
  Results based on the N\'{e}el state as CCM model state are shown in
  SUB$n$--$n$ approximations with $n=\{2,4,6,8\}$, together with the
  corresponding SUB$\infty$--$\infty$ extrapolation using Eq.\
  (\ref{Eq_exponFit}), with the data set $n=\{4,6,8\}$.  That
  portion of the curve with a thinner line indicates the respective
  unphysical region where $M<0$.  The classical result from Eq.\
  (\ref{sStiff_neel_classical}), $s=1$, is also shown for comparison.}
\label{spinstiff_s1}
\end{figure}

For further evidence of the QPT at $\kappa_{c_{1}}$ we also show in
Fig.\ \ref{spinstiff_s1} our CCM results for the spin stiffness
coefficient $\rho_{s}$ based on the twisted N\'{e}el state of Fig.\
\ref{pattern_sStiff} as our choice of model state $|\Phi\rangle$.
Once again, just as for the magnetic susceptibility, $\chi$, the
reduced symmetry of the twisted model state increases the number
$N_{f}(n)$ of fundamental CCM configurations at a given SUB$n$--$n$
level from that using its untwisted GS counterpart.  For example, for
the twisted N\'{e}el model state for the present spin-1
honeycomb-lattice model, we have $N_{f}(8) = 352515$ at the SUB8--8
level.  Just as for $\chi$, we are thus only able to perform
SUB$n$--$n$ calculations for $\rho_{s}$ with $n \leq 8$.  The results
are displayed in Fig.\ \ref{spinstiff_s1}.

While the SUB$n$--$n$ data points fit well to the extrapolation scheme
of Eq.\ (\ref{Eq_sstiff}) at or very near to the unfrustrated limit
$\kappa=0$, the curves again exhibit crossings associated with the
aforementioned $(4m-2)/4m$ staggering.  This makes extrapolation
somewhat problematic.  However, in Fig.\ \ref{spinstiff_s1} we again
use the unbiased scheme of Eq.\ (\ref{Eq_exponFit}) with the data set
$n=\{4,6,8\}$.  The corresponding SUB$\infty$--$\infty$ curve is
deemed to be reliable except in a small region around
$\kappa \approx 0.25$ where the SUB6--6 and SUB8-8 curves cross, which
is also where the order parameter $M$ vanishes.  Nevertheless, it is
apparent from both the SUB$\infty$--$\infty$ curve and the
higher-order raw SUB$n$--$n$ curves with $n=\{6,8\}$ that the exact
$\rho^{{\rm N\acute{e}el}}_{s}(\kappa)$ curve is likely to approach
zero, with a zero slope, at a critical point very close to
$\kappa = 0.25$.  Again, this value is in keeping with the QCP
$\kappa_{c_{1}}$ obtained from the vanishing of the N\'{e}el magnetic
order parameter $M^{{\rm N\acute{e}el}}$.

All of our results so far thus point towards the existence of both a
gapless and a gapped state in the intermediate regime,
$\kappa_{c_{1}} < \kappa < \kappa_{c_{2}}$, with the transition at
$\kappa_{c_{1}}$ being from a state with N\'{e}el magnetic LRO to a
gapless paramagnetic state, and that at $\kappa_{c_{2}}$ being from a
gapped paramagnetic state to one with N\'{e}el-II magnetic LRO.  The
most important open questions are (i) what is the nature of each of
these paramagnetic states, and (ii) what is the critical value $\kappa^{i}_{c}$ of
the intermediate QCP between the gapless and gapped paramagnetic
states?

\begin{figure*}[t]
\begin{center}
\mbox{
\raisebox{-8.6cm}{
\subfigure[]{\includegraphics[width=2.5cm,height=2.5cm]{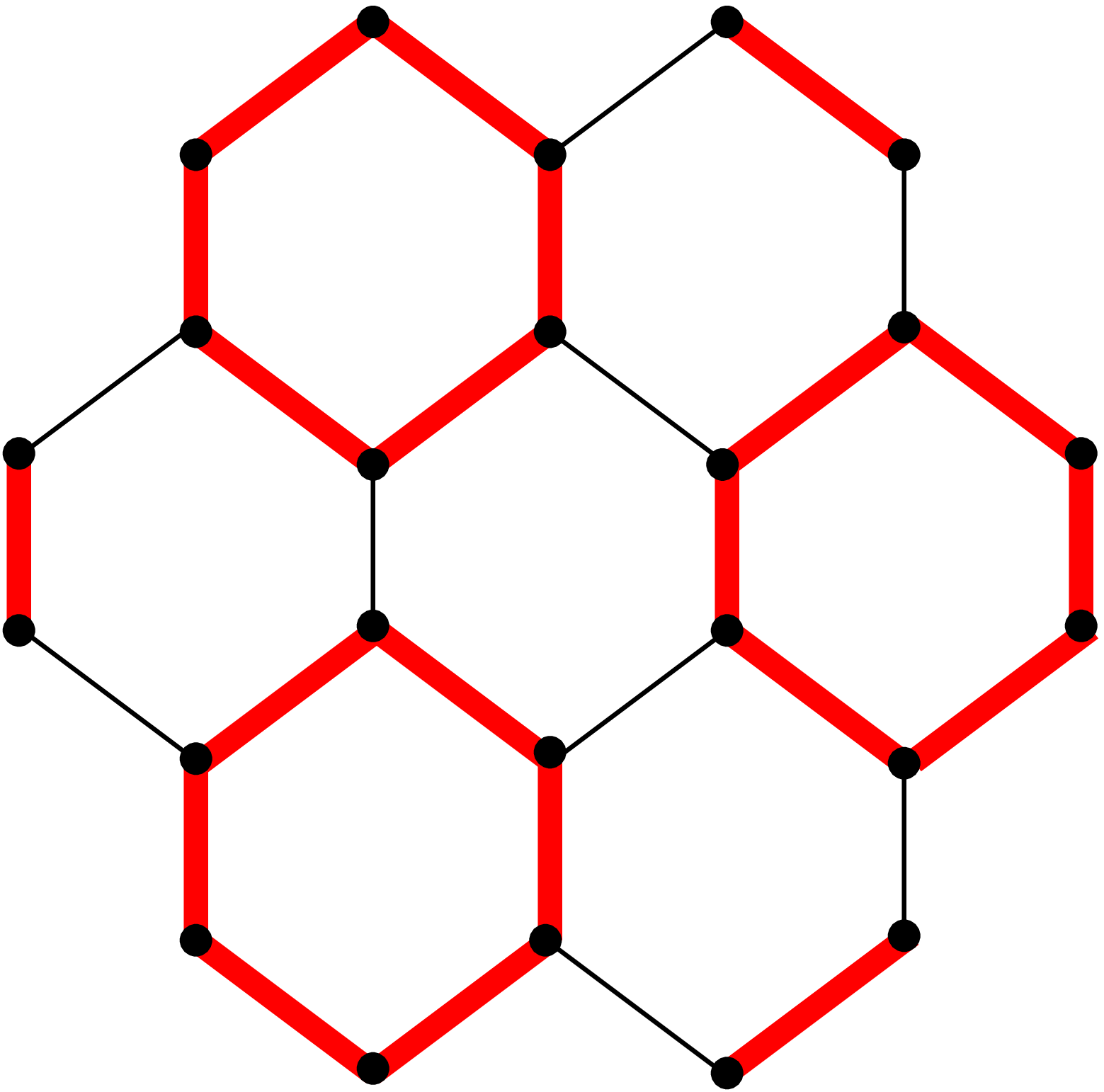}}
}
\hspace{0.5cm}\subfigure[]{\includegraphics[width=9cm,height=12cm,angle=270]{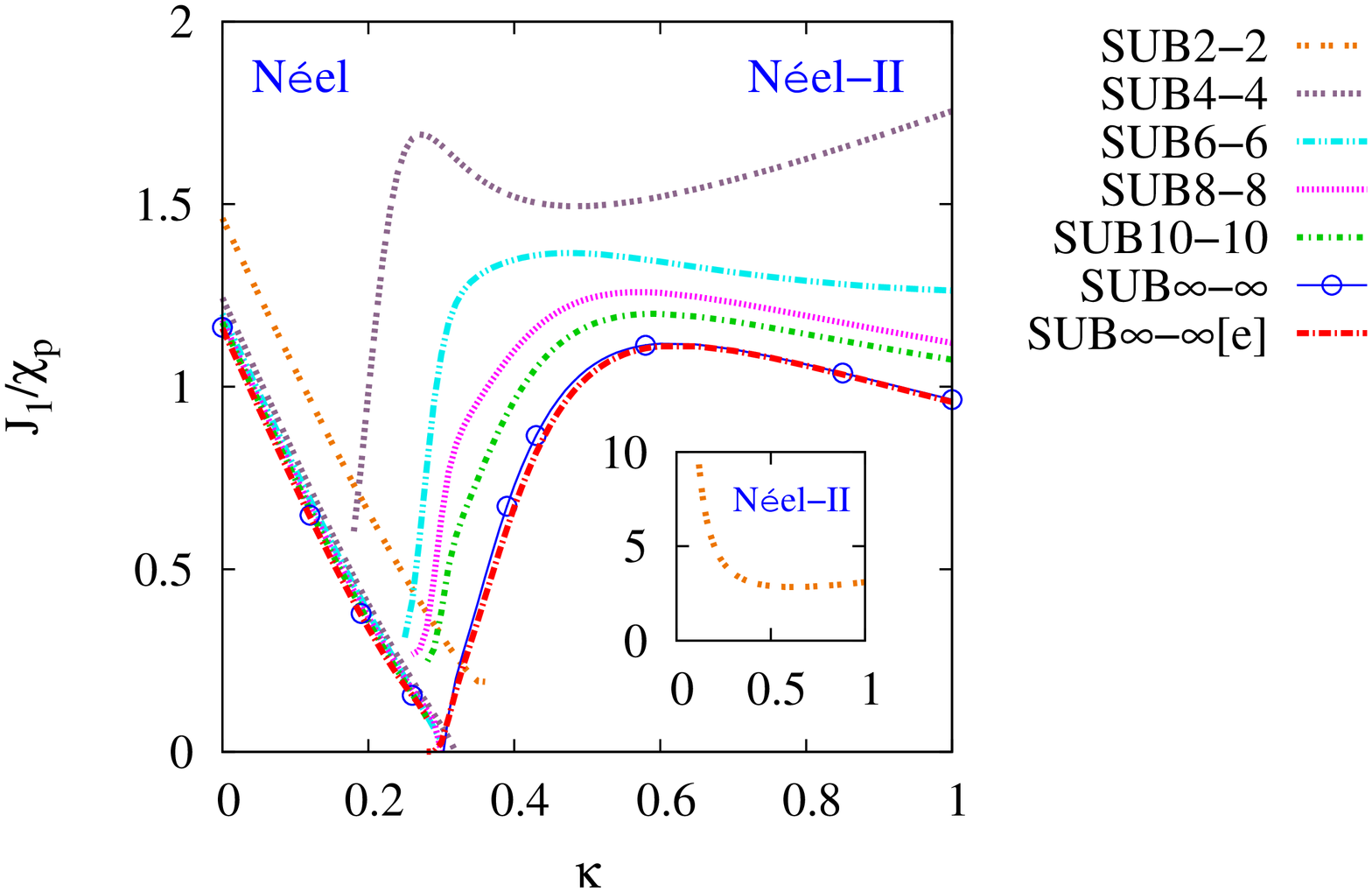}}
}
\caption{(Color online) (a) The field $F=\delta \hat{O}_{p}$ for
  the plaquette susceptibility $\chi_{p}$.  Thick (red) and thin
  (black) lines correspond respectively to strengthened and weakened
  NN exchange couplings, where
  $\hat{O}_{p} = \sum_{\langle i,j \rangle} a_{ij}
  \mathbf{s}_{i}\cdot\mathbf{s}_{j}$,
  and the sum runs over all NN bonds, with $a_{ij}=+1$ and $-1$ for
  thick (red) and thin (black) lines respectively.  (b) CCM results
  for the inverse plaquette susceptibility, $1/\chi_{p}$, (in units of
  $J^{-1}_{1}$) versus the frustration parameter
  $\kappa \equiv J_{2}/J_{1}$, for the spin-1 $J_{1}$--$J_{2}$ model
  on the honeycomb lattice (with $J_{1}>0$).  Results based on both
  the N\'{e}el and N\'{e}el-II states as CCM model states are shown
  in SUB$n$--$n$ approximations with $n=\{2,4,6,8,10\}$, together with
  the corresponding SUB$\infty$--$\infty[e]$ and SUB$\infty$--$\infty$
  extrapolations using Eqs.\ (\ref{ExponFit_Xf}) and (\ref{Extrapo-asE_inv-chi}),
  respectively, with the data sets $n=\{2,6,10\}$.}
\label{PVBC_s1}
\end{center}
\end{figure*}    

Clear candidates for the gapped state in the intermediate regime are
magnetically disordered VBC states, evidence for which we may now also
investigate within the CCM formalism.  To that end we introduce a
generalized susceptibility $\chi_{F}$, which is designed to measure
the (linear) response of our system to an imposed (infinitesimal)
external field, described by the operator $F$.  In order to do so we
thus add to our Hamiltonian $H$ of Eq.\ (\ref{Eq_H}) an extra field
term $F = \delta\hat{O}_{F}$, where $\hat{O}_{F}$ is an operator
that breaks some symmetry of $H$, and $\delta$ is simply the
(infinitesimal) strength parameter.  We will choose $F$ here to
represent various forms of VBC order, as we describe more fully below.
The energy per spin, $E(\delta)/N \equiv e(\delta)$ is then
calculated, using the same CCM technology as before and based on
either the previous N\'{e}el or N\'{e}el-II model states, for the
(infinitesimally) perturbed Hamiltonian $H + F$.  The generalized
susceptibility is then defined as
\begin{equation}
\chi_{F} \equiv -\left. \frac{\partial^2{e(\delta)}}{\partial {\delta}^2}  \label{Eq_X_partial}
\right|_{\delta=0}\,,
\end{equation}
and the energy per spin,
\begin{equation}
e(\delta) =
e_{0}-\frac{1}{2}\chi_{F}\delta^{2}+O(\delta^{4})\,,   \label{Eq_E-per-spin_Xf}
\end{equation}
is a maximum at $\delta=0$ for $\chi_{F} > 0$.  A clear signal of the
system becoming unstable against the perturbing field $F$ is then the
divergence of $\chi_{F}$ or, equivalently, the finding that
$\chi^{-1}_{F}$ become zero (and then possibly changes sign) at some
critical value of the frustration parameter $\kappa$.

Clearly our computed CCM SUB$n$--$n$ estimates for any such
susceptibility $\chi_{F}$ need, as usual, to be extrapolated to be
SUB$\infty$--$\infty$ limit.  There are various ways in practice to do
so.  The most direct way is obviously to extrapolated first our
SUB$n$--$n$ results for the perturbed energy per spin,
$e^{(n)}(\delta)$, before using them to calculate $\chi_{F}$ via Eq.\
(\ref{Eq_X_partial}) or, in practice, Eq.\ (\ref{Eq_E-per-spin_Xf}).  A
similar scheme to Eq.\ (\ref{extrapo_E}) for the GS energy, viz., one
with a leading power $1/n^{2}$, typically works well except in regions
near to a QCP.  Since we are especially interested in using $\chi_{F}$
precisely in such regions, we prefer
\cite{Li:2013_chevron,Bishop:2013_crossStripe} to use an unbiased
scheme of the form of Eq.\ (\ref{Eq_exponFit}), namely,
\begin{equation}
e^{(n)}(\delta) =
e_{0}(\delta)+e_{1}(\delta)n^{-\nu}\,,   \label{ExponFit_Xf}
\end{equation}
in which the leading exponent $\nu$ is a fitting parameter along with
the linear parameters $e_{0}(\delta)$ and $e_{1}(\delta)$.  In this
way the extrapolated, so-called SUB$\infty$--$\infty[e]$, value
$e_{0}(\delta)$ is then used to calculate $\chi_{F}$ via Eq.\
(\ref{Eq_E-per-spin_Xf}).  As an alternative method, a corresponding
direct extrapolation scheme for the SUB$n$--$n$ estimates,
$\chi^{-1}_{F}(n)$, of the inverse susceptibility, of the form
\begin{equation}
\chi^{-1}_{F}(n) = x_{0}+x_{1}n^{-2}+x_{2}n^{-4}\,,  \label{Extrapo-asE_inv-chi}
\end{equation}
has also been found previously (and see, e.g., Refs.\
\cite{DJJF:2011_honeycomb,Li:2013_chevron} to give consistently
reliable results, again with the possible exception of critical
regions where $\chi^{-1}_{F}$ becomes small or zero.  In the results
presented below we will use both the schemes of Eqs.\
(\ref{ExponFit_Xf}) and (\ref{Extrapo-asE_inv-chi}) in our plots
of $\chi^{-1}_{F}$ as a function of $\kappa$, in order to test the
reliability of the extrapolation procedures.

\begin{figure*}[t]
\begin{center}
\mbox{
\raisebox{-8.6cm}{
\subfigure[]{\includegraphics[width=2.5cm,height=2.5cm]{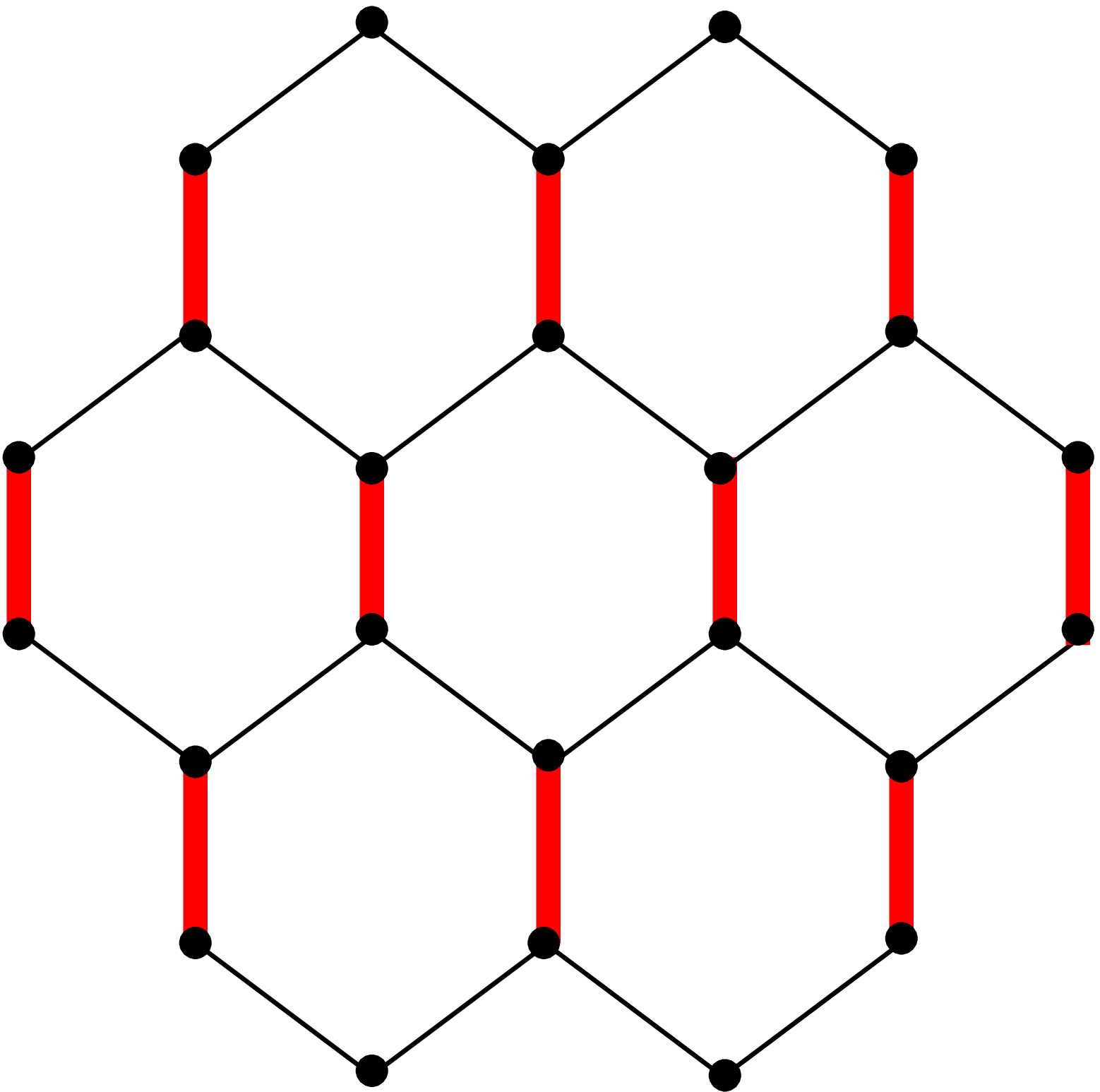}}
}
\hspace{0.5cm}\subfigure[]{\includegraphics[width=9cm,height=12cm,angle=270]{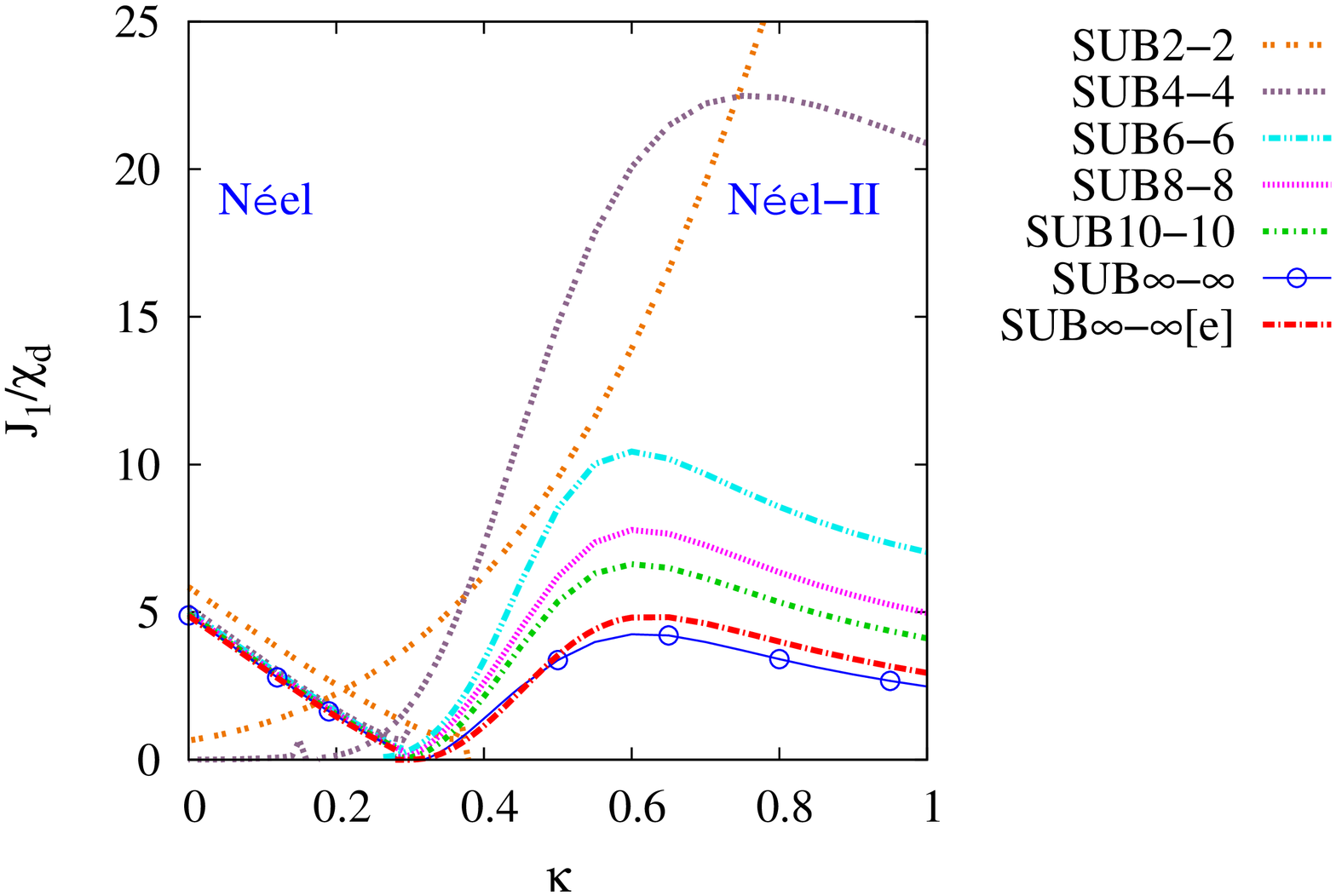}}
}
\caption{(Colour online) (a) The field $F = \delta \hat{O}_{d}$ for the staggered dimer susceptibility,
  $\chi_{d}$.  Thick (red) and thin (black) lines correspond
  respectively to strengthened and unaltered NN exchange couplings,
  where $\hat{O}_{d} = \sum_{\langle i,j \rangle} a_{ij}
  \mathbf{s}_{i}\cdot\mathbf{s}_{j}$, and the sum runs over all NN
  bonds, with $a_{ij}=+1$ and 0 for thick (red) lines and thin
  (black) lines respectively.  (b) CCM results for the inverse
  staggered dimer susceptibility, $1/\chi_{d}$, (in units of $J^{-1}_{1}$) versus the frustration parameter, $\kappa \equiv J_{2}/J_{1}$, for the
  spin-1 $J_{1}$--$J_{2}$ model on the honeycomb lattice (with $J_{1}>0$).  Results based on both the N\'{e}el and N\'{e}el-II states as CCM model states are shown in SUB$n$--$n$ approximations with $n=\{2,4,6,8,10\}$, together with
  the corresponding SUB$\infty$--$\infty[e]$ and SUB$\infty$--$\infty$
  extrapolations using Eqs.\ (\ref{ExponFit_Xf}) and (\ref{Extrapo-asE_inv-chi}),
  respectively, with the data sets $n=\{2,6,10\}$.}
\label{SDVBC_s1}
\end{center}
\end{figure*}   

Since strong evidence for a state with PVBC order has been found within the
corresponding region $\kappa_{c_{1}} < \kappa < \kappa_{c_{2}}$ of the
spin-$\frac{1}{2}$ version of the present $J_{1}$--$J_{2}$ model on the
honeycomb lattice, as reviewed in Sec.\ \ref{model_sec}, it is perhaps
natural to choose first the perturbing operator $F$ to promote PVBC
order, as illustrated in Fig.\ \ref{PVBC_s1}(a).
It clearly breaks the translational symmetry of the system.  Despite
the reduced symmetry of the perturbed Hamiltonian, $H + F$, we are
still able to perform SUB$n$--$n$ calculations for the corresponding
plaquette susceptibility $\chi_{p}$ for values of the truncation
parameter $n \leq 10$, using both the N\'{e}el and N\'{e}el-II
quasiclassical AFM states as our CCM model states.  The corresponding
SUB$n$--$n$ results for $\chi^{-1}_{p}$ as a function of $\kappa$ are
shown in Fig.\ \ref{PVBC_s1}(b).  They clearly also demonstrate a
$(4m-2)/4m$ staggering of the sort seen previously for other GS
parameters.  Hence, in Fig.\ \ref{PVBC_s1}(b) we also show
extrapolated results using our SUB$n$--$n$ data with $n=\{2,6,10\}$
and both schemes of Eqs.\ (\ref{ExponFit_Xf}) and
(\ref{Extrapo-asE_inv-chi}).

It is evident that both schemes give results that are in excellent
agreement with one another.  Furthermore, the strong collective
evidence is that $\chi^{-1}_{p}$ goes to zero only at a single point,
$\kappa \approx 0.30$, on {\it both} the N\'{e}el and N\'{e}el-II
sides.  On the N\'{e}el side the SUB10--10 data actually terminates
slightly below this value, but the two SUB$\infty$--$\infty$ curves
shown both clearly would reach $\chi^{-1}_{p} = 0$ at the value of
$\kappa \approx 0.30$ when extrapolated the small extra necessary
distance.  Clearly, on the N\'{e}el side $\chi^{-1}_{p}$ does not
vanish at the point $\kappa_{c_{1}} \approx 0.25$ at which N\'{e}el
LRO actually vanishes, as measured by
$M^{{\rm N\acute{e}el}} \rightarrow 0$.  Similarly, on the N\'{e}el-II
side the SUB$\infty$--$\infty$ curve based on Eq.\
(\ref{Extrapo-asE_inv-chi}) reaches the value zero at the value
$\kappa \approx 0.303$.  While the N\'{e}el-II
SUB$\infty$--$\infty[e]$ curve based on Eq.\ (\ref{ExponFit_Xf}) shows
a very slight tendency to flatten near the point where $\chi^{-1}_{p}$
goes to zero, it too reaches zero at $\kappa \approx 0.29$.

All of these results are in accord with our previous findings that (a)
the N\'{e}el order vanishes at a QCP $\kappa_{c_{1}}$ at which a
magnetically disordered gapless state appears; (b) this gapless state
itself disappears at a QCP $\kappa^{i}_{c}$ at which a magnetically
disordered gapped state appears; and (c) this gapped state disappears
at a QCP $\kappa_{c_{2}}$, above which a magnetically ordered state
with N\'{e}el-II LRO appears.  The additional evidence from all of the
PVBC results shown is that $\chi^{-1}_{p}$ vanishes only at a single
point, $\kappa^{i}_{c} \approx 0.30$ at which a gapped state appears,
but that this gapped state, which is expected to be stable over the
region $\kappa^{i}_{c} < \kappa < \kappa_{c_{2}}$, is not one with
PVBC order, since $1/\chi^{{\rm N\acute{e}el-II}}_{p}$ shows
essentially no tendency to vanish in this interval.

Another state with VBC order that has been associated with the
spin-$\frac{1}{2}$ version of the present model is the SDVBC (or
lattice nematic) state illustrated in Fig.\ \ref{SDVBC_s1}(a).
It is formed from the N\'{e}el-II state by replacing all of the
parallel NN spin pairs by spin-singlet dimers.  Just like the
N\'{e}el-II state, so does the SDVBC state break the lattice
rotational symmetry.  Once again, we may test for the susceptibility
of our system to form a state with SDVBC order by now choosing the
perturbing operator to promote SDVBC order,
$F \rightarrow \delta O_{d}$, as illustrated in Fig.\
\ref{SDVBC_s1}(a).  Again, we are able to perform SUB$n$--$n$
calculations for the corresponding staggered dimer susceptibility
$\chi_{d}$, for values of the truncation parameter $n \leq 10$, and
with the CCM model state chosen as either the N\'{e}el or the
N\'{e}el-II state, despite the reduced symmetry of the perturbed
Hamiltonian, $H + F$.  The corresponding SUB$n$--$n$ results for
$\chi^{-1}_{d}$ as a function of $\kappa$ are shown in Fig.\
\ref{SDVBC_s1}(b).  Unsurprisingly by now, they also illustrate a
$(4m-2)/4m$ staggering effect, and hence in Fig.\ \ref{SDVBC_s1}(b)
the two sets of extrapolations, based on Eqs.\ (\ref{ExponFit_Xf}) and
(\ref{Extrapo-asE_inv-chi}), are shown based on the SUB$n$--$n$ data
sets with $n=\{2,6,10\}$.

Just as for the corresponding extrapolated PVBC results in Fig.\
\ref{PVBC_s1}(b), so now do both extrapolation schemes for the SDVBC
results, agree very well with one another.  The SDVBC for
$\chi^{-1}_{d}$ results based on the N\'{e}el model state are
completely consistent and analogous with the corresponding PVBC
results for $\chi^{-1}_{p}$.  Again, both the SUB$\infty$--$\infty$
and SUB$\infty$--$\infty[e]$ curves shown would reach
$\chi^{-1}_{d}=0$ at the value $\kappa \approx 0.30$ when extrapolated
the small extra needed amount.  However, by contrast, the SDVBC
results based on the N\'{e}el-II state for $\chi^{-1}_{d}$ are now
qualitatively different from the corresponding PVBC results for
$\chi^{-1}_{p}$ on the N\'{e}el-II side.  Thus, all of the SDVBC
results for $\chi^{-1}_{d}$ based on the N\'{e}el-II model state, both
the ``raw'' SUB$n$--$n$ curves and the two extrapolations shown,
exhibit a clear tendency to flatten near the critical point where they
become zero, and then remain zero over a finite range of values of
$\kappa$ below the corresponding critical value.  Thus, for example,
the SUB$\infty$--$\infty$ curve for
$1/\chi^{{\rm N\acute{e}el-II}}_{d}$, based on Eq.\
(\ref{Extrapo-asE_inv-chi}), touches zero at a value
$\kappa \approx 0.32$ with a slope very close to zero, while the
corresponding SUB$\infty$--$\infty[e]$ curve, based on Eq.\
(\ref{ExponFit_Xf}), is clearly zero within small numerical
uncertainties over a range of values
$0.29 \lesssim \kappa \lesssim 0.33$.  Thus we may now identify the
gapped state as likely having SDVBC order, and the range of values of
$\kappa$ for which $\chi^{-1}_{d}=0$ as being the range
$\kappa^{i}_{c} < \kappa < \kappa_{c_{2}}$.  From both the PVBC and
SDVBC results, we estimate $\kappa^{i}_{c} \approx 0.30$, while the
SDVBC results for $1/\chi^{{\rm N\acute{e}el-II}}_{d}$ yield the estimate
$\kappa_{c_{2}} \approx 0.33$.  This latter value is in complete
agreement with the corresponding estimate,
$\kappa_{c_{2}} \approx 0.34$, at which N\'{e}el-II LRO vanishes, as
measured by the point where $M^{{\rm N\acute{e}el-II}} \rightarrow 0$.
Clearly, the latter value is intrinsically more accurate, however, than
that obtained from the point where
$1/\chi^{{\rm N\acute{e}el-II}} \rightarrow 0$, due to the totally
different slopes of the curves at their respective (vanishing)
critical points.

In the concluding section we now summarize and discuss our results.

\section{DISCUSSION AND CONCLUSIONS}
\label{summary_sec}
To summarize, we have applied the CCM in this paper, at high orders
of approximation, to the spin-1 $J_{1}$--$J_{2}$ Heisenberg
antiferromagnet on the honeycomb lattice, in the case of AFM NN bonds
($J_{1}>0$) and AFM NNN bonds ($J_{2} \equiv \kappa J_{1}>0$), in the
range $0 \leq \kappa \leq 1$ of the frustration parameter.  In
particular, our aim has been to present a comprehensive analysis of
the GS ($T=0$) quantum phase diagram of the model.  To that end we
have calculated the GS energy, the magnetic order parameter, the
zero-field transverse magnetic susceptibility, and the spin stiffness
coefficient for two quasiclassical AFM states with N\'{e}el and
N\'{e}el-II forms of LRO.  We have also calculated their generalized
susceptibilities against the formation of two forms of VBC order.  A
distinct advantage of the CCM, in which all of these quantities are
calculated within a unified framework, is that all of the calculations
are anchored from the outset in the thermodynamic limit of an infinite
lattice ($N \rightarrow \infty$).  Unlike many alternate accurate
techniques (e.g., the DMRG method), the CCM thereby obviates the need
for any finite-size scaling or extrapolation to the bulk limit, which
is often the step that is most uncontrolled in practice.

Nevertheless, it is perhaps worthwhile at this point to reflect again
on two key aspects of the CCM that are inherent to it, namely (a) the
precise role of the (reference or) model state, and (b) the fact that
while the method is certainly (bi-)variational, the lack of manifest
Hermiticity between the parametrized corresponding bra and ket states
implies that the method does not provide strict upper bounds to the GS
energy.  Thus, firstly, while the formal role of $|\Phi\rangle$ as a
cyclic vector with respect to an appropriate set of mutually
commuting, multiconfigurational creation operators $\{C^{+}_{I}\}$ has
been fully expounded in Sec.\ \ref{ccm_sec}, one may legitimately
enquire as to whether the results obtained in practice are truly
independent of the choice of model state, or whether some inherent
bias remains.  In other words, are the physical scenarios implied by
the zero-temperature, quantum phase diagram that we have obtained by
the CCM, independent of the choices of model state?

While it is perhaps difficult to be absolutely, categorically
affirmative on this point, in practice the answer has always been
found to be yes.  Thus, for example, if one considers the N\'{e}el
state as a reference state for a point in the phase diagram actually
characterized by N\'{e}el-II order (say, for $\kappa=0.4$), one will
find, at a given SUB$n$--$n$ level of approximation, either that no
solution exists or that, if one does, the corresponding N\'{e}el order
vanishes (i.e., $M \leq 0$) at either the given SUB$n$--$n$ level or
the suitably extrapolated SUB$\infty$--$\infty$ limit, as Fig.\
\ref{E_s1} clearly shows for the present model.  Results from {\it
  many} other models confirm these findings, often quite dramatically.
For example, CCM treatments of spin-$\frac{1}{2}$ $J_{1}$--$J_{2}$ models on 
both the chevron-square lattice \cite{Li:2013_chevron} and the checkerboard lattice
\cite{Bishop:2012_checkerboard} demonstrate instances where one may find SUB$n$--$n$ solutions
(for all values of $n$) using a ``wrong'' model state over extended
ranges of values of $\kappa \equiv J_{2}/J_{1}$, but for all of which
the extrapolated order parameter vanishes ($M=0$).  In both cases VBC
states have been shown to be the actual, stable GS phases.

Secondly, we reiterate that for the reasons already described, neither
our SUB$n$--$n$ estimates for the GS energy nor the
SUB$\infty$--$\infty$ extrapolation from Eq.\ (\ref{extrapo_E}) are
guaranteed to be strict upper bounds.  While our final
SUB$\infty$--$\infty$ result certainly does depend on the functional
form of the extrapolation, we have tested that the form of Eq.\
(\ref{extrapo_E}) is the appropriate one both in the present case and for
{\it many} other models, by first performing an unbiased fit of the
form of Eq.\ (\ref{Eq_exponFit}) to show that the leading exponent is
accurately given by $\nu=2$.  It is worth noting too that even for
alternative variational calculations, where strict upper bounds to the
GS energy {\it are} obtained on finite-size lattice clusters, this
bound may be lost when extrapolating to the thermodynamic
($N \rightarrow \infty$) limit.

For the usual variational approaches (i.e., in explicitly Hermitian
schemes) the GS energy values constitute a rather well-defined figure
of merit to establish which approach is more accurate.  Thus, in order
to avoid the finite-size extrapolation problem, comparison is usually
made for a given finite-sized lattice.  In this context we note that
while in principle the CCM {\it can} be implemented for finite-size
clusters, in practice it is seldom done, both since the method does
not provide energy upper bounds and since it {\it can} be applied
directly in the thermodynamic limit, which is usually of primary
interest.  Furthermore, as we point out below, results obtained only
on finite lattices (e.g., on finite cylinders for DMRG calculations)
can, and often do, as for the present model, show strong finite-size
effects that may be wholly absent in the thermodynamic limit.  For
that reason, CCM practitioners generally regard it as a strength of
the method that results are presented directly in the
$N \rightarrow \infty$ limit, with no finite-size scaling, and with the
extrapolation in the level of implementation the sole approximation
made.

Returning to our results, the wide range of low-energy parameters examined provides us with a
consistent picture of a quantum phase diagram for the model that
comprises four different stable GS phases.  For
$\kappa < \kappa_{c_{1}}$ we find a quasiclassical AFM phase with
N\'{e}el magnetic LRO, while for $\kappa > \kappa_{c_{2}}$ we find
another quasiclassical AFM phase with magnetic LRO of the N\'{e}el-II
type illustrated in Fig.\ \ref{model_pattern}(b).  For intermediate
values of the frustration parameter,
$\kappa_{c_{1}} < \kappa < \kappa_{c_{2}}$, the ground state is one of
two distinct quantum paramagnetic states, neither of which has any
discernible magnetic LRO.  We find that at the QCP $\kappa_{c_{1}}$
the phase transition is from the gapless N\'{e}el state to another
gapless state, while at the QCP $\kappa_{c_{2}}$ the transition is
from the gapless N\'{e}el-II state to a gapped state.  The latter
gapped state appears not to have PVBC order, as in the corresponding
intermediate regime for the spin-$\frac{1}{2}$ version of the model.
Strong evidence is presented that a more likely candidate gapped state
is one with SDVBC order, which breaks the same symmetries as the state
with N\'{e}el-II order.  The two paramagnetic states meet at a third
QCP $\kappa^{i}_{c}$, with $\kappa_{c_{1}} < \kappa^{i}_{c} < \kappa_{c_{2}}$.
A careful analysis of all of the CCM results yields our best estimates
for the three QCPs as $\kappa_{c_{1}} = 0.250(5)$,
$\kappa_{c_{2}} = 0.340(5)$, and $\kappa^{i}_{c} = 0.305(5)$.  The
seemingly featureless nature of the gapless GS phase in the region
$\kappa_{c_{1}} < \kappa < \kappa^{i}_{c}$, where both magnetic and
VBC orderings vanish, clearly makes it a strong candidate to be a QSL
state.

Our results may be compared to those from a recent DMRG study of the
same system \cite{Gong:2015_honey_J1J2mod_s1}, which found a
three-phase structure for the GS ($T=0$) phase diagram.  This study
also found a N\'{e}el AFM phase for $\kappa < \kappa_{c_{1}}$, and
also suggested a N\'{e}el-II AFM phase for $\kappa > \kappa_{c_{2}}$
(which was termed a stripe phase in Ref.\
\cite{Gong:2015_honey_J1J2mod_s1}).  The numerical values obtained for
these two QCPs, $\kappa_{c_{1}} \approx 0.27(1)$ and
$\kappa_{c_{2}} \approx 0.32$, are in reasonable agreement with our
own findings.  The DMRG study also identified an intermediate nonmagnetic phase
region, in which the spin gap was found to enlarge considerably on the
finite-size cylinders investigated.  However, the DMRG results differ
from ours in suggesting that in the entire intermediate region the GS
phase has PVBC order, whereas our own results give little evidence for this
form of VBC order.  It is interesting to speculate about possible
causes for this disagreement.

In the first place it is clear that the DMRG results for this system
display rather strong finite-size effects for the cylinders that are
feasible to study computationally.  Two sorts of cylinder geometries
are studied, viz., so-called AC$m$ and ZC$m$ types, with armchair and
zigzag open edges respectively, and where $m$ is the number of unit
cells (i.e., the number of sites on either sublattice ${\cal A}$ or
sublattice ${\cal B}$) in the width direction.  Both types of
cylinders are considered with lengths up to 24 unit cells.  For the
magnetically ordered phases the DMRG calculations are performed with
widths $m \leq 8$.  However, in the intermediate phase it is found
that convergence is sufficiently challenging that the calculations are
constrained to relative narrow cylinders, $m \leq 6$.

Even in the N\'{e}el-II phase region the GS energy, extracted from the
bulk bond energy on long cylinders, is different on the two types of
cylinders, with the ZC cylinders giving lower energies than the AC
cylinders.  Thus, the spin configurations on the ZC cylinders are
ordered as in the 4-sublattice N\'{e}el-II AFM phase, whereas those on
the AC cylinders order on an 8-sublattice double-N\'{e}el AFM pattern
(i.e., with an 8-site unit cell).  The double-N\'{e}el phase comprises
parallel zigzag chains of alternating pairs of parallel spins, ordered
as
$\cdots\uparrow\uparrow\downarrow\downarrow\uparrow\uparrow\downarrow\downarrow\cdots$,
and with NN spins on neighboring chains antiparallel to one another,
in such a way that each hexagonal plaquette has 3 spins pointing in
each direction.  It is degenerate in energy with the N\'{e}el-II phase
at the classical level.  When the DMRG results on long cylinders are
extrapolated to the thermodynamic limit ($m \rightarrow \infty$) using
results with $m=\{4,6,8\}$, at a value $\kappa = 0.4$, for example,
the AC cylinders give a bulk value $E/(NJ_{1}) \approx -1.267$, while
the ZC cylinders give the lower value $E/(NJ_{1}) \approx -1.274$.
Since the two geometries should presumably give the same energy in the
thermodynamic limit, these different values indicate the strong
finite-size effects on the (relatively small) systems that can be
studied with available computational resources.  By comparison, our
own extrapolated energy at $\kappa = 0.4$ is
$E/(NJ_{1}) \approx -1.2748$ based on an extrapolation using Eq.\
(\ref{extrapo_E}) from SUB$n$--$n$ calculations based on the
N\'{e}el-II model state with $n=\{2,6,10\}$.  It is in excellent
agreement with the DMRG results on ZC cylinders, which also exhibit
N\'{e}el-II ordering at $\kappa=0.4$.

Turning to the intermediate region, the DMRG calculations are even
more constrained.  On the AC4 and ZC4 cylinders the bond energies are
found to be quite uniform in the bulk of the cylinder, and in order to
detect any lattice symmetry breaking it is necessary to go to wider
systems, for which convergence can be achieved only for the AC6 and
ZC6 cylinders.  It is based only on these cylinders that PVBC ordering
is suggested, even though the truncation errors are considerably
larger than for the corresponding results in the N\'{e}el and
N\'{e}el-II regions.

From our own results shown in Figs.\ \ref{PVBC_s1} and \ref{SDVBC_s1},
it is also interesting to note that for most values of $\kappa$,
except in a small region around $\kappa = 0.3$ (near the QCP
$\kappa_{c_{2}}$), the system is actually {\it more} susceptible to
PVBC ordering than to SDVBC ordering, i.e.,
$\chi^{-1}_{p} < \chi^{-1}_{d}$.  It is only in the very narrow
region $\kappa^{i}_{c} < \kappa < \kappa_{c_{2}}$ that SDVBC
ordering clearly dominates over PVBC ordering.  Given these results,
and the strong finite-size effects observed in the DMRG results as
discussed above, it is entirely possible that the PVBC ordering
observed on the relatively narrow cylinders for which calculations are
feasible in the intermediate region
$\kappa_{c_{1}} < \kappa < \kappa_{c_{2}}$ would itself give way to
other forms of order on wider cylinders.

To conclude, it is clear that the competition between different phases
in the intermediate region $\kappa_{c_{1}} < \kappa < \kappa_{c_{2}}$
for the spin-1 $J_{1}$--$J_{2}$ model on the honeycomb lattice is
delicate and subtle.  While our own results show rather clear evidence
of two intermediate phases, viz., a gapped phase with probable SDVBC order in
the interval $\kappa^{i}_{c} < \kappa < \kappa_{c_{2}}$, and a
gapless paramagnetic phase in the interval
$\kappa_{c_{1}} < \kappa < \kappa^{i}_{c}$ (which is a possible
QSL), it would be very useful for other methods to be applied to this
system to confirm our results.

\section*{ACKNOWLEDGMENTS}
We thank the University of Minnesota Supercomputing Institute for the
grant of supercomputing facilities, on which the work reported here
was performed.  One of us (RFB) gratefully acknowledges the Leverhulme
Trust (United Kingdom) for the award of an Emeritus Fellowship (EM-2015-007).

\bibliographystyle{apsrev4-1}
\bibliography{bib_general}

\providecommand{\noopsort}[1]{}\providecommand{\singleletter}[1]{#1}%
\begin{thebibliography}{85}%
\makeatletter
\providecommand \@ifxundefined [1]{%
 \@ifx{#1\undefined}
}%
\providecommand \@ifnum [1]{%
 \ifnum #1\expandafter \@firstoftwo
 \else \expandafter \@secondoftwo
 \fi
}%
\providecommand \@ifx [1]{%
 \ifx #1\expandafter \@firstoftwo
 \else \expandafter \@secondoftwo
 \fi
}%
\providecommand \natexlab [1]{#1}%
\providecommand \enquote  [1]{``#1''}%
\providecommand \bibnamefont  [1]{#1}%
\providecommand \bibfnamefont [1]{#1}%
\providecommand \citenamefont [1]{#1}%
\providecommand \href@noop [0]{\@secondoftwo}%
\providecommand \href [0]{\begingroup \@sanitize@url \@href}%
\providecommand \@href[1]{\@@startlink{#1}\@@href}%
\providecommand \@@href[1]{\endgroup#1\@@endlink}%
\providecommand \@sanitize@url [0]{\catcode `\\12\catcode `\$12\catcode
  `\&12\catcode `\#12\catcode `\^12\catcode `\_12\catcode `\%12\relax}%
\providecommand \@@startlink[1]{}%
\providecommand \@@endlink[0]{}%
\providecommand \url  [0]{\begingroup\@sanitize@url \@url }%
\providecommand \@url [1]{\endgroup\@href {#1}{\urlprefix }}%
\providecommand \urlprefix  [0]{URL }%
\providecommand \Eprint [0]{\href }%
\providecommand \doibase [0]{http://dx.doi.org/}%
\providecommand \selectlanguage [0]{\@gobble}%
\providecommand \bibinfo  [0]{\@secondoftwo}%
\providecommand \bibfield  [0]{\@secondoftwo}%
\providecommand \translation [1]{[#1]}%
\providecommand \BibitemOpen [0]{}%
\providecommand \bibitemStop [0]{}%
\providecommand \bibitemNoStop [0]{.\EOS\space}%
\providecommand \EOS [0]{\spacefactor3000\relax}%
\providecommand \BibitemShut  [1]{\csname bibitem#1\endcsname}%
\let\auto@bib@innerbib\@empty
\bibitem [{\citenamefont {Villain}(1977)}]{Villain:1977_ordByDisord}%
  \BibitemOpen
  \bibfield  {author} {\bibinfo {author} {\bibfnamefont {J.}~\bibnamefont
  {Villain}},\ }\href@noop {} {\bibfield  {journal} {\bibinfo  {journal} {J.
  Phys. (France)}\ }\textbf {\bibinfo {volume} {38}},\ \bibinfo {pages} {385}
  (\bibinfo {year} {1977})}\BibitemShut {NoStop}%
\bibitem [{\citenamefont {Villain}\ \emph {et~al.}(1980)\citenamefont
  {Villain}, \citenamefont {Bidaux}, \citenamefont {Carton},\ and\
  \citenamefont {Conte}}]{Villain:1980_ordByDisord}%
  \BibitemOpen
  \bibfield  {author} {\bibinfo {author} {\bibfnamefont {J.}~\bibnamefont
  {Villain}}, \bibinfo {author} {\bibfnamefont {R.}~\bibnamefont {Bidaux}},
  \bibinfo {author} {\bibfnamefont {J.-P.}\ \bibnamefont {Carton}}, \ and\
  \bibinfo {author} {\bibfnamefont {R.}~\bibnamefont {Conte}},\ }\href@noop {}
  {\bibfield  {journal} {\bibinfo  {journal} {J. Phys. (France)}\ }\textbf
  {\bibinfo {volume} {41}},\ \bibinfo {pages} {1263} (\bibinfo {year}
  {1980})}\BibitemShut {NoStop}%
\bibitem [{\citenamefont {Shender}(1982)}]{Shender:1982_ordByDisord}%
  \BibitemOpen
  \bibfield  {author} {\bibinfo {author} {\bibfnamefont {E.~F.}\ \bibnamefont
  {Shender}},\ }\href@noop {} {\bibfield  {journal} {\bibinfo  {journal} {Zh.
  Eksp. Teor. Fiz.}\ }\textbf {\bibinfo {volume} {83}},\ \bibinfo {pages} {326}
  (\bibinfo {year} {1982})},\ \translation{Sov. Phys. JETP \textbf{56}, 178
  (1982)}\BibitemShut {NoStop}%
\bibitem [{\citenamefont {Jian}\ and\ \citenamefont
  {Zaletel}(2016)}]{Jiang:2016_SqLatt-honey}%
  \BibitemOpen
  \bibfield  {author} {\bibinfo {author} {\bibfnamefont {C.-M.}\ \bibnamefont
  {Jian}}\ and\ \bibinfo {author} {\bibfnamefont {M.}~\bibnamefont {Zaletel}},\
  }\href@noop {} {\bibfield  {journal} {\bibinfo  {journal} {Phys. Rev. B}\
  }\textbf {\bibinfo {volume} {93}},\ \bibinfo {pages} {035114} (\bibinfo
  {year} {2016})}\BibitemShut {NoStop}%
\bibitem [{\citenamefont {Lieb}\ \emph {et~al.}(1961)\citenamefont {Lieb},
  \citenamefont {Schultz},\ and\ \citenamefont
  {Mattis}}]{Lieb:1961_LSMH-theorem}%
  \BibitemOpen
  \bibfield  {author} {\bibinfo {author} {\bibfnamefont {E.}~\bibnamefont
  {Lieb}}, \bibinfo {author} {\bibfnamefont {T.}~\bibnamefont {Schultz}}, \
  and\ \bibinfo {author} {\bibfnamefont {D.}~\bibnamefont {Mattis}},\
  }\href@noop {} {\bibfield  {journal} {\bibinfo  {journal} {Ann. Phys. (N.
  Y.)}\ }\textbf {\bibinfo {volume} {16}},\ \bibinfo {pages} {407} (\bibinfo
  {year} {1961})}\BibitemShut {NoStop}%
\bibitem [{\citenamefont
  {Hastings}(2004)}]{Hastings:2004_Lieb-LSM-Hast-theorem}%
  \BibitemOpen
  \bibfield  {author} {\bibinfo {author} {\bibfnamefont {M.~B.}\ \bibnamefont
  {Hastings}},\ }\href@noop {} {\bibfield  {journal} {\bibinfo  {journal}
  {Phys. Rev. B}\ }\textbf {\bibinfo {volume} {69}},\ \bibinfo {pages} {104431}
  (\bibinfo {year} {2004})}\BibitemShut {NoStop}%
\bibitem [{\citenamefont {Oshikawa}(2000)}]{Oshikawa:2000_LSMN_d-gtr-1}%
  \BibitemOpen
  \bibfield  {author} {\bibinfo {author} {\bibfnamefont {M.}~\bibnamefont
  {Oshikawa}},\ }\href@noop {} {\bibfield  {journal} {\bibinfo  {journal}
  {Phys. Rev. Lett.}\ }\textbf {\bibinfo {volume} {84}},\ \bibinfo {pages}
  {1535} (\bibinfo {year} {2000})}\BibitemShut {NoStop}%
\bibitem [{\citenamefont {Haldane}(1988)}]{Haldane:1988_param-phases}%
  \BibitemOpen
  \bibfield  {author} {\bibinfo {author} {\bibfnamefont {F.~D.~M.}\
  \bibnamefont {Haldane}},\ }\href@noop {} {\bibfield  {journal} {\bibinfo
  {journal} {Phys. Rev. Lett.}\ }\textbf {\bibinfo {volume} {61}},\ \bibinfo
  {pages} {1029} (\bibinfo {year} {1988})}\BibitemShut {NoStop}%
\bibitem [{\citenamefont {Read}\ and\ \citenamefont
  {Sachdev}(1990)}]{Read:1990:param-phases}%
  \BibitemOpen
  \bibfield  {author} {\bibinfo {author} {\bibfnamefont {N.}~\bibnamefont
  {Read}}\ and\ \bibinfo {author} {\bibfnamefont {S.}~\bibnamefont {Sachdev}},\
  }\href@noop {} {\bibfield  {journal} {\bibinfo  {journal} {Phys. Rev. B}\
  }\textbf {\bibinfo {volume} {42}},\ \bibinfo {pages} {4568} (\bibinfo {year}
  {1990})}\BibitemShut {NoStop}%
\bibitem [{\citenamefont {Mermin}\ and\ \citenamefont
  {Wagner}(1966)}]{Mermin:1966}%
  \BibitemOpen
  \bibfield  {author} {\bibinfo {author} {\bibfnamefont {N.~D.}\ \bibnamefont
  {Mermin}}\ and\ \bibinfo {author} {\bibfnamefont {H.}~\bibnamefont
  {Wagner}},\ }\href@noop {} {\bibfield  {journal} {\bibinfo  {journal} {Phys.
  Rev. Lett.}\ }\textbf {\bibinfo {volume} {17}},\ \bibinfo {pages} {1133}
  (\bibinfo {year} {1966})}\BibitemShut {NoStop}%
\bibitem [{\citenamefont {Rastelli}\ \emph {et~al.}(1979)\citenamefont
  {Rastelli}, \citenamefont {Tassi},\ and\ \citenamefont
  {Reatto}}]{Rastelli:1979_honey}%
  \BibitemOpen
  \bibfield  {author} {\bibinfo {author} {\bibfnamefont {E.}~\bibnamefont
  {Rastelli}}, \bibinfo {author} {\bibfnamefont {A.}~\bibnamefont {Tassi}}, \
  and\ \bibinfo {author} {\bibfnamefont {L.}~\bibnamefont {Reatto}},\
  }\href@noop {} {\bibfield  {journal} {\bibinfo  {journal} {Physica B \& C}\
  }\textbf {\bibinfo {volume} {97}},\ \bibinfo {pages} {1} (\bibinfo {year}
  {1979})}\BibitemShut {NoStop}%
\bibitem [{\citenamefont {Mattsson}\ \emph {et~al.}(1994)\citenamefont
  {Mattsson}, \citenamefont {Fr{\"{o}}jdh},\ and\ \citenamefont
  {Einarsson}}]{Mattsson:1994_honey}%
  \BibitemOpen
  \bibfield  {author} {\bibinfo {author} {\bibfnamefont {A.}~\bibnamefont
  {Mattsson}}, \bibinfo {author} {\bibfnamefont {P.}~\bibnamefont
  {Fr{\"{o}}jdh}}, \ and\ \bibinfo {author} {\bibfnamefont {T.}~\bibnamefont
  {Einarsson}},\ }\href@noop {} {\bibfield  {journal} {\bibinfo  {journal}
  {Phys. Rev. B}\ }\textbf {\bibinfo {volume} {49}},\ \bibinfo {pages} {3997}
  (\bibinfo {year} {1994})}\BibitemShut {NoStop}%
\bibitem [{\citenamefont {Fouet}\ \emph {et~al.}(2001)\citenamefont {Fouet},
  \citenamefont {Sindzingre},\ and\ \citenamefont
  {Lhuillier}}]{Fouet:2001_honey}%
  \BibitemOpen
  \bibfield  {author} {\bibinfo {author} {\bibfnamefont {J.~B.}\ \bibnamefont
  {Fouet}}, \bibinfo {author} {\bibfnamefont {P.}~\bibnamefont {Sindzingre}}, \
  and\ \bibinfo {author} {\bibfnamefont {C.}~\bibnamefont {Lhuillier}},\
  }\href@noop {} {\bibfield  {journal} {\bibinfo  {journal} {Eur. Phys. J. B}\
  }\textbf {\bibinfo {volume} {20}},\ \bibinfo {pages} {241} (\bibinfo {year}
  {2001})}\BibitemShut {NoStop}%
\bibitem [{\citenamefont {Mulder}\ \emph {et~al.}(2010)\citenamefont {Mulder},
  \citenamefont {Ganesh}, \citenamefont {Capriotti},\ and\ \citenamefont
  {Paramekanti}}]{Mulder:2010_honey}%
  \BibitemOpen
  \bibfield  {author} {\bibinfo {author} {\bibfnamefont {A.}~\bibnamefont
  {Mulder}}, \bibinfo {author} {\bibfnamefont {R.}~\bibnamefont {Ganesh}},
  \bibinfo {author} {\bibfnamefont {L.}~\bibnamefont {Capriotti}}, \ and\
  \bibinfo {author} {\bibfnamefont {A.}~\bibnamefont {Paramekanti}},\
  }\href@noop {} {\bibfield  {journal} {\bibinfo  {journal} {Phys. Rev. B}\
  }\textbf {\bibinfo {volume} {81}},\ \bibinfo {pages} {214419} (\bibinfo
  {year} {2010})}\BibitemShut {NoStop}%
\bibitem [{\citenamefont {Wang}(2010)}]{Wang:2010_honey}%
  \BibitemOpen
  \bibfield  {author} {\bibinfo {author} {\bibfnamefont {F.}~\bibnamefont
  {Wang}},\ }\href@noop {} {\bibfield  {journal} {\bibinfo  {journal} {Phys.
  Rev. B}\ }\textbf {\bibinfo {volume} {82}},\ \bibinfo {pages} {024419}
  (\bibinfo {year} {2010})}\BibitemShut {NoStop}%
\bibitem [{\citenamefont {Cabra}\ \emph {et~al.}(2011)\citenamefont {Cabra},
  \citenamefont {Lamas},\ and\ \citenamefont {Rosales}}]{Cabra:2011_honey}%
  \BibitemOpen
  \bibfield  {author} {\bibinfo {author} {\bibfnamefont {D.~C.}\ \bibnamefont
  {Cabra}}, \bibinfo {author} {\bibfnamefont {C.~A.}\ \bibnamefont {Lamas}}, \
  and\ \bibinfo {author} {\bibfnamefont {H.~D.}\ \bibnamefont {Rosales}},\
  }\href@noop {} {\bibfield  {journal} {\bibinfo  {journal} {Phys. Rev. B}\
  }\textbf {\bibinfo {volume} {83}},\ \bibinfo {pages} {094506} (\bibinfo
  {year} {2011})}\BibitemShut {NoStop}%
\bibitem [{\citenamefont {Ganesh}\ \emph {et~al.}(2011)\citenamefont {Ganesh},
  \citenamefont {Sheng}, \citenamefont {Kim},\ and\ \citenamefont
  {Paramekanti}}]{Ganesh:2011_honey}%
  \BibitemOpen
  \bibfield  {author} {\bibinfo {author} {\bibfnamefont {R.}~\bibnamefont
  {Ganesh}}, \bibinfo {author} {\bibfnamefont {D.~N.}\ \bibnamefont {Sheng}},
  \bibinfo {author} {\bibfnamefont {Y.-J.}\ \bibnamefont {Kim}}, \ and\
  \bibinfo {author} {\bibfnamefont {A.}~\bibnamefont {Paramekanti}},\
  }\href@noop {} {\bibfield  {journal} {\bibinfo  {journal} {Phys. Rev. B}\
  }\textbf {\bibinfo {volume} {83}},\ \bibinfo {pages} {144414} (\bibinfo
  {year} {2011})}\BibitemShut {NoStop}%
\bibitem [{\citenamefont {Clark}\ \emph {et~al.}(2011)\citenamefont {Clark},
  \citenamefont {Abanin},\ and\ \citenamefont {Sondhi}}]{Clark:2011_honey}%
  \BibitemOpen
  \bibfield  {author} {\bibinfo {author} {\bibfnamefont {B.~K.}\ \bibnamefont
  {Clark}}, \bibinfo {author} {\bibfnamefont {D.~A.}\ \bibnamefont {Abanin}}, \
  and\ \bibinfo {author} {\bibfnamefont {S.~L.}\ \bibnamefont {Sondhi}},\
  }\href@noop {} {\bibfield  {journal} {\bibinfo  {journal} {Phys. Rev. Lett.}\
  }\textbf {\bibinfo {volume} {107}},\ \bibinfo {pages} {087204} (\bibinfo
  {year} {2011})}\BibitemShut {NoStop}%
\bibitem [{\citenamefont {Farnell}\ \emph {et~al.}(2011)\citenamefont
  {Farnell}, \citenamefont {Bishop}, \citenamefont {Li}, \citenamefont
  {Richter},\ and\ \citenamefont {Campbell}}]{DJJF:2011_honeycomb}%
  \BibitemOpen
  \bibfield  {author} {\bibinfo {author} {\bibfnamefont {D.~J.~J.}\
  \bibnamefont {Farnell}}, \bibinfo {author} {\bibfnamefont {R.~F.}\
  \bibnamefont {Bishop}}, \bibinfo {author} {\bibfnamefont {P.~H.~Y.}\
  \bibnamefont {Li}}, \bibinfo {author} {\bibfnamefont {J.}~\bibnamefont
  {Richter}}, \ and\ \bibinfo {author} {\bibfnamefont {C.~E.}\ \bibnamefont
  {Campbell}},\ }\href@noop {} {\bibfield  {journal} {\bibinfo  {journal}
  {Phys. Rev. B}\ }\textbf {\bibinfo {volume} {84}},\ \bibinfo {pages} {012403}
  (\bibinfo {year} {2011})}\BibitemShut {NoStop}%
\bibitem [{\citenamefont {Reuther}\ \emph {et~al.}(2011)\citenamefont
  {Reuther}, \citenamefont {Abanin},\ and\ \citenamefont
  {Thomale}}]{Reuther:2011_honey}%
  \BibitemOpen
  \bibfield  {author} {\bibinfo {author} {\bibfnamefont {J.}~\bibnamefont
  {Reuther}}, \bibinfo {author} {\bibfnamefont {D.~A.}\ \bibnamefont {Abanin}},
  \ and\ \bibinfo {author} {\bibfnamefont {R.}~\bibnamefont {Thomale}},\
  }\href@noop {} {\bibfield  {journal} {\bibinfo  {journal} {Phys. Rev. B}\
  }\textbf {\bibinfo {volume} {84}},\ \bibinfo {pages} {014417} (\bibinfo
  {year} {2011})}\BibitemShut {NoStop}%
\bibitem [{\citenamefont {Albuquerque}\ \emph {et~al.}(2011)\citenamefont
  {Albuquerque}, \citenamefont {Schwandt}, \citenamefont {Het{\'e}nyi},
  \citenamefont {Capponi}, \citenamefont {Mambrini},\ and\ \citenamefont
  {L{\"{a}}uchli}}]{Albuquerque:2011_honey}%
  \BibitemOpen
  \bibfield  {author} {\bibinfo {author} {\bibfnamefont {A.~F.}\ \bibnamefont
  {Albuquerque}}, \bibinfo {author} {\bibfnamefont {D.}~\bibnamefont
  {Schwandt}}, \bibinfo {author} {\bibfnamefont {B.}~\bibnamefont
  {Het{\'e}nyi}}, \bibinfo {author} {\bibfnamefont {S.}~\bibnamefont
  {Capponi}}, \bibinfo {author} {\bibfnamefont {M.}~\bibnamefont {Mambrini}}, \
  and\ \bibinfo {author} {\bibfnamefont {A.~M.}\ \bibnamefont
  {L{\"{a}}uchli}},\ }\href@noop {} {\bibfield  {journal} {\bibinfo  {journal}
  {Phys. Rev. B}\ }\textbf {\bibinfo {volume} {84}},\ \bibinfo {pages} {024406}
  (\bibinfo {year} {2011})}\BibitemShut {NoStop}%
\bibitem [{\citenamefont {Mosadeq}\ \emph {et~al.}(2011)\citenamefont
  {Mosadeq}, \citenamefont {Shahbazi},\ and\ \citenamefont
  {Jafari}}]{Mosadeq:2011_honey}%
  \BibitemOpen
  \bibfield  {author} {\bibinfo {author} {\bibfnamefont {H.}~\bibnamefont
  {Mosadeq}}, \bibinfo {author} {\bibfnamefont {F.}~\bibnamefont {Shahbazi}}, \
  and\ \bibinfo {author} {\bibfnamefont {S.~A.}\ \bibnamefont {Jafari}},\
  }\href@noop {} {\bibfield  {journal} {\bibinfo  {journal} {J. Phys.: Condens.
  Matter}\ }\textbf {\bibinfo {volume} {23}},\ \bibinfo {pages} {226006}
  (\bibinfo {year} {2011})}\BibitemShut {NoStop}%
\bibitem [{\citenamefont {Oitmaa}\ and\ \citenamefont
  {Singh}(2011)}]{Oitmaa:2011_honey}%
  \BibitemOpen
  \bibfield  {author} {\bibinfo {author} {\bibfnamefont {J.}~\bibnamefont
  {Oitmaa}}\ and\ \bibinfo {author} {\bibfnamefont {R.~R.~P.}\ \bibnamefont
  {Singh}},\ }\href@noop {} {\bibfield  {journal} {\bibinfo  {journal} {Phys.
  Rev. B}\ }\textbf {\bibinfo {volume} {84}},\ \bibinfo {pages} {094424}
  (\bibinfo {year} {2011})}\BibitemShut {NoStop}%
\bibitem [{\citenamefont {Mezzacapo}\ and\ \citenamefont
  {Boninsegni}(2012)}]{Mezzacapo:2012_honey}%
  \BibitemOpen
  \bibfield  {author} {\bibinfo {author} {\bibfnamefont {F.}~\bibnamefont
  {Mezzacapo}}\ and\ \bibinfo {author} {\bibfnamefont {M.}~\bibnamefont
  {Boninsegni}},\ }\href@noop {} {\bibfield  {journal} {\bibinfo  {journal}
  {Phys. Rev. B}\ }\textbf {\bibinfo {volume} {85}},\ \bibinfo {pages}
  {060402(R)} (\bibinfo {year} {2012})}\BibitemShut {NoStop}%
\bibitem [{\citenamefont {Li}\ \emph {et~al.}(2012{\natexlab{a}})\citenamefont
  {Li}, \citenamefont {Bishop}, \citenamefont {Farnell}, \citenamefont
  {Richter},\ and\ \citenamefont {Campbell}}]{PHYLi:2012_honeycomb_J1neg}%
  \BibitemOpen
  \bibfield  {author} {\bibinfo {author} {\bibfnamefont {P.~H.~Y.}\
  \bibnamefont {Li}}, \bibinfo {author} {\bibfnamefont {R.~F.}\ \bibnamefont
  {Bishop}}, \bibinfo {author} {\bibfnamefont {D.~J.~J.}\ \bibnamefont
  {Farnell}}, \bibinfo {author} {\bibfnamefont {J.}~\bibnamefont {Richter}}, \
  and\ \bibinfo {author} {\bibfnamefont {C.~E.}\ \bibnamefont {Campbell}},\
  }\href@noop {} {\bibfield  {journal} {\bibinfo  {journal} {Phys. Rev. B}\
  }\textbf {\bibinfo {volume} {85}},\ \bibinfo {pages} {085115} (\bibinfo
  {year} {2012}{\natexlab{a}})}\BibitemShut {NoStop}%
\bibitem [{\citenamefont {Bishop}\ \emph
  {et~al.}(2012{\natexlab{a}})\citenamefont {Bishop}, \citenamefont {Li},
  \citenamefont {Farnell},\ and\ \citenamefont
  {Campbell}}]{Bishop:2012_honeyJ1-J2}%
  \BibitemOpen
  \bibfield  {author} {\bibinfo {author} {\bibfnamefont {R.~F.}\ \bibnamefont
  {Bishop}}, \bibinfo {author} {\bibfnamefont {P.~H.~Y.}\ \bibnamefont {Li}},
  \bibinfo {author} {\bibfnamefont {D.~J.~J.}\ \bibnamefont {Farnell}}, \ and\
  \bibinfo {author} {\bibfnamefont {C.~E.}\ \bibnamefont {Campbell}},\
  }\href@noop {} {\bibfield  {journal} {\bibinfo  {journal} {J. Phys.: Condens.
  Matter}\ }\textbf {\bibinfo {volume} {24}},\ \bibinfo {pages} {236002}
  (\bibinfo {year} {2012}{\natexlab{a}})}\BibitemShut {NoStop}%
\bibitem [{\citenamefont {Bishop}\ and\ \citenamefont
  {Li}(2012)}]{Bishop:2012_honey_circle-phase}%
  \BibitemOpen
  \bibfield  {author} {\bibinfo {author} {\bibfnamefont {R.~F.}\ \bibnamefont
  {Bishop}}\ and\ \bibinfo {author} {\bibfnamefont {P.~H.~Y.}\ \bibnamefont
  {Li}},\ }\href@noop {} {\bibfield  {journal} {\bibinfo  {journal} {Phys. Rev.
  B}\ }\textbf {\bibinfo {volume} {85}},\ \bibinfo {pages} {155135} (\bibinfo
  {year} {2012})}\BibitemShut {NoStop}%
\bibitem [{\citenamefont {Li}\ \emph {et~al.}(2012{\natexlab{b}})\citenamefont
  {Li}, \citenamefont {Bishop}, \citenamefont {Farnell},\ and\ \citenamefont
  {Campbell}}]{Li:2012_honey_full}%
  \BibitemOpen
  \bibfield  {author} {\bibinfo {author} {\bibfnamefont {P.~H.~Y.}\
  \bibnamefont {Li}}, \bibinfo {author} {\bibfnamefont {R.~F.}\ \bibnamefont
  {Bishop}}, \bibinfo {author} {\bibfnamefont {D.~J.~J.}\ \bibnamefont
  {Farnell}}, \ and\ \bibinfo {author} {\bibfnamefont {C.~E.}\ \bibnamefont
  {Campbell}},\ }\href@noop {} {\bibfield  {journal} {\bibinfo  {journal}
  {Phys. Rev. B}\ }\textbf {\bibinfo {volume} {86}},\ \bibinfo {pages} {144404}
  (\bibinfo {year} {2012}{\natexlab{b}})}\BibitemShut {NoStop}%
\bibitem [{\citenamefont {Bishop}\ \emph
  {et~al.}(2013{\natexlab{a}})\citenamefont {Bishop}, \citenamefont {Li},\ and\
  \citenamefont {Campbell}}]{RFB:2013_hcomb_SDVBC}%
  \BibitemOpen
  \bibfield  {author} {\bibinfo {author} {\bibfnamefont {R.~F.}\ \bibnamefont
  {Bishop}}, \bibinfo {author} {\bibfnamefont {P.~H.~Y.}\ \bibnamefont {Li}}, \
  and\ \bibinfo {author} {\bibfnamefont {C.~E.}\ \bibnamefont {Campbell}},\
  }\href@noop {} {\bibfield  {journal} {\bibinfo  {journal} {J. Phys.: Condens.
  Matter}\ }\textbf {\bibinfo {volume} {25}},\ \bibinfo {pages} {306002}
  (\bibinfo {year} {2013}{\natexlab{a}})}\BibitemShut {NoStop}%
\bibitem [{\citenamefont {Ganesh}\ \emph {et~al.}(2013)\citenamefont {Ganesh},
  \citenamefont {van~den Brink},\ and\ \citenamefont
  {Nishimoto}}]{Ganesh:2013_honey_J1J2mod-XXX}%
  \BibitemOpen
  \bibfield  {author} {\bibinfo {author} {\bibfnamefont {R.}~\bibnamefont
  {Ganesh}}, \bibinfo {author} {\bibfnamefont {J.}~\bibnamefont {van~den
  Brink}}, \ and\ \bibinfo {author} {\bibfnamefont {S.}~\bibnamefont
  {Nishimoto}},\ }\href@noop {} {\bibfield  {journal} {\bibinfo  {journal}
  {Phys. Rev. Lett.}\ }\textbf {\bibinfo {volume} {110}},\ \bibinfo {pages}
  {127203} (\bibinfo {year} {2013})}\BibitemShut {NoStop}%
\bibitem [{\citenamefont {Zhu}\ \emph {et~al.}(2013)\citenamefont {Zhu},
  \citenamefont {Huse},\ and\ \citenamefont
  {White}}]{Zhu:2013_honey_J1J2mod-XXZ}%
  \BibitemOpen
  \bibfield  {author} {\bibinfo {author} {\bibfnamefont {Z.}~\bibnamefont
  {Zhu}}, \bibinfo {author} {\bibfnamefont {D.~A.}\ \bibnamefont {Huse}}, \
  and\ \bibinfo {author} {\bibfnamefont {S.~R.}\ \bibnamefont {White}},\
  }\href@noop {} {\bibfield  {journal} {\bibinfo  {journal} {Phys. Rev. Lett.}\
  }\textbf {\bibinfo {volume} {110}},\ \bibinfo {pages} {127205} (\bibinfo
  {year} {2013})}\BibitemShut {NoStop}%
\bibitem [{\citenamefont {Zhang}\ and\ \citenamefont
  {Lamas}(2013)}]{Zhang:2013_honey}%
  \BibitemOpen
  \bibfield  {author} {\bibinfo {author} {\bibfnamefont {H.}~\bibnamefont
  {Zhang}}\ and\ \bibinfo {author} {\bibfnamefont {C.~A.}\ \bibnamefont
  {Lamas}},\ }\href@noop {} {\bibfield  {journal} {\bibinfo  {journal} {Phys.
  Rev. B}\ }\textbf {\bibinfo {volume} {87}},\ \bibinfo {pages} {024415}
  (\bibinfo {year} {2013})}\BibitemShut {NoStop}%
\bibitem [{\citenamefont {Gong}\ \emph {et~al.}(2013)\citenamefont {Gong},
  \citenamefont {Sheng}, \citenamefont {Motrunich},\ and\ \citenamefont
  {Fisher}}]{Gong:2013_J1J2mod-XXX}%
  \BibitemOpen
  \bibfield  {author} {\bibinfo {author} {\bibfnamefont {S.-S.}\ \bibnamefont
  {Gong}}, \bibinfo {author} {\bibfnamefont {D.~N.}\ \bibnamefont {Sheng}},
  \bibinfo {author} {\bibfnamefont {O.~I.}\ \bibnamefont {Motrunich}}, \ and\
  \bibinfo {author} {\bibfnamefont {M.~P.~A.}\ \bibnamefont {Fisher}},\
  }\href@noop {} {\bibfield  {journal} {\bibinfo  {journal} {Phys. Rev. B}\
  }\textbf {\bibinfo {volume} {88}},\ \bibinfo {pages} {165138} (\bibinfo
  {year} {2013})}\BibitemShut {NoStop}%
\bibitem [{\citenamefont {Yu}\ \emph {et~al.}(2014)\citenamefont {Yu},
  \citenamefont {Liu}, \citenamefont {Li},\ and\ \citenamefont
  {Zou}}]{Yu:2014_honey_J1J2mod}%
  \BibitemOpen
  \bibfield  {author} {\bibinfo {author} {\bibfnamefont {X.-L.}\ \bibnamefont
  {Yu}}, \bibinfo {author} {\bibfnamefont {D.-Y.}\ \bibnamefont {Liu}},
  \bibinfo {author} {\bibfnamefont {P.}~\bibnamefont {Li}}, \ and\ \bibinfo
  {author} {\bibfnamefont {L.-J.}\ \bibnamefont {Zou}},\ }\href@noop {}
  {\bibfield  {journal} {\bibinfo  {journal} {Physica E}\ }\textbf {\bibinfo
  {volume} {59}},\ \bibinfo {pages} {41} (\bibinfo {year} {2014})}\BibitemShut
  {NoStop}%
\bibitem [{\citenamefont {Gong}\ \emph {et~al.}(2015)\citenamefont {Gong},
  \citenamefont {Zhu},\ and\ \citenamefont
  {Sheng}}]{Gong:2015_honey_J1J2mod_s1}%
  \BibitemOpen
  \bibfield  {author} {\bibinfo {author} {\bibfnamefont {S.-S.}\ \bibnamefont
  {Gong}}, \bibinfo {author} {\bibfnamefont {W.}~\bibnamefont {Zhu}}, \ and\
  \bibinfo {author} {\bibfnamefont {D.~N.}\ \bibnamefont {Sheng}},\ }\href@noop
  {} {\bibfield  {journal} {\bibinfo  {journal} {Phys. Rev. B}\ }\textbf
  {\bibinfo {volume} {92}},\ \bibinfo {pages} {195110} (\bibinfo {year}
  {2015})}\BibitemShut {NoStop}%
\bibitem [{\citenamefont {Miura}\ \emph {et~al.}(2006)\citenamefont {Miura},
  \citenamefont {Hirai}, \citenamefont {Kobayashi},\ and\ \citenamefont
  {Sato}}]{Miura:2006_honey}%
  \BibitemOpen
  \bibfield  {author} {\bibinfo {author} {\bibfnamefont {Y.}~\bibnamefont
  {Miura}}, \bibinfo {author} {\bibfnamefont {R.}~\bibnamefont {Hirai}},
  \bibinfo {author} {\bibfnamefont {Y.}~\bibnamefont {Kobayashi}}, \ and\
  \bibinfo {author} {\bibfnamefont {M.}~\bibnamefont {Sato}},\ }\href@noop {}
  {\bibfield  {journal} {\bibinfo  {journal} {J. Phys. Soc. Jpn.}\ }\textbf
  {\bibinfo {volume} {75}},\ \bibinfo {pages} {084707} (\bibinfo {year}
  {2006})}\BibitemShut {NoStop}%
\bibitem [{\citenamefont {Kataev}\ \emph {et~al.}(2005)\citenamefont {Kataev},
  \citenamefont {M{\"{o}}ller}, \citenamefont {L{\"{o}}w}, \citenamefont
  {Jung}, \citenamefont {Schittner}, \citenamefont {Kriener},\ and\
  \citenamefont {Freimuth}}]{Kataev:2005_honey}%
  \BibitemOpen
  \bibfield  {author} {\bibinfo {author} {\bibfnamefont {V.}~\bibnamefont
  {Kataev}}, \bibinfo {author} {\bibfnamefont {A.}~\bibnamefont
  {M{\"{o}}ller}}, \bibinfo {author} {\bibfnamefont {U.}~\bibnamefont
  {L{\"{o}}w}}, \bibinfo {author} {\bibfnamefont {W.}~\bibnamefont {Jung}},
  \bibinfo {author} {\bibfnamefont {N.}~\bibnamefont {Schittner}}, \bibinfo
  {author} {\bibfnamefont {M.}~\bibnamefont {Kriener}}, \ and\ \bibinfo
  {author} {\bibfnamefont {A.}~\bibnamefont {Freimuth}},\ }\href@noop {}
  {\bibfield  {journal} {\bibinfo  {journal} {J. Magn. Magn. Mater.}\ }\textbf
  {\bibinfo {volume} {290--291}},\ \bibinfo {pages} {310} (\bibinfo {year}
  {2005})}\BibitemShut {NoStop}%
\bibitem [{\citenamefont {Tsirlin}\ \emph {et~al.}(2010)\citenamefont
  {Tsirlin}, \citenamefont {Janson},\ and\ \citenamefont
  {Rosner}}]{Tsirlin:2010_honey}%
  \BibitemOpen
  \bibfield  {author} {\bibinfo {author} {\bibfnamefont {A.~A.}\ \bibnamefont
  {Tsirlin}}, \bibinfo {author} {\bibfnamefont {O.}~\bibnamefont {Janson}}, \
  and\ \bibinfo {author} {\bibfnamefont {H.}~\bibnamefont {Rosner}},\
  }\href@noop {} {\bibfield  {journal} {\bibinfo  {journal} {Phys. Rev. B}\
  }\textbf {\bibinfo {volume} {82}},\ \bibinfo {pages} {144416} (\bibinfo
  {year} {2010})}\BibitemShut {NoStop}%
\bibitem [{\citenamefont {Climent-Pascual}\ \emph {et~al.}(2012)\citenamefont
  {Climent-Pascual}, \citenamefont {Norby}, \citenamefont {Andersen},
  \citenamefont {Stephens}, \citenamefont {Zandbergen}, \citenamefont
  {Larsen},\ and\ \citenamefont {Cava}}]{Climent:2012_honey}%
  \BibitemOpen
  \bibfield  {author} {\bibinfo {author} {\bibfnamefont {E.}~\bibnamefont
  {Climent-Pascual}}, \bibinfo {author} {\bibfnamefont {P.}~\bibnamefont
  {Norby}}, \bibinfo {author} {\bibfnamefont {N.}~\bibnamefont {Andersen}},
  \bibinfo {author} {\bibfnamefont {P.}~\bibnamefont {Stephens}}, \bibinfo
  {author} {\bibfnamefont {H.}~\bibnamefont {Zandbergen}}, \bibinfo {author}
  {\bibfnamefont {J.}~\bibnamefont {Larsen}}, \ and\ \bibinfo {author}
  {\bibfnamefont {R.}~\bibnamefont {Cava}},\ }\href@noop {} {\bibfield
  {journal} {\bibinfo  {journal} {Inorg. Chem.}\ }\textbf {\bibinfo {volume}
  {51}},\ \bibinfo {pages} {557} (\bibinfo {year} {2012})}\BibitemShut
  {NoStop}%
\bibitem [{\citenamefont {Singh}\ and\ \citenamefont
  {Gegenwart}(2010)}]{Singh:2010_honey}%
  \BibitemOpen
  \bibfield  {author} {\bibinfo {author} {\bibfnamefont {Y.}~\bibnamefont
  {Singh}}\ and\ \bibinfo {author} {\bibfnamefont {P.}~\bibnamefont
  {Gegenwart}},\ }\href@noop {} {\bibfield  {journal} {\bibinfo  {journal}
  {Phys. Rev. B}\ }\textbf {\bibinfo {volume} {82}},\ \bibinfo {pages} {064412}
  (\bibinfo {year} {2010})}\BibitemShut {NoStop}%
\bibitem [{\citenamefont {Liu}\ \emph {et~al.}(2011)\citenamefont {Liu},
  \citenamefont {Berlijn}, \citenamefont {Yin}, \citenamefont {Ku},
  \citenamefont {Tsvelik}, \citenamefont {Kim}, \citenamefont {Gretarsson},
  \citenamefont {Singh}, \citenamefont {Gegenwart},\ and\ \citenamefont
  {Hill}}]{Liu:2011_honey}%
  \BibitemOpen
  \bibfield  {author} {\bibinfo {author} {\bibfnamefont {X.}~\bibnamefont
  {Liu}}, \bibinfo {author} {\bibfnamefont {T.}~\bibnamefont {Berlijn}},
  \bibinfo {author} {\bibfnamefont {W.-G.}\ \bibnamefont {Yin}}, \bibinfo
  {author} {\bibfnamefont {W.}~\bibnamefont {Ku}}, \bibinfo {author}
  {\bibfnamefont {A.}~\bibnamefont {Tsvelik}}, \bibinfo {author} {\bibfnamefont
  {Y.-J.}\ \bibnamefont {Kim}}, \bibinfo {author} {\bibfnamefont
  {H.}~\bibnamefont {Gretarsson}}, \bibinfo {author} {\bibfnamefont
  {Y.}~\bibnamefont {Singh}}, \bibinfo {author} {\bibfnamefont
  {P.}~\bibnamefont {Gegenwart}}, \ and\ \bibinfo {author} {\bibfnamefont
  {J.~P.}\ \bibnamefont {Hill}},\ }\href@noop {} {\bibfield  {journal}
  {\bibinfo  {journal} {Phys. Rev. B}\ }\textbf {\bibinfo {volume} {83}},\
  \bibinfo {pages} {220403(R)} (\bibinfo {year} {2011})}\BibitemShut {NoStop}%
\bibitem [{\citenamefont {Singh}\ \emph {et~al.}(2012)\citenamefont {Singh},
  \citenamefont {Manni}, \citenamefont {Reuther}, \citenamefont {Berlijn},
  \citenamefont {Thomale}, \citenamefont {Ku}, \citenamefont {Trebst},\ and\
  \citenamefont {Gegenwart}}]{Singh:2012_honey}%
  \BibitemOpen
  \bibfield  {author} {\bibinfo {author} {\bibfnamefont {Y.}~\bibnamefont
  {Singh}}, \bibinfo {author} {\bibfnamefont {S.}~\bibnamefont {Manni}},
  \bibinfo {author} {\bibfnamefont {J.}~\bibnamefont {Reuther}}, \bibinfo
  {author} {\bibfnamefont {T.}~\bibnamefont {Berlijn}}, \bibinfo {author}
  {\bibfnamefont {R.}~\bibnamefont {Thomale}}, \bibinfo {author} {\bibfnamefont
  {W.}~\bibnamefont {Ku}}, \bibinfo {author} {\bibfnamefont {S.}~\bibnamefont
  {Trebst}}, \ and\ \bibinfo {author} {\bibfnamefont {P.}~\bibnamefont
  {Gegenwart}},\ }\href@noop {} {\bibfield  {journal} {\bibinfo  {journal}
  {Phys. Rev. Lett.}\ }\textbf {\bibinfo {volume} {108}},\ \bibinfo {pages}
  {127203} (\bibinfo {year} {2012})}\BibitemShut {NoStop}%
\bibitem [{\citenamefont {Choi}\ \emph {et~al.}(2012)\citenamefont {Choi},
  \citenamefont {Coldea}, \citenamefont {Kolmogorov}, \citenamefont
  {Lancaster}, \citenamefont {Mazin}, \citenamefont {Blundell}, \citenamefont
  {Radaelli}, \citenamefont {Singh}, \citenamefont {Gegenwart}, \citenamefont
  {Choi}, \citenamefont {Cheong}, \citenamefont {Baker}, \citenamefont
  {Stock},\ and\ \citenamefont {Taylor}}]{Choi:2012_honey}%
  \BibitemOpen
  \bibfield  {author} {\bibinfo {author} {\bibfnamefont {S.~K.}\ \bibnamefont
  {Choi}}, \bibinfo {author} {\bibfnamefont {R.}~\bibnamefont {Coldea}},
  \bibinfo {author} {\bibfnamefont {A.~N.}\ \bibnamefont {Kolmogorov}},
  \bibinfo {author} {\bibfnamefont {T.}~\bibnamefont {Lancaster}}, \bibinfo
  {author} {\bibfnamefont {I.~I.}\ \bibnamefont {Mazin}}, \bibinfo {author}
  {\bibfnamefont {S.~J.}\ \bibnamefont {Blundell}}, \bibinfo {author}
  {\bibfnamefont {P.~G.}\ \bibnamefont {Radaelli}}, \bibinfo {author}
  {\bibfnamefont {Y.}~\bibnamefont {Singh}}, \bibinfo {author} {\bibfnamefont
  {P.}~\bibnamefont {Gegenwart}}, \bibinfo {author} {\bibfnamefont {K.~R.}\
  \bibnamefont {Choi}}, \bibinfo {author} {\bibfnamefont {S.-W.}\ \bibnamefont
  {Cheong}}, \bibinfo {author} {\bibfnamefont {P.~J.}\ \bibnamefont {Baker}},
  \bibinfo {author} {\bibfnamefont {C.}~\bibnamefont {Stock}}, \ and\ \bibinfo
  {author} {\bibfnamefont {J.}~\bibnamefont {Taylor}},\ }\href@noop {}
  {\bibfield  {journal} {\bibinfo  {journal} {Phys. Rev. Lett.}\ }\textbf
  {\bibinfo {volume} {108}},\ \bibinfo {pages} {127204} (\bibinfo {year}
  {2012})}\BibitemShut {NoStop}%
\bibitem [{\citenamefont {Regnault}\ and\ \citenamefont
  {Rossat-Mignod}(1990)}]{Regnault:1990_honey}%
  \BibitemOpen
  \bibfield  {author} {\bibinfo {author} {\bibfnamefont {L.~P.}\ \bibnamefont
  {Regnault}}\ and\ \bibinfo {author} {\bibfnamefont {J.}~\bibnamefont
  {Rossat-Mignod}},\ }in\ \href@noop {} {\emph {\bibinfo {booktitle} {Magnetic
  {P}roperties of {L}ayered {T}ransition {M}etal {C}ompounds}}},\ \bibinfo
  {editor} {edited by\ \bibinfo {editor} {\bibfnamefont {L.~J.}\ \bibnamefont
  {{De Jongh}}}}\ (\bibinfo  {publisher} {Kluwer Academic Publishers},\
  \bibinfo {address} {Dordrecht},\ \bibinfo {year} {1990})\ pp.\ \bibinfo
  {pages} {271--321}\BibitemShut {NoStop}%
\bibitem [{\citenamefont {Roudebush}\ \emph {et~al.}(2013)\citenamefont
  {Roudebush}, \citenamefont {Andersen}, \citenamefont {Ramlau}, \citenamefont
  {Garlea}, \citenamefont {Toft-Petersen}, \citenamefont {Norby}, \citenamefont
  {Schneider}, \citenamefont {Hay},\ and\ \citenamefont
  {Cava}}]{Roudebush:2013_honey}%
  \BibitemOpen
  \bibfield  {author} {\bibinfo {author} {\bibfnamefont {J.~H.}\ \bibnamefont
  {Roudebush}}, \bibinfo {author} {\bibfnamefont {N.~H.}\ \bibnamefont
  {Andersen}}, \bibinfo {author} {\bibfnamefont {R.}~\bibnamefont {Ramlau}},
  \bibinfo {author} {\bibfnamefont {V.~O.}\ \bibnamefont {Garlea}}, \bibinfo
  {author} {\bibfnamefont {R.}~\bibnamefont {Toft-Petersen}}, \bibinfo {author}
  {\bibfnamefont {P.}~\bibnamefont {Norby}}, \bibinfo {author} {\bibfnamefont
  {R.}~\bibnamefont {Schneider}}, \bibinfo {author} {\bibfnamefont {J.~N.}\
  \bibnamefont {Hay}}, \ and\ \bibinfo {author} {\bibfnamefont {R.~J.}\
  \bibnamefont {Cava}},\ }\href@noop {} {\bibfield  {journal} {\bibinfo
  {journal} {Inorg. Chem.}\ }\textbf {\bibinfo {volume} {52}},\ \bibinfo
  {pages} {6083} (\bibinfo {year} {2013})}\BibitemShut {NoStop}%
\bibitem [{\citenamefont {Senthil}\ \emph
  {et~al.}(2004{\natexlab{a}})\citenamefont {Senthil}, \citenamefont
  {Vishwanath}, \citenamefont {Balents}, \citenamefont {Sachdev},\ and\
  \citenamefont {Fisher}}]{Senthil:2004_Science_deconfinedQC}%
  \BibitemOpen
  \bibfield  {author} {\bibinfo {author} {\bibfnamefont {T.}~\bibnamefont
  {Senthil}}, \bibinfo {author} {\bibfnamefont {A.}~\bibnamefont {Vishwanath}},
  \bibinfo {author} {\bibfnamefont {L.}~\bibnamefont {Balents}}, \bibinfo
  {author} {\bibfnamefont {S.}~\bibnamefont {Sachdev}}, \ and\ \bibinfo
  {author} {\bibfnamefont {M.~P.~A.}\ \bibnamefont {Fisher}},\ }\href@noop {}
  {\bibfield  {journal} {\bibinfo  {journal} {Science}\ }\textbf {\bibinfo
  {volume} {303}},\ \bibinfo {pages} {1490} (\bibinfo {year}
  {2004}{\natexlab{a}})}\BibitemShut {NoStop}%
\bibitem [{\citenamefont {Senthil}\ \emph
  {et~al.}(2004{\natexlab{b}})\citenamefont {Senthil}, \citenamefont {Balents},
  \citenamefont {Sachdev}, \citenamefont {Vishwanath},\ and\ \citenamefont
  {Fisher}}]{Senthil:2004_PRB_deconfinedQC}%
  \BibitemOpen
  \bibfield  {author} {\bibinfo {author} {\bibfnamefont {T.}~\bibnamefont
  {Senthil}}, \bibinfo {author} {\bibfnamefont {L.}~\bibnamefont {Balents}},
  \bibinfo {author} {\bibfnamefont {S.}~\bibnamefont {Sachdev}}, \bibinfo
  {author} {\bibfnamefont {A.}~\bibnamefont {Vishwanath}}, \ and\ \bibinfo
  {author} {\bibfnamefont {M.~P.~A.}\ \bibnamefont {Fisher}},\ }\href@noop {}
  {\bibfield  {journal} {\bibinfo  {journal} {Phys. Rev. B}\ }\textbf {\bibinfo
  {volume} {70}},\ \bibinfo {pages} {144407} (\bibinfo {year}
  {2004}{\natexlab{b}})}\BibitemShut {NoStop}%
\bibitem [{\citenamefont {K{\"{u}}mmel}\ \emph {et~al.}(1978)\citenamefont
  {K{\"{u}}mmel}, \citenamefont {L{\"{u}}hrmann},\ and\ \citenamefont
  {Zabolitzky}}]{Kummel:1978_ccm}%
  \BibitemOpen
  \bibfield  {author} {\bibinfo {author} {\bibfnamefont {H.}~\bibnamefont
  {K{\"{u}}mmel}}, \bibinfo {author} {\bibfnamefont {K.~H.}\ \bibnamefont
  {L{\"{u}}hrmann}}, \ and\ \bibinfo {author} {\bibfnamefont {J.~G.}\
  \bibnamefont {Zabolitzky}},\ }\href@noop {} {\bibfield  {journal} {\bibinfo
  {journal} {Phys Rep.}\ }\textbf {\bibinfo {volume} {36C}},\ \bibinfo {pages}
  {1} (\bibinfo {year} {1978})}\BibitemShut {NoStop}%
\bibitem [{\citenamefont {Bishop}\ and\ \citenamefont
  {L{\"{u}}hrmann}(1978)}]{Bishop:1978_ccm}%
  \BibitemOpen
  \bibfield  {author} {\bibinfo {author} {\bibfnamefont {R.~F.}\ \bibnamefont
  {Bishop}}\ and\ \bibinfo {author} {\bibfnamefont {K.~H.}\ \bibnamefont
  {L{\"{u}}hrmann}},\ }\href@noop {} {\bibfield  {journal} {\bibinfo  {journal}
  {Phys. Rev. B}\ }\textbf {\bibinfo {volume} {17}},\ \bibinfo {pages} {3757}
  (\bibinfo {year} {1978})}\BibitemShut {NoStop}%
\bibitem [{\citenamefont {Bishop}\ and\ \citenamefont
  {L{\"{u}}hrmann}(1982)}]{Bishop:1982_ccm}%
  \BibitemOpen
  \bibfield  {author} {\bibinfo {author} {\bibfnamefont {R.~F.}\ \bibnamefont
  {Bishop}}\ and\ \bibinfo {author} {\bibfnamefont {K.~H.}\ \bibnamefont
  {L{\"{u}}hrmann}},\ }\href@noop {} {\bibfield  {journal} {\bibinfo  {journal}
  {Phys. Rev. B}\ }\textbf {\bibinfo {volume} {26}},\ \bibinfo {pages} {5523}
  (\bibinfo {year} {1982})}\BibitemShut {NoStop}%
\bibitem [{\citenamefont {Arponen}(1983)}]{Arponen:1983_ccm}%
  \BibitemOpen
  \bibfield  {author} {\bibinfo {author} {\bibfnamefont {J.}~\bibnamefont
  {Arponen}},\ }\href@noop {} {\bibfield  {journal} {\bibinfo  {journal} {Ann.
  Phys. (N.Y.)}\ }\textbf {\bibinfo {volume} {151}},\ \bibinfo {pages} {311}
  (\bibinfo {year} {1983})}\BibitemShut {NoStop}%
\bibitem [{\citenamefont {Bishop}\ and\ \citenamefont
  {K{\"{u}}mmel}(1987)}]{Bishop:1987_ccm}%
  \BibitemOpen
  \bibfield  {author} {\bibinfo {author} {\bibfnamefont {R.~F.}\ \bibnamefont
  {Bishop}}\ and\ \bibinfo {author} {\bibfnamefont {H.~G.}\ \bibnamefont
  {K{\"{u}}mmel}},\ }\href@noop {} {\bibfield  {journal} {\bibinfo  {journal}
  {Phys. Today}\ }\textbf {\bibinfo {volume} {40(3)}},\ \bibinfo {pages} {52}
  (\bibinfo {year} {1987})}\BibitemShut {NoStop}%
\bibitem [{\citenamefont {Bartlett}(1989)}]{Bartlett:1989_ccm}%
  \BibitemOpen
  \bibfield  {author} {\bibinfo {author} {\bibfnamefont {R.~J.}\ \bibnamefont
  {Bartlett}},\ }\href@noop {} {\bibfield  {journal} {\bibinfo  {journal} {J.
  Phys. Chem.}\ }\textbf {\bibinfo {volume} {93}},\ \bibinfo {pages} {1697}
  (\bibinfo {year} {1989})}\BibitemShut {NoStop}%
\bibitem [{\citenamefont {Arponen}\ and\ \citenamefont
  {Bishop}(1991)}]{Arponen:1991_ccm}%
  \BibitemOpen
  \bibfield  {author} {\bibinfo {author} {\bibfnamefont {J.~S.}\ \bibnamefont
  {Arponen}}\ and\ \bibinfo {author} {\bibfnamefont {R.~F.}\ \bibnamefont
  {Bishop}},\ }\href@noop {} {\bibfield  {journal} {\bibinfo  {journal} {Ann.
  Phys. (N.Y.)}\ }\textbf {\bibinfo {volume} {207}},\ \bibinfo {pages} {171}
  (\bibinfo {year} {1991})}\BibitemShut {NoStop}%
\bibitem [{\citenamefont {Bishop}(1991)}]{Bishop:1991_TheorChimActa_QMBT}%
  \BibitemOpen
  \bibfield  {author} {\bibinfo {author} {\bibfnamefont {R.~F.}\ \bibnamefont
  {Bishop}},\ }\href@noop {} {\bibfield  {journal} {\bibinfo  {journal} {Theor.
  Chim. Acta}\ }\textbf {\bibinfo {volume} {80}},\ \bibinfo {pages} {95}
  (\bibinfo {year} {1991})}\BibitemShut {NoStop}%
\bibitem [{\citenamefont {Bishop}(1998)}]{Bishop:1998_QMBT_coll}%
  \BibitemOpen
  \bibfield  {author} {\bibinfo {author} {\bibfnamefont {R.~F.}\ \bibnamefont
  {Bishop}},\ }in\ \href@noop {} {\emph {\bibinfo {booktitle} {Microscopic
  Quantum Many-Body Theories and Their Applications}}},\ \bibinfo {series and
  number} {Lecture Notes in Physics Vol. 510},\ \bibinfo {editor} {edited by\
  \bibinfo {editor} {\bibfnamefont {J.}~\bibnamefont {Navarro}}\ and\ \bibinfo
  {editor} {\bibfnamefont {A.}~\bibnamefont {Polls}}}\ (\bibinfo  {publisher}
  {Springer-Verlag},\ \bibinfo {address} {Berlin},\ \bibinfo {year} {1998})\
  p.~\bibinfo {pages} {1}\BibitemShut {NoStop}%
\bibitem [{\citenamefont {Zeng}\ \emph {et~al.}(1998)\citenamefont {Zeng},
  \citenamefont {Farnell},\ and\ \citenamefont
  {Bishop}}]{Zeng:1998_SqLatt_TrianLatt}%
  \BibitemOpen
  \bibfield  {author} {\bibinfo {author} {\bibfnamefont {C.}~\bibnamefont
  {Zeng}}, \bibinfo {author} {\bibfnamefont {D.~J.~J.}\ \bibnamefont
  {Farnell}}, \ and\ \bibinfo {author} {\bibfnamefont {R.~F.}\ \bibnamefont
  {Bishop}},\ }\href@noop {} {\bibfield  {journal} {\bibinfo  {journal} {J.
  Stat. Phys.}\ }\textbf {\bibinfo {volume} {90}},\ \bibinfo {pages} {327}
  (\bibinfo {year} {1998})}\BibitemShut {NoStop}%
\bibitem [{\citenamefont {Farnell}\ and\ \citenamefont
  {Bishop}(2004)}]{Fa:2004_QM-coll}%
  \BibitemOpen
  \bibfield  {author} {\bibinfo {author} {\bibfnamefont {D.~J.~J.}\
  \bibnamefont {Farnell}}\ and\ \bibinfo {author} {\bibfnamefont {R.~F.}\
  \bibnamefont {Bishop}},\ }in\ \href@noop {} {\emph {\bibinfo {booktitle}
  {Quantum Magnetism}}},\ \bibinfo {series and number} {Lecture Notes in
  Physics Vol. 645},\ \bibinfo {editor} {edited by\ \bibinfo {editor}
  {\bibfnamefont {U.}~\bibnamefont {Schollw{\"{o}}ck}}, \bibinfo {editor}
  {\bibfnamefont {J.}~\bibnamefont {Richter}}, \bibinfo {editor} {\bibfnamefont
  {D.~J.~J.}\ \bibnamefont {Farnell}}, \ and\ \bibinfo {editor} {\bibfnamefont
  {R.~F.}\ \bibnamefont {Bishop}}}\ (\bibinfo  {publisher} {Springer-Verlag},\
  \bibinfo {address} {Berlin},\ \bibinfo {year} {2004})\ p.\ \bibinfo {pages}
  {307}\BibitemShut {NoStop}%
\bibitem [{ccm()}]{ccm_code}%
  \BibitemOpen
  \href@noop {} {}\bibinfo {note} {We use the program package CCCM of D.~J.~J.
  Farnell and J.~Schulenburg, see
  http://www-e.uni-magdeburg.de/jschulen/ccm/index.html}\BibitemShut {NoStop}%
\bibitem [{\citenamefont {Mila}(2000)}]{Mila:2000_M-Xcpty_spinGap}%
  \BibitemOpen
  \bibfield  {author} {\bibinfo {author} {\bibfnamefont {F.}~\bibnamefont
  {Mila}},\ }\href@noop {} {\bibfield  {journal} {\bibinfo  {journal} {Eur. J.
  Phys.}\ }\textbf {\bibinfo {volume} {21}},\ \bibinfo {pages} {499} (\bibinfo
  {year} {2000})}\BibitemShut {NoStop}%
\bibitem [{\citenamefont {Bernu}\ and\ \citenamefont
  {Lhuillier}(2015)}]{Bernu:2015_M-Xcpty_spinGap}%
  \BibitemOpen
  \bibfield  {author} {\bibinfo {author} {\bibfnamefont {B.}~\bibnamefont
  {Bernu}}\ and\ \bibinfo {author} {\bibfnamefont {C.}~\bibnamefont
  {Lhuillier}},\ }\href@noop {} {\bibfield  {journal} {\bibinfo  {journal}
  {Phys. Rev. Lett.}\ }\textbf {\bibinfo {volume} {114}},\ \bibinfo {pages}
  {057201} (\bibinfo {year} {2015})}\BibitemShut {NoStop}%
\bibitem [{\citenamefont {Bishop}\ \emph {et~al.}(2000)\citenamefont {Bishop},
  \citenamefont {Farnell}, \citenamefont {Kr{\"{u}}ger}, \citenamefont
  {Parkinson}, \citenamefont {Richter},\ and\ \citenamefont
  {Zeng}}]{Bishop:2000_XXZ}%
  \BibitemOpen
  \bibfield  {author} {\bibinfo {author} {\bibfnamefont {R.~F.}\ \bibnamefont
  {Bishop}}, \bibinfo {author} {\bibfnamefont {D.~J.~J.}\ \bibnamefont
  {Farnell}}, \bibinfo {author} {\bibfnamefont {S.~E.}\ \bibnamefont
  {Kr{\"{u}}ger}}, \bibinfo {author} {\bibfnamefont {J.~B.}\ \bibnamefont
  {Parkinson}}, \bibinfo {author} {\bibfnamefont {J.}~\bibnamefont {Richter}},
  \ and\ \bibinfo {author} {\bibfnamefont {C.}~\bibnamefont {Zeng}},\
  }\href@noop {} {\bibfield  {journal} {\bibinfo  {journal} {J. Phys.: Condens.
  Matter}\ }\textbf {\bibinfo {volume} {12}},\ \bibinfo {pages} {6887}
  (\bibinfo {year} {2000})}\BibitemShut {NoStop}%
\bibitem [{\citenamefont {Kr{\"{u}}ger}\ \emph {et~al.}(2000)\citenamefont
  {Kr{\"{u}}ger}, \citenamefont {Richter}, \citenamefont {Schulenburg},
  \citenamefont {Farnell},\ and\ \citenamefont {Bishop}}]{Kruger:2000_JJprime}%
  \BibitemOpen
  \bibfield  {author} {\bibinfo {author} {\bibfnamefont {S.~E.}\ \bibnamefont
  {Kr{\"{u}}ger}}, \bibinfo {author} {\bibfnamefont {J.}~\bibnamefont
  {Richter}}, \bibinfo {author} {\bibfnamefont {J.}~\bibnamefont
  {Schulenburg}}, \bibinfo {author} {\bibfnamefont {D.~J.~J.}\ \bibnamefont
  {Farnell}}, \ and\ \bibinfo {author} {\bibfnamefont {R.~F.}\ \bibnamefont
  {Bishop}},\ }\href@noop {} {\bibfield  {journal} {\bibinfo  {journal} {Phys.
  Rev. B}\ }\textbf {\bibinfo {volume} {61}},\ \bibinfo {pages} {14607}
  (\bibinfo {year} {2000})}\BibitemShut {NoStop}%
\bibitem [{\citenamefont {Farnell}\ \emph {et~al.}(2001)\citenamefont
  {Farnell}, \citenamefont {Gernoth},\ and\ \citenamefont
  {Bishop}}]{Fa:2001_SqLatt_s1}%
  \BibitemOpen
  \bibfield  {author} {\bibinfo {author} {\bibfnamefont {D.~J.~J.}\
  \bibnamefont {Farnell}}, \bibinfo {author} {\bibfnamefont {K.~A.}\
  \bibnamefont {Gernoth}}, \ and\ \bibinfo {author} {\bibfnamefont {R.~F.}\
  \bibnamefont {Bishop}},\ }\href@noop {} {\bibfield  {journal} {\bibinfo
  {journal} {Phys. Rev. B}\ }\textbf {\bibinfo {volume} {64}},\ \bibinfo
  {pages} {172409} (\bibinfo {year} {2001})}\BibitemShut {NoStop}%
\bibitem [{\citenamefont {Darradi}\ \emph {et~al.}(2005)\citenamefont
  {Darradi}, \citenamefont {Richter},\ and\ \citenamefont
  {Farnell}}]{Darradi:2005_Shastry-Sutherland}%
  \BibitemOpen
  \bibfield  {author} {\bibinfo {author} {\bibfnamefont {R.}~\bibnamefont
  {Darradi}}, \bibinfo {author} {\bibfnamefont {J.}~\bibnamefont {Richter}}, \
  and\ \bibinfo {author} {\bibfnamefont {D.~J.~J.}\ \bibnamefont {Farnell}},\
  }\href@noop {} {\bibfield  {journal} {\bibinfo  {journal} {Phys. Rev. B}\
  }\textbf {\bibinfo {volume} {72}},\ \bibinfo {pages} {104425} (\bibinfo
  {year} {2005})}\BibitemShut {NoStop}%
\bibitem [{\citenamefont {Darradi}\ \emph {et~al.}(2008)\citenamefont
  {Darradi}, \citenamefont {Derzhko}, \citenamefont {Zinke}, \citenamefont
  {Schulenburg}, \citenamefont {Kr{\"{u}}ger},\ and\ \citenamefont
  {Richter}}]{Darradi:2008_J1J2mod}%
  \BibitemOpen
  \bibfield  {author} {\bibinfo {author} {\bibfnamefont {R.}~\bibnamefont
  {Darradi}}, \bibinfo {author} {\bibfnamefont {O.}~\bibnamefont {Derzhko}},
  \bibinfo {author} {\bibfnamefont {R.}~\bibnamefont {Zinke}}, \bibinfo
  {author} {\bibfnamefont {J.}~\bibnamefont {Schulenburg}}, \bibinfo {author}
  {\bibfnamefont {S.~E.}\ \bibnamefont {Kr{\"{u}}ger}}, \ and\ \bibinfo
  {author} {\bibfnamefont {J.}~\bibnamefont {Richter}},\ }\href@noop {}
  {\bibfield  {journal} {\bibinfo  {journal} {Phys. Rev. B}\ }\textbf {\bibinfo
  {volume} {78}},\ \bibinfo {pages} {214415} (\bibinfo {year}
  {2008})}\BibitemShut {NoStop}%
\bibitem [{\citenamefont {Bishop}\ \emph
  {et~al.}(2008{\natexlab{a}})\citenamefont {Bishop}, \citenamefont {Li},
  \citenamefont {Darradi},\ and\ \citenamefont
  {Richter}}]{Bi:2008_EPL_J1J1primeJ2_s1}%
  \BibitemOpen
  \bibfield  {author} {\bibinfo {author} {\bibfnamefont {R.~F.}\ \bibnamefont
  {Bishop}}, \bibinfo {author} {\bibfnamefont {P.~H.~Y.}\ \bibnamefont {Li}},
  \bibinfo {author} {\bibfnamefont {R.}~\bibnamefont {Darradi}}, \ and\
  \bibinfo {author} {\bibfnamefont {J.}~\bibnamefont {Richter}},\ }\href@noop
  {} {\bibfield  {journal} {\bibinfo  {journal} {EPL}\ }\textbf {\bibinfo
  {volume} {83}},\ \bibinfo {pages} {47004} (\bibinfo {year}
  {2008}{\natexlab{a}})}\BibitemShut {NoStop}%
\bibitem [{\citenamefont {Bishop}\ \emph
  {et~al.}(2008{\natexlab{b}})\citenamefont {Bishop}, \citenamefont {Li},
  \citenamefont {Darradi}, \citenamefont {Richter},\ and\ \citenamefont
  {Campbell}}]{Bi:2008_JPCM_J1xxzJ2xxz_s1}%
  \BibitemOpen
  \bibfield  {author} {\bibinfo {author} {\bibfnamefont {R.~F.}\ \bibnamefont
  {Bishop}}, \bibinfo {author} {\bibfnamefont {P.~H.~Y.}\ \bibnamefont {Li}},
  \bibinfo {author} {\bibfnamefont {R.}~\bibnamefont {Darradi}}, \bibinfo
  {author} {\bibfnamefont {J.}~\bibnamefont {Richter}}, \ and\ \bibinfo
  {author} {\bibfnamefont {C.~E.}\ \bibnamefont {Campbell}},\ }\href@noop {}
  {\bibfield  {journal} {\bibinfo  {journal} {J. Phys.: Condens. Matter}\
  }\textbf {\bibinfo {volume} {20}},\ \bibinfo {pages} {415213} (\bibinfo
  {year} {2008}{\natexlab{b}})}\BibitemShut {NoStop}%
\bibitem [{\citenamefont {Bishop}\ \emph {et~al.}(2009)\citenamefont {Bishop},
  \citenamefont {Li}, \citenamefont {Farnell},\ and\ \citenamefont
  {Campbell}}]{Bi:2009_SqTriangle}%
  \BibitemOpen
  \bibfield  {author} {\bibinfo {author} {\bibfnamefont {R.~F.}\ \bibnamefont
  {Bishop}}, \bibinfo {author} {\bibfnamefont {P.~H.~Y.}\ \bibnamefont {Li}},
  \bibinfo {author} {\bibfnamefont {D.~J.~J.}\ \bibnamefont {Farnell}}, \ and\
  \bibinfo {author} {\bibfnamefont {C.~E.}\ \bibnamefont {Campbell}},\
  }\href@noop {} {\bibfield  {journal} {\bibinfo  {journal} {Phys. Rev. B}\
  }\textbf {\bibinfo {volume} {79}},\ \bibinfo {pages} {174405} (\bibinfo
  {year} {2009})}\BibitemShut {NoStop}%
\bibitem [{\citenamefont {Bishop}\ \emph
  {et~al.}(2010{\natexlab{a}})\citenamefont {Bishop}, \citenamefont {Li},
  \citenamefont {Farnell},\ and\ \citenamefont {Campbell}}]{Bishop:2010_UJack}%
  \BibitemOpen
  \bibfield  {author} {\bibinfo {author} {\bibfnamefont {R.~F.}\ \bibnamefont
  {Bishop}}, \bibinfo {author} {\bibfnamefont {P.~H.~Y.}\ \bibnamefont {Li}},
  \bibinfo {author} {\bibfnamefont {D.~J.~J.}\ \bibnamefont {Farnell}}, \ and\
  \bibinfo {author} {\bibfnamefont {C.~E.}\ \bibnamefont {Campbell}},\
  }\href@noop {} {\bibfield  {journal} {\bibinfo  {journal} {Phys. Rev. B}\
  }\textbf {\bibinfo {volume} {82}},\ \bibinfo {pages} {024416} (\bibinfo
  {year} {2010}{\natexlab{a}})}\BibitemShut {NoStop}%
\bibitem [{\citenamefont {Bishop}\ \emph
  {et~al.}(2010{\natexlab{b}})\citenamefont {Bishop}, \citenamefont {Li},
  \citenamefont {Farnell},\ and\ \citenamefont
  {Campbell}}]{Bishop:2010_KagomeSq}%
  \BibitemOpen
  \bibfield  {author} {\bibinfo {author} {\bibfnamefont {R.~F.}\ \bibnamefont
  {Bishop}}, \bibinfo {author} {\bibfnamefont {P.~H.~Y.}\ \bibnamefont {Li}},
  \bibinfo {author} {\bibfnamefont {D.~J.~J.}\ \bibnamefont {Farnell}}, \ and\
  \bibinfo {author} {\bibfnamefont {C.~E.}\ \bibnamefont {Campbell}},\
  }\href@noop {} {\bibfield  {journal} {\bibinfo  {journal} {Phys. Rev. B}\
  }\textbf {\bibinfo {volume} {82}},\ \bibinfo {pages} {104406} (\bibinfo
  {year} {2010}{\natexlab{b}})}\BibitemShut {NoStop}%
\bibitem [{\citenamefont {Bishop}\ and\ \citenamefont
  {Li}(2011)}]{Bishop:2011_UJack_GrtSpins}%
  \BibitemOpen
  \bibfield  {author} {\bibinfo {author} {\bibfnamefont {R.~F.}\ \bibnamefont
  {Bishop}}\ and\ \bibinfo {author} {\bibfnamefont {P.~H.~Y.}\ \bibnamefont
  {Li}},\ }\href@noop {} {\bibfield  {journal} {\bibinfo  {journal} {Eur. Phys.
  J. B}\ }\textbf {\bibinfo {volume} {81}},\ \bibinfo {pages} {37} (\bibinfo
  {year} {2011})}\BibitemShut {NoStop}%
\bibitem [{\citenamefont {Li}\ and\ \citenamefont
  {Bishop}(2012)}]{PHYLi:2012_SqTriangle_grtSpins}%
  \BibitemOpen
  \bibfield  {author} {\bibinfo {author} {\bibfnamefont {P.~H.~Y.}\
  \bibnamefont {Li}}\ and\ \bibinfo {author} {\bibfnamefont {R.~F.}\
  \bibnamefont {Bishop}},\ }\href@noop {} {\bibfield  {journal} {\bibinfo
  {journal} {Eur. Phys. J. B}\ }\textbf {\bibinfo {volume} {85}},\ \bibinfo
  {pages} {25} (\bibinfo {year} {2012})}\BibitemShut {NoStop}%
\bibitem [{\citenamefont {Li}\ \emph {et~al.}(2012{\natexlab{c}})\citenamefont
  {Li}, \citenamefont {Bishop}, \citenamefont {Campbell}, \citenamefont
  {Farnell}, \citenamefont {G{\"{o}}tze},\ and\ \citenamefont
  {Richter}}]{Li:2012_anisotropic_kagomeSq}%
  \BibitemOpen
  \bibfield  {author} {\bibinfo {author} {\bibfnamefont {P.~H.~Y.}\
  \bibnamefont {Li}}, \bibinfo {author} {\bibfnamefont {R.~F.}\ \bibnamefont
  {Bishop}}, \bibinfo {author} {\bibfnamefont {C.~E.}\ \bibnamefont
  {Campbell}}, \bibinfo {author} {\bibfnamefont {D.~J.~J.}\ \bibnamefont
  {Farnell}}, \bibinfo {author} {\bibfnamefont {O.}~\bibnamefont
  {G{\"{o}}tze}}, \ and\ \bibinfo {author} {\bibfnamefont {J.}~\bibnamefont
  {Richter}},\ }\href@noop {} {\bibfield  {journal} {\bibinfo  {journal}
  {Phys.\ Rev.\ B}\ }\textbf {\bibinfo {volume} {86}},\ \bibinfo {pages}
  {214403} (\bibinfo {year} {2012}{\natexlab{c}})}\BibitemShut {NoStop}%
\bibitem [{\citenamefont {Richter}\ \emph {et~al.}(2015)\citenamefont
  {Richter}, \citenamefont {Zinke},\ and\ \citenamefont
  {Farnell}}]{Richter:2015_ccm_J1J2sq_spinGap}%
  \BibitemOpen
  \bibfield  {author} {\bibinfo {author} {\bibfnamefont {J.}~\bibnamefont
  {Richter}}, \bibinfo {author} {\bibfnamefont {R.}~\bibnamefont {Zinke}}, \
  and\ \bibinfo {author} {\bibfnamefont {D.~J.~J.}\ \bibnamefont {Farnell}},\
  }\href@noop {} {\bibfield  {journal} {\bibinfo  {journal} {Eur. Phys. J. B}\
  }\textbf {\bibinfo {volume} {88}},\ \bibinfo {pages} {2} (\bibinfo {year}
  {2015})}\BibitemShut {NoStop}%
\bibitem [{\citenamefont {Bishop}\ \emph {et~al.}(2015)\citenamefont {Bishop},
  \citenamefont {Li}, \citenamefont {G{\"{o}}tze}, \citenamefont {Richter},\
  and\ \citenamefont {Campbell}}]{Bishop:2015_honey_low-E-param}%
  \BibitemOpen
  \bibfield  {author} {\bibinfo {author} {\bibfnamefont {R.~F.}\ \bibnamefont
  {Bishop}}, \bibinfo {author} {\bibfnamefont {P.~H.~Y.}\ \bibnamefont {Li}},
  \bibinfo {author} {\bibfnamefont {O.}~\bibnamefont {G{\"{o}}tze}}, \bibinfo
  {author} {\bibfnamefont {J.}~\bibnamefont {Richter}}, \ and\ \bibinfo
  {author} {\bibfnamefont {C.~E.}\ \bibnamefont {Campbell}},\ }\href@noop {}
  {\bibfield  {journal} {\bibinfo  {journal} {Phys. Rev. B}\ }\textbf {\bibinfo
  {volume} {92}},\ \bibinfo {pages} {224434} (\bibinfo {year}
  {2015})}\BibitemShut {NoStop}%
\bibitem [{\citenamefont {Bishop}\ and\ \citenamefont
  {Li}(2015)}]{Bishop:2015_J1J2-triang_spinGap}%
  \BibitemOpen
  \bibfield  {author} {\bibinfo {author} {\bibfnamefont {R.~F.}\ \bibnamefont
  {Bishop}}\ and\ \bibinfo {author} {\bibfnamefont {P.~H.~Y.}\ \bibnamefont
  {Li}},\ }\href@noop {} {\bibfield  {journal} {\bibinfo  {journal} {EPL}\
  }\textbf {\bibinfo {volume} {112}},\ \bibinfo {pages} {67002} (\bibinfo
  {year} {2015})}\BibitemShut {NoStop}%
\bibitem [{\citenamefont {Kr{\"{u}}ger}\ \emph {et~al.}(2006)\citenamefont
  {Kr{\"{u}}ger}, \citenamefont {Darradi}, \citenamefont {Richter},\ and\
  \citenamefont {Farnell}}]{SEKruger:2006_spinStiff}%
  \BibitemOpen
  \bibfield  {author} {\bibinfo {author} {\bibfnamefont {S.~E.}\ \bibnamefont
  {Kr{\"{u}}ger}}, \bibinfo {author} {\bibfnamefont {R.}~\bibnamefont
  {Darradi}}, \bibinfo {author} {\bibfnamefont {J.}~\bibnamefont {Richter}}, \
  and\ \bibinfo {author} {\bibfnamefont {D.~J.~J.}\ \bibnamefont {Farnell}},\
  }\href@noop {} {\bibfield  {journal} {\bibinfo  {journal} {Phys. Rev. B}\
  }\textbf {\bibinfo {volume} {73}},\ \bibinfo {pages} {094404} (\bibinfo
  {year} {2006})}\BibitemShut {NoStop}%
\bibitem [{\citenamefont {G{\"{o}}tze}\ \emph {et~al.}(2016)\citenamefont
  {G{\"{o}}tze}, \citenamefont {Richter}, \citenamefont {Zinke},\ and\
  \citenamefont {Farnell}}]{Gotze:2016_triang}%
  \BibitemOpen
  \bibfield  {author} {\bibinfo {author} {\bibfnamefont {O.}~\bibnamefont
  {G{\"{o}}tze}}, \bibinfo {author} {\bibfnamefont {J.}~\bibnamefont
  {Richter}}, \bibinfo {author} {\bibfnamefont {R.}~\bibnamefont {Zinke}}, \
  and\ \bibinfo {author} {\bibfnamefont {D.~J.~J.}\ \bibnamefont {Farnell}},\
  }\href@noop {} {\bibfield  {journal} {\bibinfo  {journal} {J. Magn. Magn.
  Mater.}\ }\textbf {\bibinfo {volume} {397}},\ \bibinfo {pages} {333}
  (\bibinfo {year} {2016})}\BibitemShut {NoStop}%
\bibitem [{\citenamefont {Bishop}\ and\ \citenamefont
  {Li}(2016)}]{Bishop:2016_honey_grtSpins}%
  \BibitemOpen
  \bibfield  {author} {\bibinfo {author} {\bibfnamefont {R.~F.}\ \bibnamefont
  {Bishop}}\ and\ \bibinfo {author} {\bibfnamefont {P.~H.~Y.}\ \bibnamefont
  {Li}},\ }\href@noop {} {\bibfield  {journal} {\bibinfo  {journal} {J. Magn.
  Magn. Mater.}\ }\textbf {\bibinfo {volume} {407}},\ \bibinfo {pages} {348}
  (\bibinfo {year} {2016})}\BibitemShut {NoStop}%
\bibitem [{\citenamefont {Farnell}\ \emph {et~al.}(2009)\citenamefont
  {Farnell}, \citenamefont {Zinke}, \citenamefont {Schulenburg},\ and\
  \citenamefont {Richter}}]{Farnell:2009_Xcpty_ExtMagField}%
  \BibitemOpen
  \bibfield  {author} {\bibinfo {author} {\bibfnamefont {D.~J.~J.}\
  \bibnamefont {Farnell}}, \bibinfo {author} {\bibfnamefont {R.}~\bibnamefont
  {Zinke}}, \bibinfo {author} {\bibfnamefont {J.}~\bibnamefont {Schulenburg}},
  \ and\ \bibinfo {author} {\bibfnamefont {J.}~\bibnamefont {Richter}},\
  }\href@noop {} {\bibfield  {journal} {\bibinfo  {journal} {J. Phys.: Condens.
  Matter}\ }\textbf {\bibinfo {volume} {21}},\ \bibinfo {pages} {406002}
  (\bibinfo {year} {2009})}\BibitemShut {NoStop}%
\bibitem [{\citenamefont {Li}\ \emph {et~al.}(2015)\citenamefont {Li},
  \citenamefont {Bishop},\ and\ \citenamefont
  {Campbell}}]{Li:2015_j1j2-triang}%
  \BibitemOpen
  \bibfield  {author} {\bibinfo {author} {\bibfnamefont {P.~H.~Y.}\
  \bibnamefont {Li}}, \bibinfo {author} {\bibfnamefont {R.~F.}\ \bibnamefont
  {Bishop}}, \ and\ \bibinfo {author} {\bibfnamefont {C.~E.}\ \bibnamefont
  {Campbell}},\ }\href@noop {} {\bibfield  {journal} {\bibinfo  {journal}
  {Phys. Rev. B}\ }\textbf {\bibinfo {volume} {91}},\ \bibinfo {pages} {014426}
  (\bibinfo {year} {2015})}\BibitemShut {NoStop}%
\bibitem [{\citenamefont {Li}\ \emph {et~al.}(2013)\citenamefont {Li},
  \citenamefont {Bishop},\ and\ \citenamefont {Campbell}}]{Li:2013_chevron}%
  \BibitemOpen
  \bibfield  {author} {\bibinfo {author} {\bibfnamefont {P.~H.~Y.}\
  \bibnamefont {Li}}, \bibinfo {author} {\bibfnamefont {R.~F.}\ \bibnamefont
  {Bishop}}, \ and\ \bibinfo {author} {\bibfnamefont {C.~E.}\ \bibnamefont
  {Campbell}},\ }\href@noop {} {\bibfield  {journal} {\bibinfo  {journal}
  {Phys. Rev. B}\ }\textbf {\bibinfo {volume} {88}},\ \bibinfo {pages} {144423}
  (\bibinfo {year} {2013})}\BibitemShut {NoStop}%
\bibitem [{\citenamefont {Bishop}\ \emph
  {et~al.}(2013{\natexlab{b}})\citenamefont {Bishop}, \citenamefont {Li},\ and\
  \citenamefont {Campbell}}]{Bishop:2013_crossStripe}%
  \BibitemOpen
  \bibfield  {author} {\bibinfo {author} {\bibfnamefont {R.~F.}\ \bibnamefont
  {Bishop}}, \bibinfo {author} {\bibfnamefont {P.~H.~Y.}\ \bibnamefont {Li}}, \
  and\ \bibinfo {author} {\bibfnamefont {C.~E.}\ \bibnamefont {Campbell}},\
  }\href@noop {} {\bibfield  {journal} {\bibinfo  {journal} {Phys. Rev. B}\
  }\textbf {\bibinfo {volume} {88}},\ \bibinfo {pages} {214418} (\bibinfo
  {year} {2013}{\natexlab{b}})}\BibitemShut {NoStop}%
\bibitem [{\citenamefont {Bishop}\ \emph
  {et~al.}(2012{\natexlab{b}})\citenamefont {Bishop}, \citenamefont {Li},
  \citenamefont {Farnell}, \citenamefont {Richter},\ and\ \citenamefont
  {Campbell}}]{Bishop:2012_checkerboard}%
  \BibitemOpen
  \bibfield  {author} {\bibinfo {author} {\bibfnamefont {R.~F.}\ \bibnamefont
  {Bishop}}, \bibinfo {author} {\bibfnamefont {P.~H.~Y.}\ \bibnamefont {Li}},
  \bibinfo {author} {\bibfnamefont {D.~J.~J.}\ \bibnamefont {Farnell}},
  \bibinfo {author} {\bibfnamefont {J.}~\bibnamefont {Richter}}, \ and\
  \bibinfo {author} {\bibfnamefont {C.~E.}\ \bibnamefont {Campbell}},\
  }\href@noop {} {\bibfield  {journal} {\bibinfo  {journal} {Phys. Rev. B}\
  }\textbf {\bibinfo {volume} {85}},\ \bibinfo {pages} {205122} (\bibinfo
  {year} {2012}{\natexlab{b}})}\BibitemShut {NoStop}%
\end{thebibliography}%

\end{document}